\documentclass[aps,prx,showpacs,twocolumn,reprint,superscriptaddress]{revtex4-2}
\usepackage[dvips]{graphicx}
\usepackage{amsmath,amssymb,amsthm,mathrsfs,amsfonts,dsfont,mathtools}
\usepackage{epsfig}
\usepackage{braket}
\usepackage{hyperref}
\usepackage{bm}
\usepackage{enumerate}
\usepackage{color}
\usepackage{graphicx}
\usepackage{pgf}
\usepackage{pgfplots}
\pgfplotsset{compat=1.18}
\usepackage{tikz}
\usepackage{enumitem}
\usepackage[capitalize]{cleveref}
\usepackage[normalem]{ulem}
\usepackage{yquant, quantikz}
\useyquantlanguage{groups}
\usetikzlibrary{calc}

\usepackage{amsthm}

\newtheorem{remark}{Remark}

\newtheorem{statement}{Statement}
\newtheorem{theorem}{Theorem}
\newtheorem{lemma}{Lemma}

\Crefname{theorem}{Theorem}{Theorems}
\theoremstyle{remark}

\usepackage{xcolor}

\begin{document}

\title{TE-PAI: Exact Time Evolution by Sampling Random Circuits}
\author{Chusei Kiumi}
\email{c.kiumi.qiqb@osaka-u.ac.jp}
\affiliation{Mathematical Institute, University of Oxford, Woodstock Road, Oxford OX2 6GG, United Kingdom}
\affiliation{Center for Quantum Information and Quantum Biology, Osaka University, 1-2 Machikaneyama, Toyonaka 560-0043, Japan}
\author{B\'alint Koczor}
\email{balint.koczor@maths.ox.ac.uk}
\affiliation{Mathematical Institute, University of Oxford, Woodstock Road, Oxford OX2 6GG, United Kingdom}

\begin{abstract}
	Simulating time evolution under quantum Hamiltonians is one of the most natural applications of quantum computers. We introduce TE-PAI, which simulates time evolution exactly by sampling random quantum circuits for the purpose of estimating observable expectation values at the cost of an increased circuit repetition. The approach builds on the Probabilistic Angle Interpolation (PAI) technique and we prove that it simulates time evolution without discretisation or algorithmic error while achieving shallow circuit depths with optimal scaling that saturates the Lieb-Robinson bound. Another significant advantage of TE-PAI is that it only requires executing random circuits that consist of Pauli rotation gates of only two kinds of rotation angles $\pm\Delta$ and $\pi$, along with measurements. While TE-PAI is highly beneficial for NISQ devices, we additionally develop an optimised early fault-tolerant implementation using catalyst circuits and repeat-until-success teleportation, concluding that the approach requires orders of magnitude fewer T-states than conventional techniques, such as Trotterization -- we estimate $3 \times 10^{5}$ T states are sufficient for the fault-tolerant simulation of a $100$-qubit Heisenberg spin Hamiltonian. Furthermore, TE-PAI allows for a highly configurable trade-off between circuit depth and measurement overhead by adjusting the rotation angle $\Delta$ arbitrarily. We expect that the approach will be a major enabler in the late NISQ and early fault-tolerant periods as it can compensate circuit-depth and qubit-number limitations through an increased circuit repetition.
\end{abstract}

\maketitle

\section{Introduction}
Accurately modelling the time evolution of quantum systems is an important task but
presents a significant challenge in classical computing.
Thus, simulating quantum dynamics is regarded as one of the most promising applications
of quantum computers \cite{Feynman,Lloyd} and may provide an exponential speedup
over classical computers.
The simplest such approach, the Trotter-Suzuki decomposition \cite{Suzuki1,Suzuki2},
approximates the time evolution through a relatively simple circuit that contains evolutions under
the individual Hamiltonian terms. A drawback of the approach is that circuits may need to be quite deep
to sufficiently suppress approximation errors.
This discretisation error, also called the Trotter error, is inevitable with finite circuit
depth and can be particularly daunting in, e.g., quantum chemistry applications~\cite{chemistry1,chemistry2,chemistry3,chemistry4}.
These issues are further exacerbated when the aim is to simulate dynamics under
time-dependent Hamiltonians, as we demonstrate below.
Indeed, sophisticated quantum algorithms, such as linear combination of unitaries (LCU)
\cite{LCU1,LCU2,LCU3}, quantum signal processing \cite{QSP1,QSP2} or quantum walks \cite{qw1,qw2},
can achieve a fundamentally improved circuit-depth scaling compared to Trotterisation; however, they require significant
overheads in quantum resources.

We make significant progress and develop TE-PAI, which:
a) simulates ``effectively exact'' time evolution on average;
b) requires executing only very simple circuits and performing measurements on them, i.e., it does not require advanced quantum resources such as ancillary qubits or controlled evolutions;
c) can naturally simulate time evolution under time-dependent Hamiltonians;
d) achieves shallow circuit depths with optimal scaling that saturates the Lieb-Robinson bound.
TE-PAI is ``effectively exact'' in the sense that one needs to choose a priori a
desired precision $\epsilon$ to which the expectation value of the time-evolved observable is estimated, i.e.,
the level of statistical uncertainty (see \cref{app:classical_cost} for details).
While algorithmic errors associated with time evolution are finite,
they can be suppressed arbitrarily below $\epsilon$ without increasing circuit depths, i.e.,
by committing $O(\epsilon^{-1})$ classical pre-processing resources.

TE-PAI proceeds by constructing an unbiased estimator for the entire exact time-evolution
superoperator -- this allows us to estimate expectation values of time-evolved observables
by sampling the output of quantum circuits and is thus compatible with
advanced measurement techniques, such as classical shadows \cite{mitigation_shadow,shadow} or Pauli grouping techniques~\cite{Crawford2019,jena2019pauli}.
This immediately enables a wide range of applications, such
as shadow spectroscopy for estimating energy gaps in the problem Hamiltonian~\cite{chan2022algorithmic}.
TE-PAI can be combined straightforwardly with a broad range of further techniques that
require the estimation of time-dependent correlators.
Specifically, one can estimate the Loschmidt echoes $\langle \psi(0)|\psi(t) \rangle$  and $\langle \psi(t)|O|\psi(t') \rangle$
of fundamental importance, by estimating the Pauli $Z$ expectation value on an ancilla qubit.
This enables the estimation of ground and excited state energies via statistical phase estimation~\cite{Wang_2023, Lin_2022, Ding2023simultaneous},
the estimation of the density of states~\cite{goh2024direct}, as well as the estimation of expectation values of observables $O$ in
eigenstates of the problem Hamiltonian~\cite{PRXQuantum.2.040361}.
These techniques are expected to unlock some of the most important application areas of early fault-tolerant quantum computers, e.g., obtaining ground-state energies, correlation functions and molecular configurations, etc.
However, we note that our approach is based on expected value estimation and is thus
not compatible with conventional phase estimation or with other ``single-shot" techniques.

Previous work, such as qDRIFT~\cite{qDRIFT1,qDRIFT2,qDRIFT3}, similarly use randomisation
but the circuit depths are not independent of the approximation error.
Prior works~\cite{dyson-base, similar} achieve some of the advantages of TE-PAI
(exact evolution, comparable circuit depths), however, require more complex controlled circuits and are not compatible with
advanced measurement techniques, such as classical shadows \cite{mitigation_shadow,shadow} or Pauli grouping techniques~\cite{Crawford2019,jena2019pauli}. For instance, ~\cite{similar}  requires estimating observables one-by-one using Hadamard tests and thus requires random circuits to be controlled on an ancillary qubit,
as we detail in Appendix~\ref{app:comparison}.
In contrast, TE-PAI only requires executing simple random circuits and measuring their outputs,
and thus benefits from almost unlimited compatibility
with a broad range of applications and can be combined naturally with all error mitigation techniques,
classical shadows, and randomised protocols that use time evolution as a subroutine,
such as spectroscopy or estimating the density of states and beyond~\cite{mitigation_shadow, chan2022algorithmic, goh2024direct}.

\begin{figure*}[tb]
	\centering
	\includegraphics[width=1\textwidth]{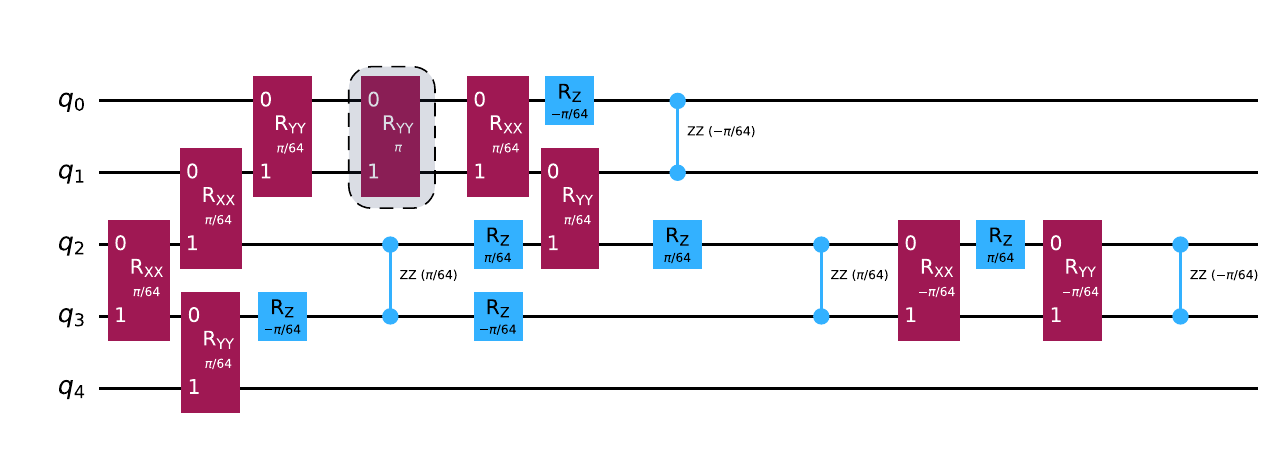}
	\caption{A single random circuit instance of TE-PAI -- by executing multiple such random
	circuits and post-processing their measurement outcomes, one can implement effectively exact time evolution on average via \cref{main_statement}. In the present example, we consider a 5-qubit Hamiltonian
	defined in \cref{eq:spinring} and a rotation angle  $\Delta = \pi/2^{6} = \pi/64$.
	TE-PAI then uses the Pauli gates $RXX, RYY, RZZ$ and $RZ$ only with rotation angles $\pm\Delta = \pm\pi/64$ and only rarely with
	$\pi$ (gate highlighted by dotted rectangle) --
	when the angle $\pi$ is chosen then all measurement outcomes are multiplied by a factor $-1$.
	The example considers a short time evolution of $T=0.05$ which is the reason for obtaining a shallow circuit
	with an expected number of gates $\nu_\infty \approx 25$.
	We note that existing compilation techniques, including ones that were specifically developed for
	Trotterised circuit structures~\cite{2qan}, can be applied immediately to reduce the circuit depth.
	}
	\label{fig:example}
\end{figure*}

The main technical tool we exploit is Probabilistic Angle Interpolation (PAI) \cite{PAI}, which is particularly relevant for near-term and early fault-tolerant applications, where averaging over many circuit executions is required for expected-value measurements. While PAI increases the sampling cost, it provably achieves the least possible overhead.
TE-PAI applies this mathematical formalism to time evolution circuits, generating random circuits with only two kinds of rotation angles, $\pm\Delta$ and $\pi$ as illustrated in \cref{fig:example}; post-processing their outputs yields exact time-evolved expected values. Conventional trotterisation requires small gate angle settings that are proportional to the individual Hamiltonian terms. However, in
real experiments the gate angles always have a finite precision as well as added coherent over or under rotations,
which effectively leads to implementing the correct trotter evolution of a different
Hamiltonian with slightly different Pauli coefficients -- the effect of this may be radical near phase transitions. Our randomised approach effectively overcomes this limitation as the rotation angle $\Delta$ can be chosen arbitrarily in our approach and can be updated according to
our best knowledge of the true, experimentally calibrated rotation angle.

We prove that TE-PAI requires a number of gates proportional to the total simulation time $T$ and the system size, thus saturating the Lieb-Robinson bound~\cite{LR, Haah2021}.
A significant advantage of TE-PAI is that it offers a trade-off between
sampling overhead and circuit depth, i.e., one can use NISQ-friendly
shallow circuits at the cost of an increased sampling overhead.
While these features are particularly important in the  NISQ era,
we construct optimised implementations for early fault-tolerant
quantum computers (FTQC), whereby circuit-depth and width limitations
will be similarly crucial.

Furthermore, we exploit that our approach works even if we use the same rotation angle in
every gate in our time evolution circuits
and develop an explicit, heavily optimised  fault-tolerant implementation using catalyst circuits and repeat-until-success teleportation.
We explicitly estimate fault-tolerant resource requirements for a benchmark problem and demonstrate that it
may require significantly less magic states than prior techniques, e.g., we achieve orders of  magnitude
lower T-counts than conventional Trotterisation.
As a result, our method imposes substantially lower fault-tolerance overheads, potentially enabling early practical demonstrations of quantum advantage in early FTQC.

This manuscript is structured as follows: We begin with a detailed description of the time-independent and time-dependent Trotter decompositions. In \cref{sec:results}, we derive the unbiased estimator of TE-PAI and prove that the sampling overhead and the expected number of gates in the circuit are finite for simulating exact time evolution. In \cref{sec:numeric}, we demonstrate the superiority of TE-PAI through numerical simulations of practically motivated quantum simulation tasks under time-dependent Hamiltonians. We also demonstrate the benefit of executing shallow circuits with TE-PAI by simulating a noisy NISQ device. In \cref{sec:impl}, we detail explicit fault-tolerant implementations and provide cost estimations of TE-PAI, concluding that it can achieve orders of magnitude more T-cost-efficient implementations than conventional Trotterisation. Finally, we conclude our work in \cref{sec:summary}.

\subsection{Quantum simulation with product formulas}
\subsubsection{Time-independent Hamiltonians}
We start by briefly reviewing product formulas used for simulating time evolution
under a time-independent quantum Hamiltonian $H$ which is typically specified as a linear combination
$H=\sum _{k=1}^{L} c_{k} h_{k}$,
where $c_{k}$ are real coefficients, and  $h_k \in \{  X,  Y, Z, \openone \}^n$ are Pauli strings.
We also define the $\ell _{1}$ norm of the coefficients as it will
determine the complexity of simulating such quantum systems as
$\| c\| _{1} =\sum _{k=1}^{L}| c_{k}|$.

Then, the first-order Trotter-Suzuki decomposition provides a way to approximate the evolution
operator as a product of exponentials of each term in the Hamiltonian as
\begin{equation}\label{eq:simple_trotter}
	e^{-iHT} \approx \left(\prod _{k=1}^{L} e^{-ic_{k} h_{k}\frac{T}{N}}\right)^{N}.
\end{equation}
Each term represents the evolution under one component of the Hamiltonian $h_{k}$
for a short time interval $\frac{T}{N}$ and the approximation becomes increasingly more accurate
as $N$ increases. While in the present work we are focusing on the above first-order Trotter decomposition we note that
our results can immediately be applied to higher-order Trotter decompositions~\cite{trotter_err,oscillate}.

We now briefly summarise the error analysis for time-independent Trotter decompositions following ref.~\cite{trotter_err}.
\begin{statement}\label{prop:trotter_error}
	The additive approximation error of the first-order Trotter decomposition
	can be bounded as
	\begin{equation*}
		\epsilon_T:=\left\Vert \left( \prod _{k =1}^{L} e^{-ic_{k} h_{k} \frac{T}{N}}\right) ^N-e^{-iHT}\right\Vert
		\leq
		\frac{T^{2}}{2N} \lVert c \rVert^2_T,
	\end{equation*}
	where the Trotterisation error norm was defined in ref.~\cite{trotter_err} as
	\begin{equation}\label{eq:commutator_norm}
		\lVert c \rVert^2_T	:=\sum _{\gamma _{1} =1}^{L}\left\Vert \left[\sum _{\gamma _{2} =\gamma _{1} +1}^{L} c_{\gamma _{2}} h_{\gamma _{2}} ,c_{\gamma _{1}} h_{\gamma _{1}}\right]\right\Vert.
	\end{equation}
	It follows that achieving a precision $\epsilon$ requires the following number of
	quantum gates in a Trotter circuit
	\begin{equation}\label{eq:trotter_bound}
		\nu \leq \frac{1}{2} L T^2 \lVert c \rVert^2_T \epsilon^{-1}.
	\end{equation}
\end{statement}
Indeed, the additive Trotter error, $\epsilon$, can be reduced at the expense of proportionally increasing the circuit depth. In the next section we describe our algorithm which utilizes a probabilistic approach to generate random circuits from these Trotter-Suzuki circuit templates. In stark contrast, the circuit depth in our approach is independent of the precision, and $N$ is a parameter that only affects the complexity of classical pre-processing as we detail in \cref{app:classical_cost}.

\subsubsection{Time-dependent Hamiltonians}

Building on the previous time-independent case, we extend the formalism to time-dependent Hamiltonians, which are crucial for accurately simulating complex quantum systems in practice
that evolve under time-dependent interactions, such as in quantum control~\cite{koch2022quantum}.
In particular, we consider a Hamiltonian $H(t)$ whose decomposition coefficients $c_k(t)$
are time-dependent as
\begin{equation*}
	H(t)=\sum _{k=1}^{L} c_{k} (t)h_{k}.
\end{equation*}
We will assume that $c_{k} (t)$ are absolutely continuous functions of time. This assumption is necessary to bound the convergence rate in our proofs in \cref{app:proof1,app:proof2}.
We then define their average $\ell _{1}$-norm as
\begin{equation}
	\label{eq:l1-norm}
	\overline{\| c\| _{1}} :=\frac{1}{T}\int _{0}^{T}\sum _{k=1}^{L}| c_{k} (t)| \, \mathrm{d} t
\end{equation}
In this work, we consider the following discretised product approximation for the unitary evolution operator
$U(T)$ for a time-dependent Hamiltonian as
\begin{equation}\label{eq:timedep_trotter}
	U(T)\approx e^{-iH(t_{N} )\tfrac{T}{N}} \cdots e^{-iH(t_{1} )\tfrac{T}{N}} ,
\end{equation}
where $H(t_{j} )$ represents the Hamiltonian at discrete time points
via $t_{j} =\frac{T}{N} j$ for $j=1\dotsc N$ and $\tfrac{T}{N}$.

Finally, we define the Trottererised circuit for our time-dependent Hamiltonian as
\begin{equation*}
	U(T)\approx \prod _{j=1}^{N}\left(\prod _{k=1}^{L} e^{-ic_{k} (t_{j} )h_{k}\frac{T}{N}}\right).
\end{equation*}
The approximation assumes that the Hamiltonian remains constant
within each small interval $\tfrac{T}{N}$, this approximation improves as $N$ is increased,
and error bounds have been reported in ref.~\cite{oscillate}.
We also note that a generalisation of the above product formula can be found in ref.~\cite{oscillate}
which has been derived by truncating the Magnus expansion~\cite{magnus} to the first order.

\section{Unbiased estimators via TE-PAI\label{sec:results}}
\subsection{Unbiased estimator for general product formulas}
In this section, we present details of our protocol that samples and post-processes
measurement outcomes of random, shallow-depth quantum circuits in
order to simulate effectively exact time evolution.
By comparing \cref{eq:simple_trotter} and \cref{eq:timedep_trotter} it is apparent that
both time-dependent and time-independent product formulas are generally of the form (and similarly higher order product formulas can be written in this form)
as
\begin{equation} \label{eq:general_product_formula}
	U =\prod _{j=1}^{N}\left(\prod _{k=1}^{L} R_{k} (\theta _{kj}) \right) .
\end{equation}
The gates above are Pauli rotation gates, $R_k(\theta) := e^{-ih_k \theta/2}$
and for time-dependent Hamiltonians their rotation angles are set to $\theta_{kj} = 2 c_k(t_j) \frac{T}{N}$,
which simplify to $\theta_{kj} := \theta_k = 2 c_k\frac{T}{N} $ for time-independent Hamiltonians. The parameter $N$ controls the number of Trotter steps, and by increasing $N$, we can approximate the desired time evolution operator with arbitrary accuracy via \cref{prop:trotter_error}.

We use the \textit{Probabilistic Angle Interpolation} (PAI) technique~\cite{PAI},
which we summarise in \cref{app:PAI}, and which builds on the observation that estimating an expected value of
an observable requires a quantum circuit to be run and measured many times.
At each circuit run, PAI randomly replaces the angle $\theta _{kj}$ in the rotation gate $R_{k}(\theta _{kj})$
with one of only three discrete rotation angles $0$, $\operatorname{sign}( \theta_{kj}) \Delta $, or $\pi $.
We will denote these three  gate variants as
\begin{equation} \label{eq:3_gates}
	A =I,\ \ B_{kj} = R_{k} (\operatorname{sign}( \theta_{kj}) \Delta ),\ \ C_{k} =R_{k} (\pi ).
\end{equation}
We note that one can choose a uniquely different $\Delta_{kj}$ specifically for
each rotation angle $\theta _{kj}$, however, for ease of notation we assume that one global $\Delta$ is chosen such that
$|\theta_{kj}| \leq \Delta < \pi$ for all $k \in {1, \dotsc, L}$ -- this choice will significantly reduce resources required
for fault-tolerant implementations as we will detail below.

The crucial observation that TE-PAI exploits is that in the limit $N \rightarrow \infty$, where the Trotter circuit approaches exact evolution, the rotation angles $\theta_{kj}$ become vanishingly small. Thus, with probability nearly equal to 1, we almost always choose the first gate variant, the identity operation (see Appendix \ref{app:proof2}).
As we prove below, this ultimately guarantees that the total circuit depth remains finite even in the limit $N \rightarrow \infty$ and thus the parameter $N$ only influences the complexity of classical pre-computation.

We review in detail in \cref{app:PAI}
that the PAI approach builds on the fact that the superoperator representation $\mathcal{R}_{k} (\theta_{kj} )$ of each unitary gate $R_{k} (\theta_{kj} )$ can be decomposed analytically as
\begin{equation} \label{eq:linear_comb}
	\mathcal{R}_{k} (\theta _{kj} )=\gamma _{1} ( |\theta _{kj}| ) \mathcal{A} + \gamma _{2} (| \theta _{kj} | ) \mathcal{B}_{kj}+\gamma _{3} (|\theta _{kj} |) \mathcal{C}_k,
\end{equation}
where $\mathcal{A}$, $\mathcal{B}_{kj}$ and $\mathcal{C}_k$ are superoperator representations of the unitary gates in
\cref{eq:3_gates} which generally act isomorphically via
conjugation, e.g., $\mathcal{R}_k \mathrm{vec}[\rho] =  \mathrm{vec}[R_k \rho R_k^\dagger]$.
The coefficients $\gamma _{l} (\theta)$ are provided explicitly as trigonometric functions in \cref{app:PAI}.

Focusing on a single gate element $\hat{\mathcal{R}}_{k}( \theta _{kj})$,
our classical pre-processing algorithm randomly selects one of
the three discrete gate variants as $\hat{\mathcal{D}}_{l} \in \{ \mathcal{A} , \mathcal{B}_{kj}, \mathcal{C}_k \}$
according to the probabilities
$p_l = |\gamma _{l} (\theta_{kj} )|/\lVert\gamma (\theta _{kj} )\rVert_{1}$ for $l\in \{1,2,3\}$
in order to sample the unbiased estimator
of the desired, continuous-angle gate $\mathcal{R}_{k}( \theta _{kj})$ as
\begin{equation} \label{eq:PAI}
	\hat{\mathcal{R}}_{k}( \theta _{kj}) =\| \gamma (|\theta_{kj}| )\| _{1}\operatorname{sign}[ \gamma _{l} (|\theta_{kj}| )] \hat{\mathcal{D}}_{l}.
\end{equation}
Ref.~\cite{PAI} analytically proved that this choice of the three discrete angle settings
minimises the measurement overhead characterised by $\| \gamma (|\theta_{kj}| )\| _{1}$,
confirming numerical optimisation results of ref.~\cite{koczor2024sparse}.

Replacing each gate in the product formula in \cref{eq:general_product_formula}
with the above unbiased estimator
then allows us to construct an unbiased estimator for the entire time evolution operator.
As we now summarise, this is a direct consequence of Statement~2 of ref.~\cite{PAI}.
\begin{statement}\label{main_statement}
	We obtain an unbiased
	estimator of the superoperator representation of the entire
	product formula in \cref{eq:general_product_formula} as
	\begin{equation}
		\hat{\mathcal{U}} = \prod _{j=1}^{N}\left(\prod _{k=1}^{L} \hat{\mathcal{R}}_{k} (\theta _{kj}) \right),
	\end{equation}
	using the unbiased estimators $\hat{\mathcal{R}}_{k} (\theta _{kj})$ of the individual
	continuous-angle rotations from \cref{eq:PAI}. Specifically, the mean value of the estimator is
	$\mathds{E}[\hat{\mathcal{U}}] = \mathcal{U}$.
	The classical computational complexity of generating $N_s$ random circuits is $O(NLN_{s} )$.
\end{statement}
Please refer to \cref{app:PAI} for a detailed derivation. \cref{fig:example} shows an example of a random circuit generated using the PAI approach that simulates the time evolution
under the Hamiltonian \cref{eq:spinring} for 5-qubit.
Finally our protocol is summarised as follows.
\begin{itemize}
	\item Take a quantum circuit $\mathcal{U}$ of the form of \cref{eq:general_product_formula} which implements a product formula for simulating the time evolution under the input Hamiltonian
	      $\mathcal{H}$ with parameters $N$ and $T$.
	\item Generate $N_s$ random circuits by randomly replacing gates in the circuit $\mathcal{U}$
	      with fixed rotation angles of $\pm \Delta$ and $\pi$
	      according to the PAI protocol in  \cref{main_statement}.

	\item Execute all random circuits and in post-processing multiply all measurement outcomes
	      with their corresponding prefactor $\prod_{j=1}^{N}\prod _{k=1}^{L} \| \gamma (|\theta|_{kj} )\|_{1}
		      \operatorname{sign}[ \gamma _{l} (|\theta_{kj}| )]
	      $ where the index $l := l_{kj}$ is chosen randomly for each gate in the circuit as
	      $l \in \{1,2,3\}$.
	      When classical shadows are estimated then the
	      expected value from each snapshot  needs to be multiplied by this factor as detailed in
	      ref.~\cite{mitigation_shadow}.
\end{itemize}

\begin{figure}
	\centering
	\includegraphics[width=0.5\textwidth]{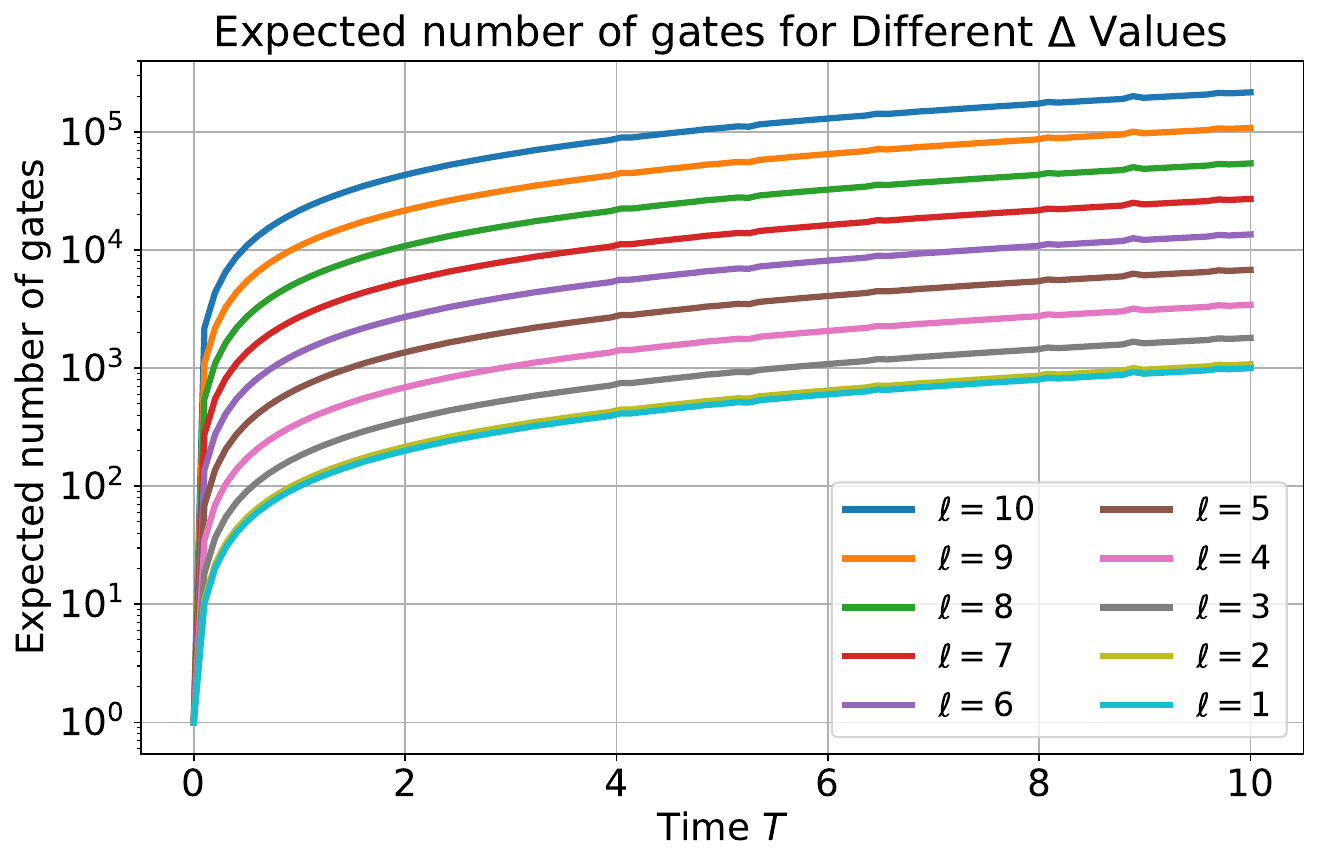}
	\caption{
	Expected number of gates when simulating the time evolution under the Hamiltonian in \cref{eq:spinring} for 14 qubits
	using different rotation angle settings as $\Delta = \pi/2^{\ell},\ \ell = 1, 2, \dots , 10$.
	While the number of gates grows linearly with the total time $T$,
	the slope is determined by the angle $\Delta$ --  decreasing $\Delta$ increases the circuit depth, however, can exponentially reduce the measurement overhead as we detail below.
	}
	\label{fig:num_gate_t}
\end{figure}

\subsection{Gate count in the random circuits}
For ease of notation, in the following we specifically consider first-order
Trotter circuits in \cref{eq:simple_trotter} and \cref{eq:timedep_trotter}, but our proofs
apply to any higher-order product formula.
In the standard first-order Trotter approach, the number of gates $\nu = NL$ is directly
proportional to $N$.
In contrast, TE-PAI generates circuits randomly,
and the number of gates is thus formally a random variable.
We now prove that, as we increase $N$, the mean value $\mathds{E}[\nu]$ asymptotically approaches a constant.

\begin{theorem}\label{thm:num_gate2}
	The expected number of gates $\mathbb{E} (\nu )$ can be approximated  in terms of
	\begin{equation*}
		\nu_{\infty } :=\lim _{N\rightarrow \infty }\mathbb{E} (\nu )=\csc (\Delta )(3-\cos \Delta )\overline{\| c\| _{1}} T
	\end{equation*}
	up to an error term as $\mathbb{E} (\nu )  = \nu_{\infty } + O(N^{-1})$. Furthermore, the variance of the gate count satisfies the same scaling: $\operatorname{Var} [\nu ] = \nu_{\infty } + O(N^{-1})$. The asymptotic gate count is lower bounded as
	\begin{equation}\label{eq:min}
		\nu_{\infty } \geq \overline{\| c\| _{1}} T \,2\sqrt{2}.
	\end{equation}
	This bound is saturated
	when using the large angle $\Delta  = 2\arctan\left(1/\sqrt{2}\right) \approx 0.392 \pi$.
	In the special case of time-independent Hamiltonians the same result holds
	up to formally replacing $\overline{\| c\| _{1}} \equiv \| c\| _{1}$.
\end{theorem}
We explicitly bound the leading constant factors in the error term in $\mathbb{E} (\nu )  = \nu_{\infty } + O(N^{-1})$
in our proof in \cref{app:proof2}.
We thus find that for a constant rotation angle $\Delta$, TE-PAI saturates the Lieb-Robinson bound \cite{LR, Haah2021}, which determines a fundamental bound on the speed at which local information can spread due to time evolution under local interactions. However, we will later prove that in practice, $\Delta$ needs to scale with both $T$ and the system size to avoid an exponential increase in the sample complexity.

In addition to the above mean value, we also characterise the
distribution of the number of gates in the circuit.
\begin{lemma}\label{lemma:distribution}
	The distribution of the number of gates in the circuit approaches a normal distribution $\mathcal{N} (\nu _{\infty },\nu _{\infty } )$ as $N\rightarrow \infty $, where $\nu_{\infty}$
	is the mean value from \cref{thm:num_gate2}.
\end{lemma}

As illustrated in Figure \ref{fig:num_gate_t}, the expected number of gates for the time-independent
Trotter circuit grows linearly with the total time $T$.

Additionally, the figure demonstrates that decreasing $\Delta$ results
in an increased number of gates, however, as we will see below,
it also decreases the measurement overhead exponentially.

\begin{figure}[tb]
	\centering
	\includegraphics[width=0.5\textwidth]{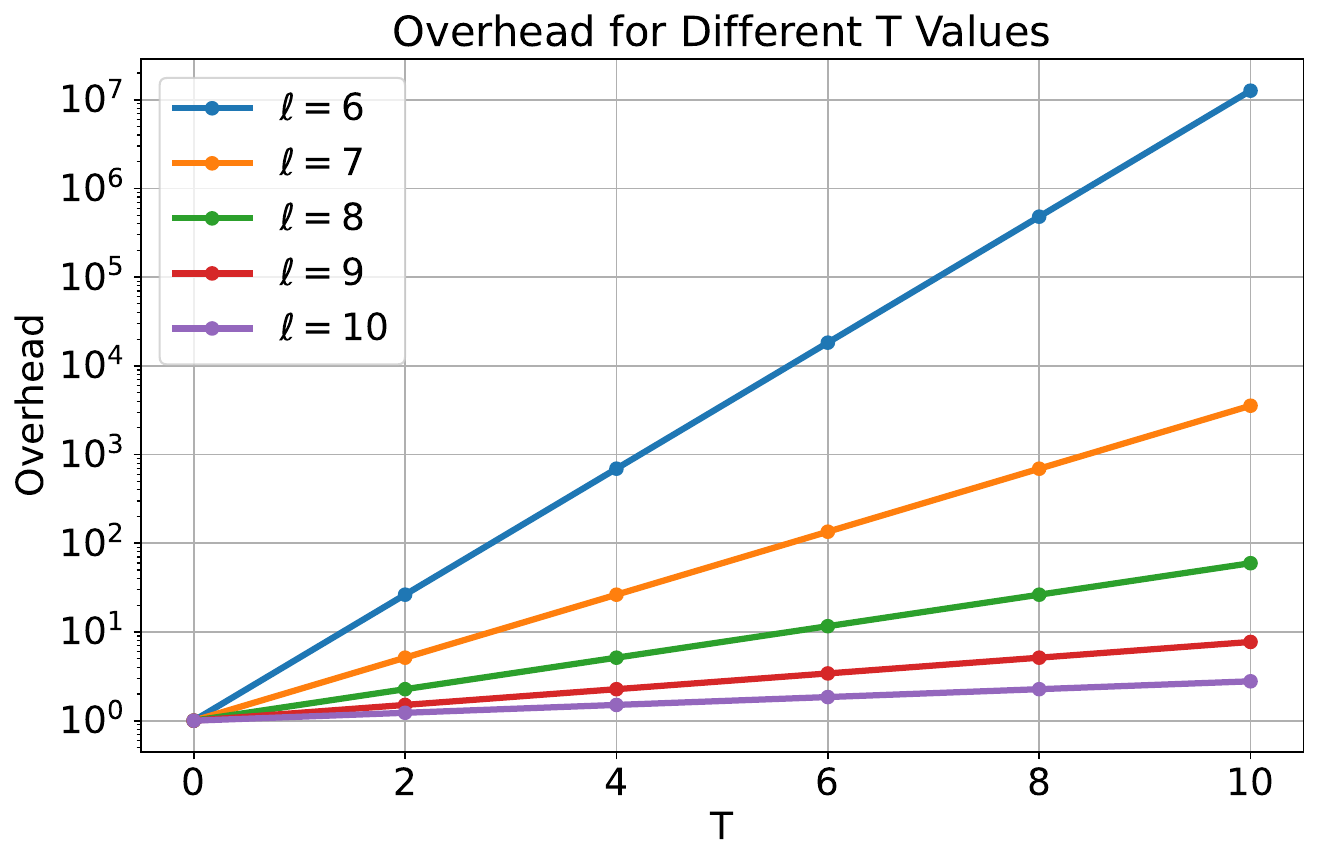}
	\caption{Measurement overhead for the time-dependent Trotter circuit with different $\Delta = \pi/2^{\ell},\ \ell = 6, 7, 8, 9, 10$. We consider the Hamiltonian in \cref{eq:spinring} for 14 qubits. We observe that the overhead grows exponentially with the total time $T$. Since $\Delta$ directly affects the exponent, a smaller $\Delta$ results in a slower exponential blowup.}
	\label{fig:overhead_T}
\end{figure}

\begin{figure}[tb]
	\centering
	\includegraphics[width=0.48\textwidth]{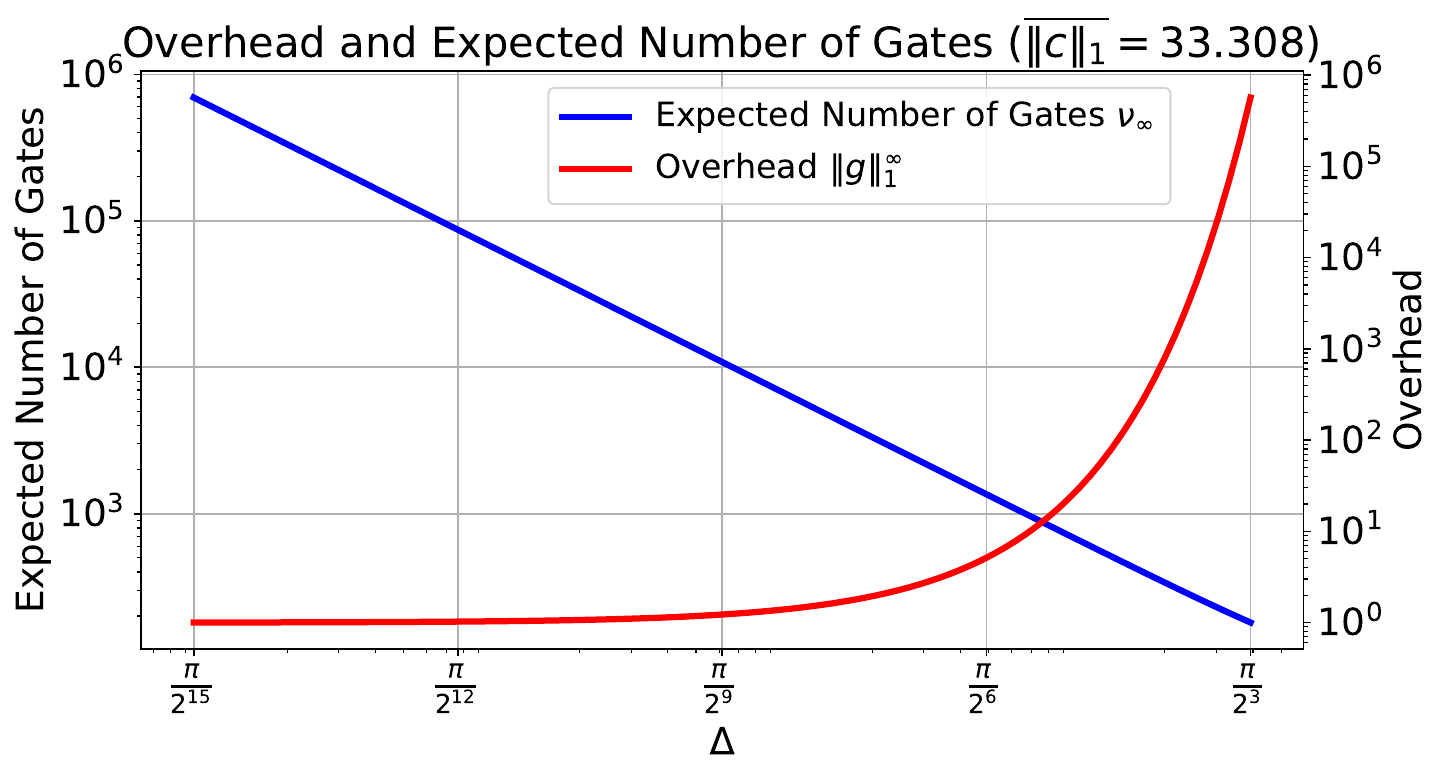}
	\caption{Trade-off between the expected number of gates $\nu_\infty$ (left axis) and measurement overhead (right axis) as a function of the rotation angle $\Delta = \pi/2^{\ell},\ \ell = 3, 6, 9, 12, 15$ for the time-dependent Hamiltonian in \cref{eq:spinring} for 14 qubits and $T=1$.}
	\label{fig:overhead}
\end{figure}

\subsection{Measurement overhead}
In TE-PAI, we randomly replace the continuous-angle gates with
three discrete gate variants: the third gate variant has a very low associated probability for small $\Delta$, however,
when it
does get selected then any measured observable is multiplied with a negative sign.
This negative sign leads to an increase in the variance of the expectation value of the observable
being estimated. Thus, in order to estimate the observable expected value to
the same precision as with an infinitely deep Trotter circuit,
one needs to perform an increased number of measurements.
We detail in \cref{app:PAI} that this measurement overhead is upper bounded by the following factor as
\begin{equation}
	\| g\| _{1} :=\prod _{j=1}^{N} \prod _{k=1}^{L} \| \gamma (|\theta_{kj}| )\| _{1} .
\end{equation}
Now we prove that the measurement overhead $\| g\| _{1}$ converges to a constant as we increase $N$.
\begin{theorem}\label{thm:overhead}
	We bound the number of shots $N_{s}$ required to achieve a specified precision $\epsilon $
	in estimating time-evolved expectation values.
	The number of circuit repetitions in TE-PAI is upper bounded as
	\begin{equation*}
		N_{s} \leq \| g\| _{1}^{2} /\epsilon ^{2},
	\end{equation*}
	whereas having access to an infinitely deep Trotter circuit results in the upper bound $N_{s} \leq \epsilon ^{-2}$.
	The overhead determined by $\| g\|^2_{1}$ can be approximated via
	\begin{equation*}
		\| g\| _{1}^{\infty }:=\exp\left[ 2\overline{\| c\| _{1}} \, T \,  \tan\left(\frac{\Delta }{2}\right)\right]
	\end{equation*}
	up to an term as $\| g\| _{1} = \| g\| _{1}^{\infty } + O(N^{-1})$.
	In the special case of time-independent Hamiltonians the measurement overhead simplifies via $\overline{\| c\| _{1}} \equiv \| c\| _{1}$.
\end{theorem}
We explicitly derive and bound the leading constant factors in the above error term in our proof
in \cref{app:proof1}.
We note that a variant of PAI was developed in ref~\cite{koczor2024sparse} that
introduces a trade-off parameter $\lambda$ that allows to continuously interpolate
between the unbiased PAI estimator (exact estimator with measurement overhead $\| g\| _{1}$)
and an approximate, biased estimator which has no measurement overhead at all.
The numerical approach of ref~\cite{koczor2024sparse} can be used immediately
for reducing the above measurement overhead at the cost of introducing a bias, however,
for ease of notation in the present work we focus on the exact, unbiased implementation.

We illustrate in \cref{fig:overhead_T}, that the measurement
overhead of our unbiased estimator grows exponentially with the total time $T$
but decreasing $\Delta$ results in a slower exponential blowup.
Furthermore, in \cref{fig:overhead}, we illustrate the trade-off between
the measurement overhead and the expected number of gates
at different rotation angles $\Delta$. As we increase $\Delta$, we decrease the circuit depth but also increase the measurement overhead. While the expected number of gates $\nu_\infty$ is
a constant that is independent of $N$, decreasing the
rotation angle $\Delta$ increases $\nu_\infty$ and ultimately can yield
to divergence in the limit $\lim _{\Delta \rightarrow 0}\mathbb{E} (\nu )=\infty $. In a quantum computer where the quantum gates are not perfect---for example, noisy physical operations or logical operations in an early fault-tolerant machine---the circuit depth needs to be chosen such that $p_{\mathrm{err}}^{-1} \propto \nu$ to ensure that error-mitigation techniques can be used effectively. This desired circuit depth can be achieved by setting the corresponding $\Delta$ angle.

Both the measurement overhead and the circuit depth of TE-PAI depends implicitly on the number of qubits, through the dependence of the norm $\overline{\lVert c\rVert}_1$ on the number of qubits $n$. For example, for the Heisenberg spin model considered later in this manuscript, $\overline{\lVert c\rVert}_1 \in O(n)$, while in quantum chemistry a pessimistic worst-case bound is $\overline{\lVert c\rVert}_1 \in O(n^4)$. However, advanced techniques such as tensor hypercontraction (THC) can significantly reduce this scaling~\cite{hyper_contraction}. The  complexity of other quantum simulation methods, such as qubitization and qDRIFT, similarly depends implicitly on the number of qubits via the norm $\overline{\lVert c\rVert}_1$.

Finally, we consider fixing the measurement overhead as a constant and establish how the rotation angle $\Delta$ scales with the system size and total simulation time.
\begin{remark}\label{cor:Q_control_param}
	We introduce a trade-off parameter $ Q $ that governs the trade-off between circuit depth and the measurement overhead. By using a rotation angle $ \Delta = 2\arctan\left(\frac{Q}{2\overline{\| c\| _{1}} T}\right) $, we achieve a constant overhead of $ \exp(Q) $ and obtain the number of gates as
	\begin{equation*}
		\nu _{\infty }=\frac{2\left(\overline{\| c\| _{1}} T\right)^{2}}{Q} + Q
		\leq \frac{4\left(\overline{\| c\| _{1}} T\right)^{2}}{Q}.
	\end{equation*}
	The upper bound above is due to the fact that \( Q \leq \overline{\| c\| _{1}} T \sqrt{2} \),
	via the lower bound on \( \nu_\infty \) in \cref{eq:min}.
\end{remark}
In practice one would choose $Q\geq 1$ given the measurement overhead $\exp(1)$ is still very reasonable.
Let us now compare the number of gates and the total time complexity of TE-PAI to similar techniques.

\textbf{Number of gates:}
The parameter $Q$ allows us to have a constant measurement overhead
at the cost of the number of gates increasing quadratically with the system size and with the
time depth $T$.
This scaling is the same as in the case of first-order Trotterisation whereby
$\nu \in O( T^{2} /\epsilon) $ in \cref{eq:trotter_bound},
however, the crucial difference is that the constant factor $\epsilon^{-1}$ in Trotterisation
is replaced here with a controllable hyperparameter $Q^{-1} \leq 1$ which in practice is many orders of magnitude
smaller. Thus, we expect TE-PAI to require orders of magnitude fewer gates than first-order Trotterisation.

\begin{figure}[tb]
	\centering
	\includegraphics[width=0.5\textwidth]{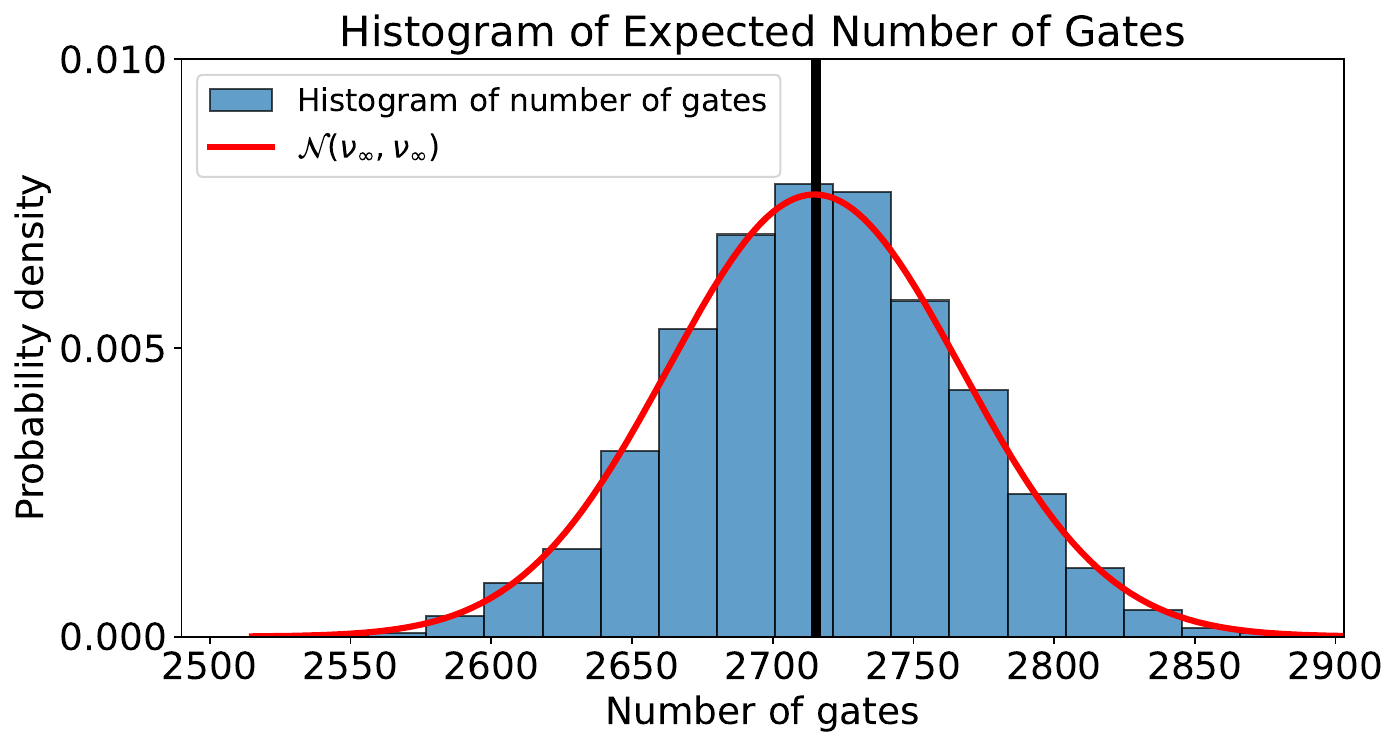}
	\caption{Histogram of the number of gates in the randomly generated TE-PAI circuits  for the time-dependent Hamiltonian in \cref{eq:spinring} for 14 qubits and $\Delta=2^{-7}\pi,\ T=1$.
	The expected number of gates in \cref{thm:num_gate2} is $\nu _{\infty} \approx 2715$ which is in good agreement
	with the empirical mean (black line).
	Furthermore, \cref{lemma:distribution} guarantees that the distribution is well approximated by
	a Gaussian distribution $\mathcal{N}(\nu_{\infty}, \nu_{\infty})$ (red line)
	which is in good agreement with the histogram.
	}
	\label{fig:num_gate_hist}
\end{figure}

\textbf{Time complexity:} We now estimate the end-to-end time complexity of TE-PAI. The number of gates in a single circuit is $O(T^2 Q^{-1})$, and estimating an observable to precision $\epsilon$ requires $O(\epsilon^{-2} \exp{Q} )$ repetitions (we assume the circuits are repeated serially and therefore TE-PAI does not require a space overhead).
The end-to-end time complexity of TE-PAI (or equivalently its space-time volume) therefore scales as $O( T^2 \epsilon^{-2} Q^{-1} \exp[Q]  )$.
We can compare this to a first-order Trotter circuit from which the expected value
is extracted using amplitude estimation in which case
the Trotter circuit is repeated coherently $O(\epsilon^{-1})$ times
leading to a total time complexity
$O(T^{2} \epsilon ^{-2})$ (the space-time volume of this approach scales similarly given amplitude estimation requires little space overhead).

In conclusion, the time complexity of extracting a time-evolved expected value in TE-PAI is comparable to using first-order Trotterisation in combination with amplitude estimation. However, TE-PAI has a number of significant advantages.
First, TE-PAI requires only shallow circuits, making it feasible to run when
coherence time or code distance is limited, whereas amplitude amplification requires significantly deeper circuits.
Second, while amplitude estimation estimates observables one-by-one, we detail below that TE-PAI is compatible
with advanced measurement techniques, including classical shadows and thus allows simultaneous estimation of many observables.
Third, certain platforms, such as silicon qubit technologies, enable the fabrication of many independent QPUs on a single chip with no or limited communication between QPUs. TE-PAI can be fully parallelized, with the sampling task distributed across many QPUs,
leading to a proportional reduction of the runtime.
In contrast, some amplitude estimation variants can be parallelised, but only to a more limited extent~\cite{labib2024quantum}.
Finally, TE-PAI only requires the implementation of a single type of non-Clifford rotation,
specifically a single-qubit rotation with angle $\Delta$, which allows us to design a particularly efficient fault-tolerant implementation
below.

Note that, in terms of gate complexity, truncated Taylor--series LCU methods scale as $\mathcal{O}\!\left(\|c\|_1\,T \,\frac{\log(1/\epsilon)}{\log\log(1/\epsilon)}\right)$~\cite{LCU1,LCU2,LCU3}, while qubitization, QSP, and QSVT scale as $\mathcal{O}\!\left(\|c\|_1\,T + \log(1/\epsilon)\right)$~\cite{QSP1,QSP2,QSVT}, but these techniques are fundamentally different and require deep coherent circuits that are not feasible to execute on current or early-FTQC devices unlike TE-PAI.

\begin{figure*}[tb]
	\centering
	\begin{minipage}[b]{0.51\linewidth}
		\centering
		\includegraphics[width=\textwidth]{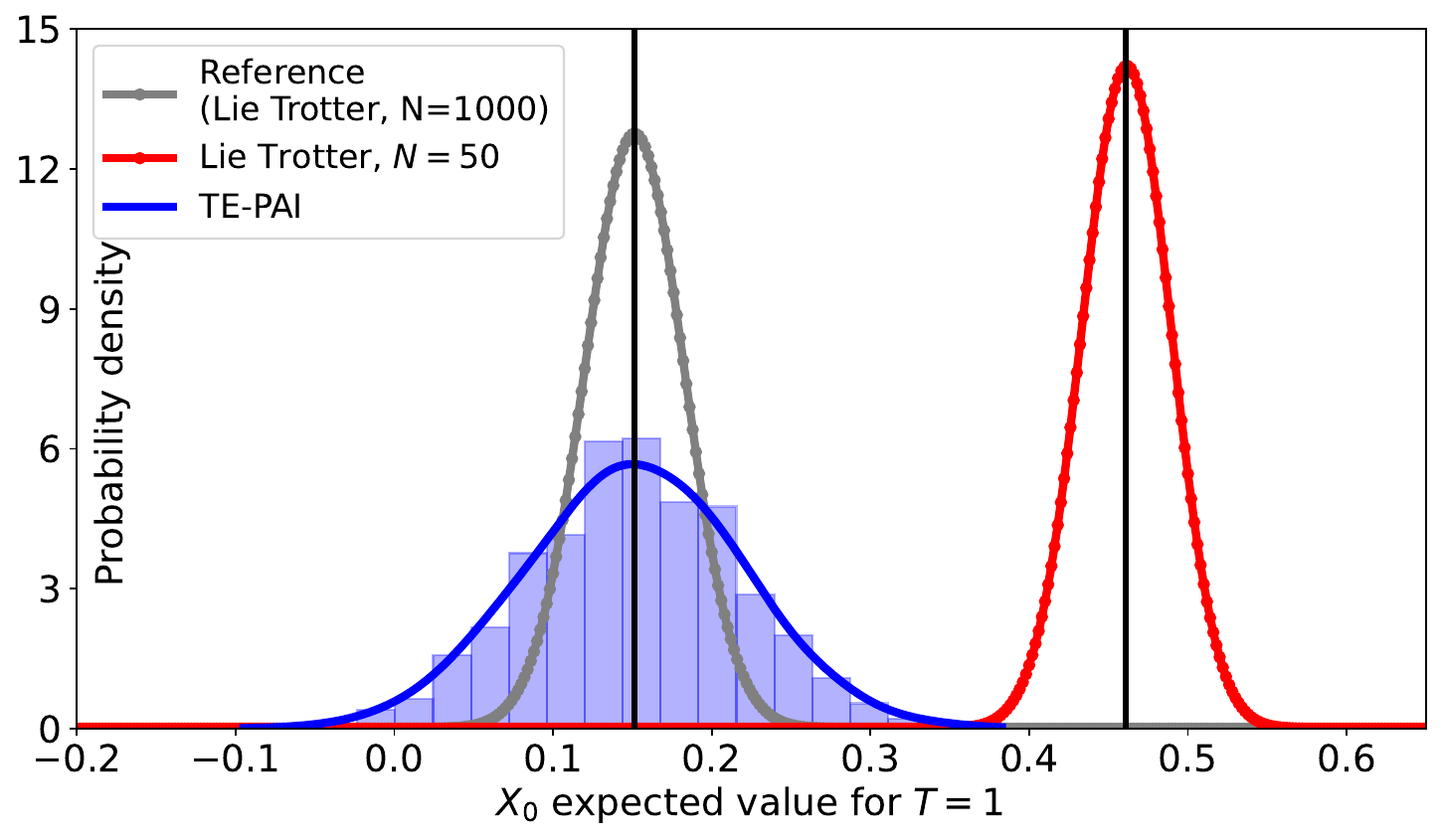}
	\end{minipage}
	\hfill
	\begin{minipage}[b]{0.48\linewidth}
		\centering
		\includegraphics[width=\textwidth]{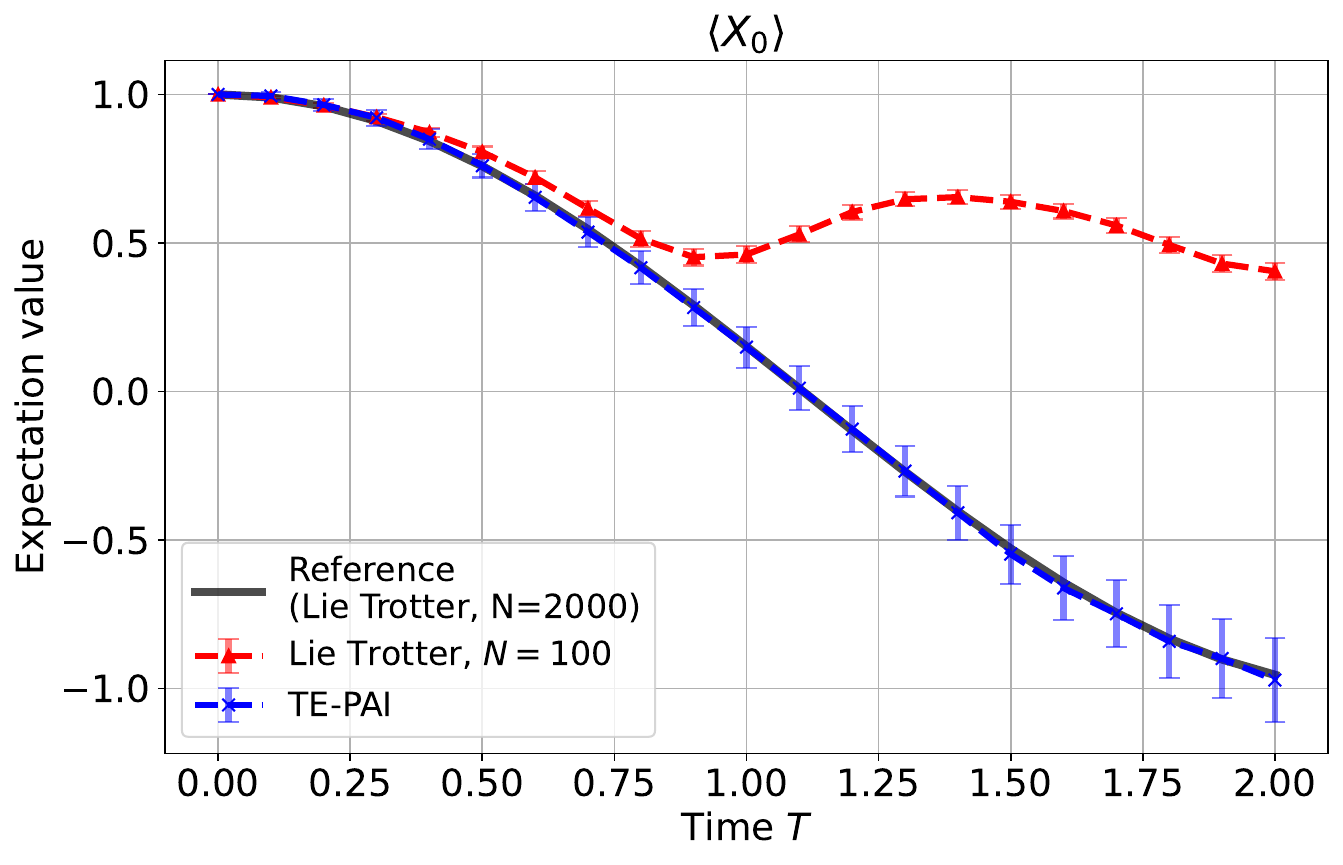}
	\end{minipage}
	\caption{(left) Distribution of estimated expected values $\langle X_{0} \rangle $ from time evolution
		circuits that implement evolution under the 14-qubit spin Hamiltonian in \cref{eq:spinring} for $T=1$ and using $N_s=1000$ circuit repetitions (shots).
		(left, grey) distribution of $\langle X_{0} \rangle $ from a deep Trotter circuit with $N=1000$ layers, consisting of $\nu =56000$ continuous-angle rotations. (left, red) distribution of $\langle X_{0} \rangle $
		from a shallow Trotter circuit consisting of $N=50$ layers and $\nu =2800$ continuous-angle rotations. (left, blue) Histogram and estimated distribution of $\langle X_{0} \rangle $ using TE-PAI using only two kinds of discrete rotation angles $\Delta = 2^{-7} \pi$ and $\pi$.
		The distribution width of TE-PAI is slightly increased but its mean is identical to that of the deep Trotter circuit (blue) while its gate count is very close to that of the shallow Trotter circuit as $\nu_\infty = 2715$. (right) Expectation values $\langle X_{0} \rangle $ as the simulation time $T$ increases from $0$ to $2$.
		Due to its shallow depth, the $N=100$ Trotter circuit (right, red)
		introduces a bias to the expected value measurement, albeit its standard deviation remains unchanged
		as we increase $T$. In contrast, TE-PAI (right, blue) recovers the same mean value as
		the $N=2000$ deep Trotter circuit (right, grey), however, its standard deviation increases as $T$ increases.
		\label{fig:simulation}
	}
\end{figure*}

\section{Numerical demonstrations\label{sec:numeric}}
\subsection{Heisenberg spin model benchmark}
We consider the benchmarking task of simulating the spin-ring Hamiltonian as
\begin{equation} \label{eq:spinring}
	\mathcal{H} =\sum _{k\in \text{ring} (N)} \omega _{k} Z_{k} +J(t)\vec{\sigma }_{k} \cdot \vec{\sigma }_{k+1},
\end{equation}
where we choose time-dependent coupling terms $J(t)=\cos (99\pi t)$ and the parameters $\omega _{k}$ are chosen uniformly randomly within the range $[-1,1]$.
This model is representative of problems considered in condensed matter physics
for studying many-body localisation. These problems could be effectively explored
using early quantum computers and may be hard to simulate classically
for large numbers of qubits~\cite{PhysRevB.91.081103, Childs_2018}.
For the present demonstration, we use $n=14$ qubits and due to the periodicity of the Hamiltonian,
$\overline{\| c\| _{1}} \approx 33.30$ is
constant for integer evolution times $T=1,2\dotsc $, and we choose $T=1$ and $\Delta = 2^{-7} \pi$.
We generate random circuits using TE-PAI for a
product formula of $N=1000$ layers and a total evolution time of $1$ -- we chose a relatively
low value of $N$ as in our demonstrations we will use a relatively low number of shots
$N_s=1000$ and therefore shot noise will dominate over residual algorithmic errors
(while indeed $N$ can be increased without requiring more quantum resources).

First, we present a histogram in \cref{fig:num_gate_hist} that estimates the distribution of the
number of gates from $N_s=1000$ different randomly generated TE-PAI circuits.
The expected number of gates via \cref{thm:num_gate2} is $\nu _{\infty} \approx 2715$
which shows good agreement with the empirical mean in our histogram (black line).
Furthermore, as predicted by \cref{lemma:distribution}, the histogram agrees well with a Gaussian distribution $\mathcal{N}\left( \nu _{\infty } ,\nu _{\infty }\right)$.

Second, we execute the TE-PAI circuits to estimate expected values
$\langle X_{0} \rangle $ and compare them to conventional Lie-Trotter circuits.
In \cref{fig:simulation} (left) we report the distribution of expected values estimated using $N_s=1000$ shots:
while TE-PAI in \cref{fig:simulation} (left, blue) has a slightly increased distribution width, its mean value matches exactly the mean value of an
arbitrarily deep Trotter circuit in \cref{fig:simulation} (left, grey). Furthermore, with small measurement overhead of $\|g\|_1^\infty \approx 2.15$, TE-PAI
requires only a small number of gates comparable to a shallow Trotter circuit with $N=50$ --  which shallow circuit
introduces a significant bias due to significant algorithmic errors, as shown in \cref{fig:simulation} (left, red).

Finally, we generate a family of TE-PAI circuits for an increasing total simulation time $T$ and compare
the expected values $\langle X_{0} \rangle (T)$ to shallow and deep Trotter circuits in \cref{fig:simulation} (right). Since this plot considers $T = 0 \ldots 2$, we doubled the Trotter step numbers to $N = 2000$ for the deep Trotter circuit and $N = 100$ for the shallow Trotter circuit.
Expected values estimated from shallow Trotter circuits have
relatively low statistical uncertainty \cref{fig:simulation} (right, red error bars) throughout
the evolution, however, suffer from a significant bias, i.e., red curve is significantly off from the reference deep
time evolution circuit \cref{fig:simulation} (right, grey). In contrast, TE-PAI (blue line) achieves the same mean value as
the deep Trotter circuit, however, its statistical uncertainty increases with the simulation time.

\subsection{Quantum chemistry benchmark}
The complexity of TE-PAI depends on the coefficient $\ell_1$-norm $\overline{\|c\|_1}$ rather than on the number of Hamiltonian terms $L$. This suggests that TE-PAI could be advantageous for quantum chemistry problems with large $L$, where standard Trotter methods require prohibitively large gate counts per Trotter step. An important example is the simulation of the FeMoCo active-space Hamiltonian. TE-PAI is therefore expected to play a role in enabling quantum advantage in challenging quantum chemistry applications.

As a simple demonstration, we consider a time-independent chemistry simulation of a linear chain of $N_{\mathrm{H}}=6$ hydrogen atoms in the STO-6G minimal basis with an interatomic spacing of $2.0$~Bohr, described by the corresponding electronic Hamiltonian:
\begin{equation}
	H = \sum_{p q} h_{p q}\, a^\dagger_p a_q
	+ \frac{1}{2}\sum_{p q r s} h_{p q r s}\, a^\dagger_p a^\dagger_q a_r a_s,
\end{equation}
where $a_p^\dagger$ and $a_q$ are fermionic creation and annihilation operators, and
$h_{pq}$, $h_{pqrs}$ denote one- and two-electron integrals obtained from a restricted Hartree-Fock calculation.
Mapping to qubits is performed using the Jordan-Wigner transformation, which yields a Pauli expansion acting on twelve qubits (corresponding to twelve spin orbitals).

Since orbital occupation numbers,
\begin{equation*}
	n_p = a^\dagger_p a_p = \tfrac{1 - Z_p}{2},
\end{equation*}
are an important observable in quantum chemistry, we compute the single-qubit expectation value $\langle Z_{0}(t)\rangle$, which is directly related to the occupation of the first spin orbital.

\begin{figure}[tb]
	\centering
	\includegraphics[width=0.5\textwidth]{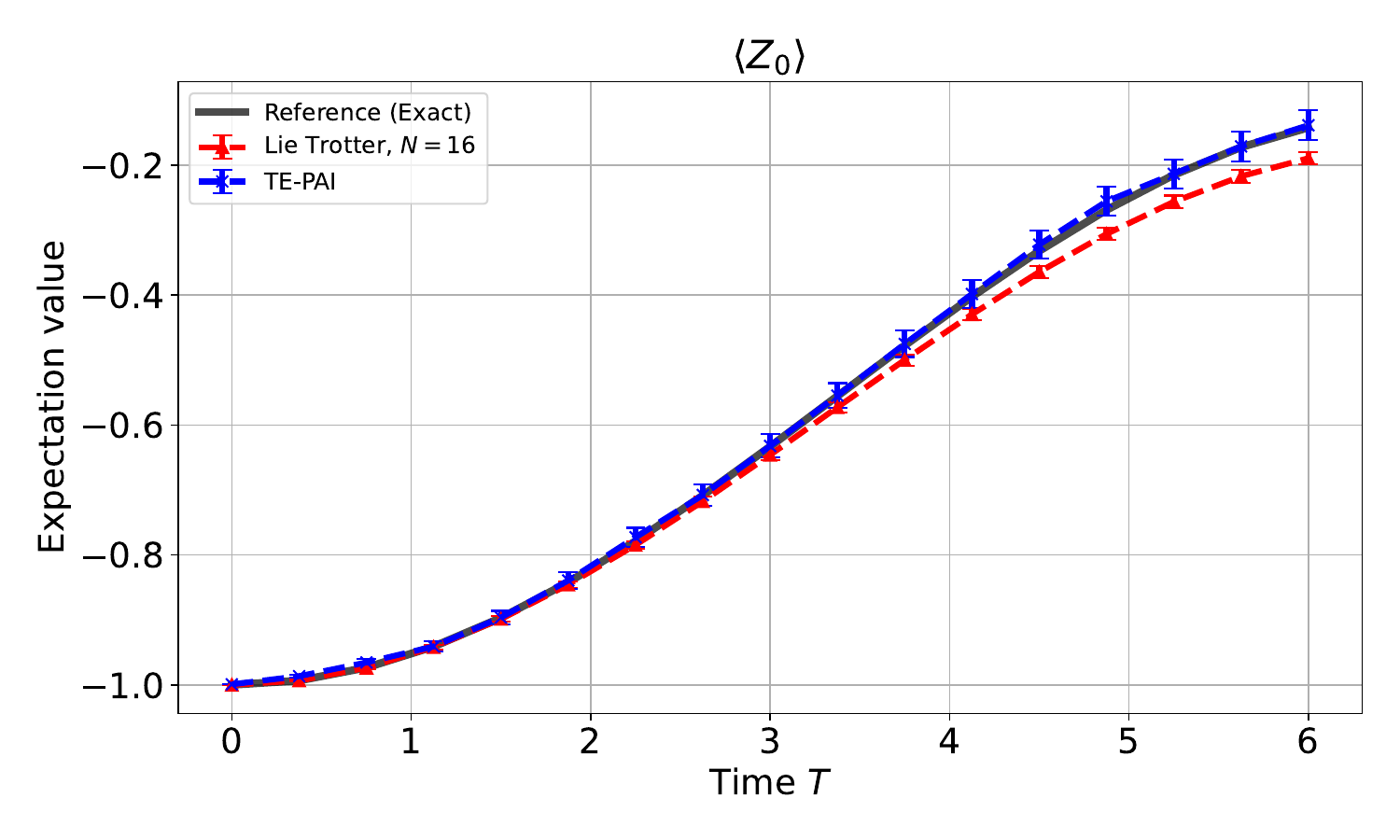}
	\caption{Expectation values $\langle Z_{0} \rangle$ from time evolution under the H$_6$ Hamiltonian with $919$ Pauli terms, starting from the initial state $\ket{101001010101}$ and using $N_s = 10000$ circuit repetitions (shots), as the simulation time $T$ increases from $0$ to $6$. (grey) Exact simulation obtained by directly exponentiating the Hamiltonian. (red) Shallow Trotter circuit with $N=16$ layers, consisting of $\nu = 14704$ continuous-angle rotations, which introduces a systematic bias in the estimated values. (blue) TE-PAI simulation using discrete rotation angles with $\Delta = \pi/2^{8}$, which recovers the same mean as the exact reference while requiring only $\nu_\infty = 14138$ gates; however, its standard deviation increases with $T$.}

	\label{fig:chem_snap}
\end{figure}

\cref{fig:chem_snap} shows the results of simulating time evolution under the H$_6$ Hamiltonian with $919$ Pauli terms, starting from the initial state $\ket{101001010101}$ and using $N_s = 10000$ circuit repetitions (shots), as the simulation time $T$ increases from $0$ to $6$. We compare the estimated expectation values $\langle Z_{0}(t)\rangle$ with an exact simulation obtained by directly exponentiating the Hamiltonian, a shallow Trotter circuit with $N=16$ layers requiring $\nu = 14704$ continuous-angle rotations that introduces systematic bias, and TE-PAI with $\nu_\infty = 14138$ gates and fixed rotation angle $\Delta = 2^{-8}\pi$, implemented using random circuits corresponding to a product formula with $N=1000$ layers. In this setting, TE-PAI achieves the same unbiased result as the exact simulation with a gate cost comparable to that of the shallow Trotter circuit.

In chemistry simulations typically a relatively high precision, the chemical accuracy $\propto10^{-3}$ is required which is challenging to meet using shallow Trotter circuits. TE-PAI, on the other hand, reproduces the exact result, with the only remaining error coming from sampling noise, which can be reduced by increasing the number of shots.

\vspace{2mm}
\subsection{Noisy Implementation\label{sec:nisq}}
Before fault tolerance is achieved, one needs to resort to noisy physical gates
to execute circuits in the NISQ era. This poses  limitations on the achievable circuit
depths, as the total number of noisy gates is typically restricted to a small constant multiple of the inverse average
gate error rate.
Compared to Trotterisation, TE-PAI has the significant advantage that the circuit depth can be
reduced without introducing bias. Here we demonstrate the improved robustness of TE-PAI against gate
noise and consider a noise model where each gate is followed by depolarisation that acts on the same qubit(s)
as the gate itself. We assume a typical two-qubit gate error probability of $p_2 = 10^{-3}$
and for single-qubit gates we assume an order of magnitude lower error probability of $p_1 = 10^{-4}$.

\begin{figure}[tb]
	\centering
	\includegraphics[width=0.5\textwidth]{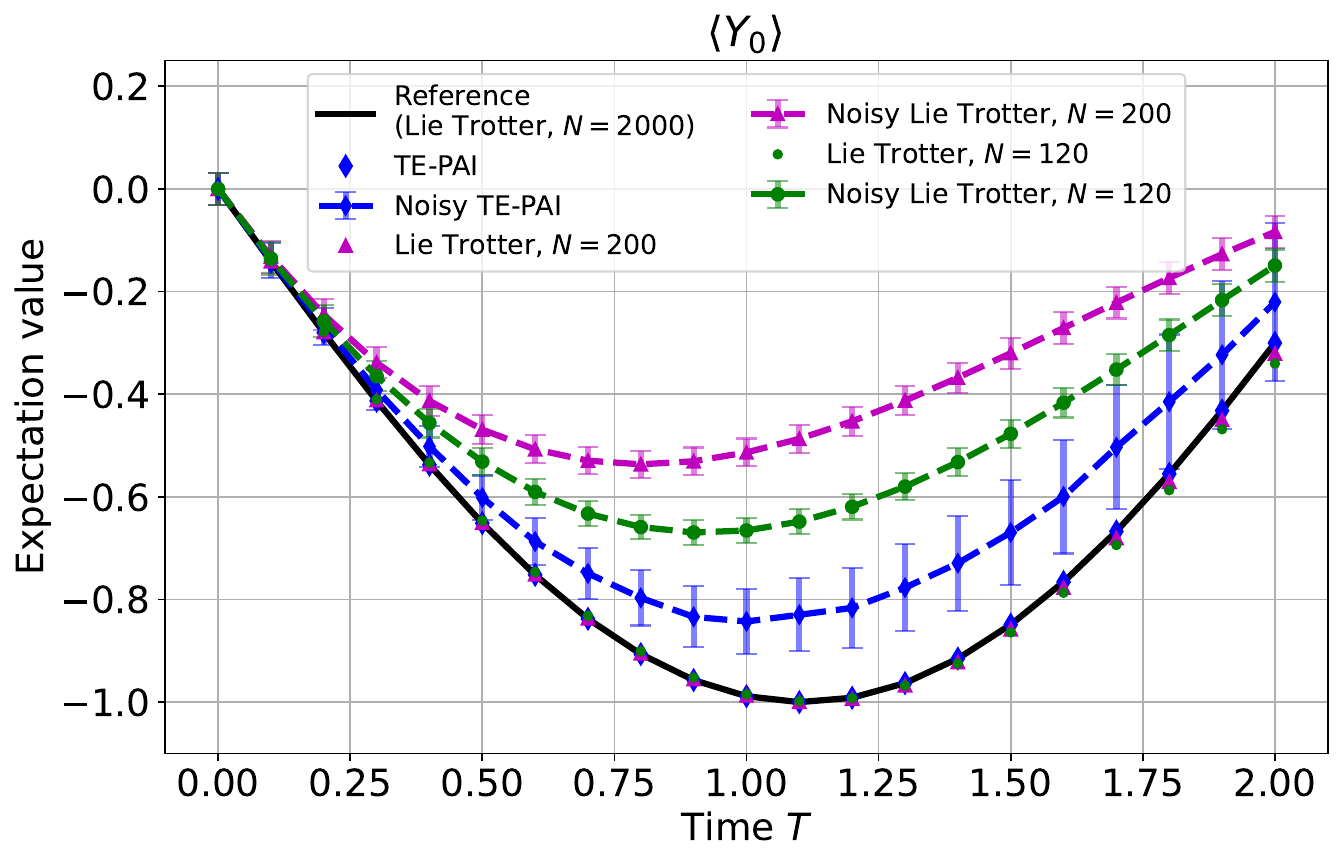}
	\caption{
		Expected value $\langle Y_{0} \rangle$ after time evolution using 1000 circuit repetitions (shots)
		in a 7-qubit spin-ring model from \cref{sec:numeric} using noisy quantum gates.
		Our reference is a noise-free Lie-Trotter simulation with $N=2000$ Trotter steps (grey) consisting of
		$\nu = 56000$ parameterized gates;
		Using a shallower Trotter circuit as $N=120$ layers consisting of $\nu = 3360$ noisy parameterized gates (green)
		achieves a smaller bias than using $N=200$ layers (magenta) consisting of $\nu = 5600$ noisy parameterized gates.
		However, TE-PAI (blue) achieves the smallest bias as it uses a fewer number of noisy quantum gates at
		the expense of an increased statistical uncertainty (increasing error bars).
	}
	\label{fig:noisy_snap}
\end{figure}

In \cref{fig:noisy_snap}, we repeat simulations of the spin-ring model defined in \cref{sec:numeric}
but assuming a noisy 7-qubit system and a rotation angle $\Delta = \frac{\pi}{2^{6}}$
which yields an expected number of gates of $1364$ at $T=2$.
We consider Trotterisation using $N=100$ (green) and $N=200$ layers (magenta). For both $N = 100$ and $N = 200$, the Trotter step sizes are larger than those of the shallow Trotter circuit in \cref{fig:simulation}, but they are still small enough that the Trotter error cannot be ignored in our noiseless simulations. Despite the small number of steps, the resulting circuits remain highly susceptible to noise. We also observe that increasing the number of Trotter layers introduces a more significant bias due to the increased number of noisy gates. In contrast, TE-PAI achieves a smaller bias, i.e., blue is closer to the reference simulation (grey), as it requires fewer noisy quantum gates. This configuration of $\Delta$ enables us to flexibly reduce the circuit depth, thereby enhancing its resilience to noise compared with Trotter circuits.

Additionally, to further reduce the effect of gate noise, TE-PAI can naturally be combined with quantum error mitigation~\cite{RevModPhys.95.045005} and can also be combined with classical shadows as
detailed in ref.~\cite{mitigation_shadow}. Furthermore, TE-PAI estimates an expectation value by executing a large number of
structurally radically different random circuits; such randomisation protocols have been shown to
scramble local gate noise to global depolarising noise---with theoretical proofs for global random circuits~\cite{dalzell2024random}
and numerical evidence for shallow structured circuits~\cite{white_noise}---which allows for very simple
and effective error mitigation by global rescaling.

\section{Fault-tolerant resource estimation \label{sec:impl}}
TE-PAI implements time evolution exactly by averaging outputs of random circuits.
A significant advantage compared to, e.g., Trotterisation, is that
our circuits only have two kinds of rotation angles as $\pm \Delta$ and $\pi$,
and, as we demonstrate in the following, this significantly reduces overall non-Clifford resources required
for fault-tolerant implementations.

In particular, early generations of fault-tolerant quantum computers will likely
be limited by the number of logical qubits and by the achievable circuit depths.
Here, we perform resource estimation for a typical example of simulating the time evolution under a 100-qubit spin Hamiltonian
(which we introduced in \cref{sec:numeric}).
This Hamiltonian has $L=400$ terms, and we fix $T= 1$, $\overline{\| c \|_{1}} = 241.3$.
The expected number of rotation gates using $\Delta = \frac{\pi}{2^8}$ is approximately $39,328$
and the measurement overhead is then about $19.32$. We set the Trotter step to $N=10^{4}$ because, by \cref{eq:trotter_bound} in \cref{app:classical_cost}, this choice guarantees a Trotter error to be $\epsilon_T \le 0.1$. However, such worst-case bounds are typically loose by about an order of magnitude~\cite{trotter_err}, so we expect a error on the order of $10^{-2}$ in practice.

In \cref{app:resource}, we detail how our random circuits can be compiled into a sequence of Clifford gates and
discrete angle, single-qubit $Z$ rotation gates $R_{Z} (\Delta )$. Thus the only non-Clifford
resource we require are single-qubit $Z$ rotation gates, all with the same rotation angle.
In fault-tolerant quantum computing (FTQC), Clifford gates are relatively cheaper
and less error prone than non-Clifford resources. Thus, our focus in this section is to minimize
the implementation cost of our method by efficiently implementing the $R_{Z} (\Delta )$ rotations.

Fault-tolerant machines are emerging; however, there is a range of different hardware platforms with radically different qubit technologies. For example, solid-state platforms (superconducting qubits and quantum dots) have fast gates but relatively noisier qubits, and the leading error-correction approach is the surface code. On the other hand, ion traps and neutral atoms have significantly higher fidelities but slower gate times, and therefore the requirements for error correction can be quite different. Nevertheless, most platforms aim to implement fault-tolerant Clifford gates augmented with magic state teleportation, and therefore we expect the results in this section to apply to a broad range of hardware platforms.

\begin{table*}[tb]
	\centering
	\begin{tabular}{lc@{\hspace{6mm}}c@{\hspace{6mm}}c@{\hspace{6mm}}c}
		\hline
		\hline
		               & Trotter circuit & TE-PAI, direct synthesis & TE-PAI, Hamming weight phasing & TE-PAI, towers \\
		\hline                                                                                                        \\[-2mm]
		T-gates        & $356,000,000$   & $2,438,336 $             & $1,880,980$                    & $298,647$      \\
		storage qubits & --              & --                       & 63                             & 63             \\
		ancilla qubits & --              & --                       & $56$                           & $60$           \\
		\hline
		\hline
	\end{tabular}
	\caption{Resource estimates for the fault-tolerant simulation of a 100-qubit spin-ring Hamiltonian in \cref{eq:spinring}, with $ L = 400 $ Hamiltonian terms, $ T = 1 $, $ \overline{\| c \|_1} = 241.3 $, and $ \Delta = \frac{\pi}{2^8} $. The expected number of $ R_\sigma (\pm \Delta) $ gates is approximately $39,328$, with a measurement overhead of $19.32$. For the Trotter circuit, we use the direct synthesis method assuming $N=10000$ trotter steps.
		While we assumed limited storage space, the efficiency of the Hamming phasing approach would
		asymptotically for large number of storage qubits approach that of the catalyst towers.
	}
\end{table*}

\subsection{Method 1: direct gate synthesis}

The most straightforward approach is a direct gate synthesis whereby the non-Clifford rotation is
decomposed into a sequence of Clifford gates and typically T-gates.
Here we consider approximating $R_{Z} (\Delta )$ to a precision $\epsilon$ deterministically without using ancilla qubits as in~\cite{synthesis0},
which requires $\approx 3.02\log_{2} (1/\epsilon )+1.77$ T-gates on average. Thus, in our
example we estimate approximately $62$ T-gates are required to synthesize
each rotation gate to a precision $\epsilon = 10^{-6}$ which adds up to a total of $2,438,336$ T-gates.
In contrast, performing Trotterisation with $N=10000$ rounds
requires $4,000,000$ rotation gates each with a precision $\epsilon = 10^{-8}$
which adds up to a total of $356,000,000$ T-gates.
As a remark, by using approximate or non-deterministic synthesis, e.g., repeat-until-success synthesis,
the costs can be reduced to $\approx 1.03\log_{2} (1/\epsilon )+5.75$~\cite{synthesis2,synthesis, synthesis_summary}.
In contrast, exact synthesis can also be achieved by randomly choosing from a library of
short T-depth approximate rotations~\cite{koczor2024sparse}.

\subsection{Repeat-Until-Success Methods}
The repeat-until-success approach implements rotation gates by iteratively teleporting the
following resource states as
\begin{equation*}
	\ket{\theta } \equiv R_{Z} (\theta )|+\rangle =\frac{1}{\sqrt{2}}\left( e^{-i\theta /2} |0\rangle +e^{+i\theta /2} |1\rangle \right).
\end{equation*}
With probability $1/2$ the teleportation circuit yields a measurement  outcome $+1$ which
indicates that the qubit is correctly rotated as $R_{Z} (\theta )|\psi \rangle $.
However, with equal chance, it yields a $-1$ measurement outcome which indicates an inverse rotation $R_{Z} (-\theta )|\psi \rangle $.
In the latter case, one needs to apply a rotation gate with twice the angle $R_{Z} (2\theta )$
in order to obtain the desired $R_{Z} (\theta )|\psi \rangle $ net effect.
The approach is therefore repeated iteratively using resource states with angle settings $2^k \theta $
until a $+1$ outcome is achieved, which in general requires on average the following
number of trials as
\begin{equation*}
	1\times \frac{1}{2} +2\times \frac{1}{4} +3\times \frac{1}{8} +\cdots =\sum _{i=1}^{\infty }\frac{i}{2^{i}} =2.
\end{equation*}

Using the Clifford hierarchy, which we define in \cref{app:resource}, is particularly beneficial for our purposes.
In particular, in the following we assume that our rotation angles are of the form
$\Delta _{\ell } =\pi 2^{-\ell +1}$, thus the rotation gate $R_{Z} (\Delta _{\ell } )$
is in the $\ell $-th Clifford hierarchy. The significant advantage is that the repeat until success approach can terminate successfully after
$\ell{-}3$ unsuccessful trials given the $\ell=3$ case is a T-gate which can be applied deterministically
using T-state teleportation.
Since we consider $\Delta\equiv\Delta_{\ell_0=9}$ in the present
example, the probability of termination with a T state teleportation is relatively high as $2^{-6}$.

We note that we can directly prepare Clifford hierarchy states $\ket{\Delta_\ell}$ by distillation using
Reed-Mueller codes~\cite{reed}. This may be beneficial when $\ell$ is sufficiently low,
however for large $\ell$, the cost of directly distilling a resource state $\ket{\Delta_\ell}$ to high precision
will likely exceed the cost of distilling a multiple T-states that can produce $\ket{\Delta_\ell}$
using the techniques we now introduce.
Therefore, we consider alternative approaches that enable the preparation of these resource
states by consuming relatively few T-states~\cite{reed}.

\subsubsection{Method 2: Hamming weight phasing}

Using the Hamming weight phasing method introduced in \cite{hamming-Gidney} and Appendix~A
of \cite{hamming}, we can efficiently apply $n$ equal-angle $R_z(\theta)$ rotations
simultaneously to $n$ qubits.
The technique still uses $\lfloor \log_{2} n+1\rfloor $ directly synthesised, arbitrary-angle
$R_z(\cdot)$ rotations but only uses in addition at most $4( n-1)$ T-gates
(or $n-1$ Toffoli gates) and $n-1$ ancilla qubits. The total required number of T-gates is therefore
\begin{equation}\label{eq:hamming_cost}
	h(n) := C_{\text{syn}}( \epsilon )\left\lfloor\log _2 n+1\right\rfloor+4(n-1),
\end{equation}
where $C_{\text{syn}}( \epsilon )$ is number of T-gates required to synthesise an arbitrary-angle
$R_z(\cdot)$ rotation to precision $\epsilon$ and we assume that $C_{\text{syn }} = 62$
as in Method 1.

While we could directly apply these rotations to the computational qubits, instead we
apply them to $n$ storage qubits to prepare resource states for the repeat-until-success approach
-- then the more storage qubits
are available, the more T-gate savings this approach can yield and in the limit of infinite storage space
the approach could produce $K$ resource states using $4(K-1)$ T-states.

We assume that in the present example we store $n_\ell = 2^{\ell - 4}$ number of $\ket{\Delta_{\ell}}$ resource states, i.e., we use $n_9 = 32$ qubits to store the states $\ket{\Delta_{\ell = 9}}$ and store only one of the $\ket{\Delta_{\ell = 4}}$. We do not assume additional storage qubits for the $\ell=3$ case given this is a T-state which we assume is provided
natively on demand in the fault-tolerant machine.
The total storage space then adds up to $n_9 +n_8+\dots n_4 =  \sum_{\ell=4}^{9} 2^{\ell-4} = 63$ qubits, whereas the total number of ancillary qubits is $(n_9-1) +(n_8-1)+\dots+ (n_4-1) = 57$.

Ultimately, our aim is to power a total of $K=39,328$ rotations using the repeat-until-success approach, thus we need to run the procedure $R = \lceil{K/n_9}\rceil$ times which in total costs $R [ h(n_9) + h(n_8)+ \cdots h(n_4) + n_3 ] $ T-states, where $h(n)$ is defined in \cref{eq:hamming_cost} and for the 3rd level in the hierarchy we use $n_3$ T-states. In total we thus need $1,880,980$ T-gates and $63$ storage qubits.

\subsubsection{Method 3: Catalyst generation of resource states}
A catalyst tower construction was proposed in ref.~\cite{sun2024low}, which builds on refs.~\cite{hamming, gidney2019efficient}.
The central object is a so-called catalyst circuit that
we define in \cref{fig:ct} and which consumes two $\ket{+}$ states, a resource state $\ket{\Delta}$ and a rotation gate $R_z(2\Delta)$, and outputs three resource states $\ket{\Delta}$, thus, in effect applies two $R_z(\Delta)$ rotations at the cost of consuming one $R_z(2\Delta)$ rotation and 4 T-states. Ref.~\cite{sun2024low} then stacked these catalyst circuits so that the overall circuit prepares
a family of resource states $\ket{2^k \Delta}$ by catalysis consuming only
a single rotation $R_z(2^{h} \Delta)$ where $h$ is the height of the tower.

Using our Clifford hierarchy construction with angles $\Delta \equiv \Delta_{\ell_0}$, the tower height can be set to $h=\ell_0-3$, given the rotation angle $2^h \Delta_{\ell_0} = \Delta_3$
can be applied directly by T-state teleportation. Thus, the catalyst towers can be initialised by the family of resource states $\ket{\Delta_{\ell=4\dots\ell_0}}$ (negligible initial cost) and can then continuously produce the required  resource states. Thus, we prepare and store in total $n_\ell =  2^{\ell-4}$ resource states
as in the previous subsection.

In \cref{app:tower}, we explicitly
construct catalyst towers that branch out a relatively large number of
catalysts at the top level to produce $n_\ell$ resource states $\ket{\Delta_{\ell_0}}$
and branch out fewer and fewer at the lower levels, in order to precisely output the desired exponential distribution
$n_{\ell}$ of the resource states $\ket{\Delta_{\ell}}$
required for the repeat-until-success method. Then the total number of ancilla qubits required to
produce the desired distribution $n_\ell = 2^{\ell - 4}$ is
\begin{equation*}
	\left\lceil\frac{2^{\ell_0 -2} -\ell_0 +1}{2}\right\rceil
\end{equation*}
and the total number of T-gates required is
\begin{equation*}
	\begin{cases}
		\left( 2^{\ell_0 } -3\ell_0 +1\right) /2 & \ell_0\text{ is odd}
		\\
		\left( 2^{\ell_0 } -3\ell_0 +6\right) /2 & \ell_0\text{ is even}
	\end{cases}.
\end{equation*}

As in the previous section, our aim is to power a total of $K=39,328$ rotations via the repeat-until-success approach.
Thus, we prepare the $63$ storage qubits in the resource states using
$\lceil{\left( 2^{\ell_0 } -3\ell_0 +1\right) /2}\rceil=243$ T-states in a single round,
and we repeat this procedure $ \lceil{K/n_9}\rceil=1229$ times as in the previous section.
Thus in total, we estimate $298,647$ T-gates are required.



\begin{statement}\normalfont
	Given an arbitrarily chosen $\ell_0$,
	the expected number of rotation gates in \cref{thm:num_gate2} is
	$\nu_{\infty } = \csc (\Delta_{\ell_0} )(3-\cos \Delta_{\ell_0} )\overline{\| c\| _{1}} T$
	and thus
	the expected T cost of implementing a time evolution using our catalyst approach is
	\[N_{\text{Tgate}} = \begin{cases}
			\left( 2^{\ell_0 } -3\ell_0 +1\right) /2^{\ell_0-3} \cdot \nu_{\infty } & \ell_0\text{ is odd,}
			\\
			\left( 2^{\ell_0 } -3\ell_0 +6\right) /2^{\ell_0-3}\cdot \nu_{\infty }  & \ell_0\text{ is even,}
		\end{cases}\]
	using the repeat-until-success approach with
	$\sum_{\ell=4}^{\ell_0} 2^{\ell-4}$   storage qubits that store Clifford hierarchy states.
	It follows that the T cost can be upper bounded given $\ell_0 > 2$ as
	$N_{\text{Tgate}} \leq 8 \nu_\infty$.
\end{statement}

\section{Conclusion and Discussion\label{sec:summary}}

We introduced TE-PAI to estimate observable expected values from effectively exactly time-evolved quantum states.
The approach proceeds by generating a number of random circuits in classical pre-processing, whose outputs
are post-processed to yield on average exact time evolution.
A significant advantage of the approach is that the random circuits are built entirely of Pauli rotations $R_\sigma(\cdot)$ using the Pauli operators $\sigma$ in the system Hamiltonian and using only two kinds of rotation
angles $\Delta$ and $\pi$, which is particularly well-suited for fault-tolerant implementations.
Furthermore, another significant advantage of TE-PAI is that it allows for a highly configurable trade-off
between circuit depth and measurement overhead
by adjusting a single parameter, $\Delta$, offering flexibility to fine-tune.
This feature is particularly useful in NISQ and early-FTQC devices, where circuit depth and
qubit coherence are the primary limitations. In the limit $\Delta \to \pi/2$, TE-PAI reduces to a Clifford circuit that is classically simulable---and Clifford circuits can be trivially exectued in fault tolerant quantum computers---but requires infinite sampling overhead. In contrast, in the limit $\Delta \to 0$ the measurement overhead vanishes at the cost of an infinitely deep quantum circuit. TE-PAI thus provides a continuous interpolation between these extremes and trades off coherent resources for incorherent repetition, i.e., by tuning $\Delta$, one effectively trades classical resources (sampling) against quantum resources (circuit depth) as also relevant in quantum resource theory. Furthermore, we proved that our circuits saturate the Lieb-Robinson bound in the sense that the number of gates required for simulating a total time $T$ is directly proportional to $T$. By construction, TE-PAI respects the Lieb-Robinson bound on information propagation in expectation, so that quantities such as entanglement growth and state-transfer fidelity are reproduced on average as in the ideal evolution. These observations indicate the potential for TE-PAI's randomized structure to be compatible with error-resilient strategies in quantum communication and computation.

Compared to other time-evolution algorithms, the key advantage of this approach is its ability to simulate time evolution without discretisation or algorithmic errors in the sense that finite Trotterisation error can be suppressed efficiently in classical pre-processing to arbitrarily low levels without affecting circuit depth.
This is a particularly powerful feature when the aim is to simulate the evolution under time-dependent Hamiltonians.
Furthermore, we require no ancillary qubits or advanced quantum resources, only the ability
to execute random circuits with rotation angles $\Delta$ and $\pi$, and perform measurements.
The main limitation of the approach is that its measurement overhead potentially grows exponentially
unless the circuit depth is increased with growing system size.
Nonetheless, an approximate version of TE-PAI can interpolate between
the edge cases of exact time evolution with measurement overhead and approximate time evolution
but no measurement overhead
by tuning a continuous trade-off parameter $\lambda$ from~\cite{koczor2024sparse}. Building on this flexibility, future work could explore variance reduction techniques to further reduce sampling cost.

While TE-PAI is well-suited for NISQ applications due to its shallow depth,
we develop particularly efficient fault-tolerant implementations building on the
observation that the only non-Clifford
gates we require are single-qubit $R_z(\Delta)$ rotations with identical rotation angles.
Our architecture prepares resource states $R_z(\Delta)\ket{+}$
using less than 4 T-states via catalyst towers~\cite{sun2024low,hamming-Gidney}
and applies the desired rotations via repeat until success teleportation~\cite{synthesis}.

TE-PAI can also naturally be combined with other randomised quantum protocols. First, our random circuits could be distributed via circuit cutting algorithms and phase estimation algorithms.
For instance, Ref~\cite{circuit_cutting} employs circuit-cutting techniques using the quasiprobabilistic decomposition framework of our method to remove entangling gates rather than to reduce overall circuit depth. Moreover, ~\cite{statistical-phase-estimation} proposes a randomized algorithm for phase estimation that balances gate complexity and shot noise, aiming to eliminate dependence on the sparsity of the Hamiltonian.
Second, TE-PAI can be combined immediately with classical shadows by appending a layer of random measurement-basis
transformation gates~\cite{mitigation_shadow,shadow}. This opens up powerful applications, such as shadow spectroscopy~\cite{chan2022algorithmic}
or subspace expansion using time-evolved trial states~\cite{boyd2024high}. Alternatively, one can also combine the
approach with Pauli grouping techniques~\cite{Crawford2019,jena2019pauli}.
Third, the approach can also be combined with algorithms whereby random initial states are time evolved, such as
when estimating the density of states~\cite{goh2024direct}.
As many of these randomised protocols treat the evolution time $T$ a random variable,
TE-PAI can be used naturally to implement queries to random evolution times.
Furthermore, our random circuits are composed of Pauli operators that appear in the Hamiltonian
and can thus present opportunities for further optimisation through advanced compilation and transpilation tools,
such as 2QAN, which was specifically designed for structurally similar Trotter circuits~\cite{2qan} and can significantly
reduce the circuit depth by parallel execution of non-overlapping gates.

Given its simplicity, our approach is immediately deployable to a broad range of problems with applications
in, e.g., quantum chemistry, materials science, combinatorial optimisation, high-energy physics, optimisation, quantum machine learning etc.,
given time evolution is one of the most important quantum algorithmic subroutines.
\\[2mm]

\noindent \textbf{Data availability:} The simulation code used in this work is available online at \url{https://github.com/CKiumi/te_pai}.

\section*{Acknowledgements}
The authors thank Hayata Morisaki, Simon Benjamin, Gergory Boyd and Zhu Sun for helpful technical conversations.
This work was supported by JST ASPIRE Japan Grant Number JPMJAP2319. BK thanks the University of Oxford for a Glasstone Research Fellowship and Lady Margaret Hall, Oxford for a Research Fellowship. The numerical modelling involved in this study made use of the Quantum Exact Simulation Toolkit (QuEST) \cite{quest} via the QuESTlink\,\cite{questlink} frontend. BK thanks UKRI for the Future Leaders Fellowship Theory to Enable Practical Quantum Advantage (MR/Y015843/1).
The authors also acknowledge funding from the EPSRC projects Robust and Reliable Quantum Computing (RoaRQ, EP/W032635/1) and Software Enabling Early Quantum Advantage (SEEQA, EP/Y004655/1).

\appendix
\section{Comparison to Prior Work\label{app:comparison}}
The \emph{qDRIFT} algorithm~\cite{qDRIFT1} is a randomized Hamiltonian simulation method based on Hamiltonian sparsification. It was originally proposed for a time-independent Hamiltonian $H = \sum_{j=1}^L c_j P_j$ with $\|c\|_1 = \sum_j |c_j|$. At each step, a term $P_j$ is sampled with probability $p_j = |c_j| / \|c\|_1$, and the unitary $e^{-i\,\mathrm{sgn}(c_j) P_j\,\tau}$ is applied with $\tau = T\|c\|_1 / N$, where $T$ is the total simulation time and $N$ is the number of steps. Averaging over the $N$ independent random steps yields a quantum channel whose expectation matches $e^{-iHT}$ up to an additive error $\epsilon$, with gate complexity $\mathcal{O}(\|c\|_1^2 T^2 / \epsilon)$, independent of the number of Hamiltonian terms $L$. The rotation angle $\tau$ is fixed across all steps of the algorithm but depends on the target accuracy $\epsilon$ through the choice of $N$. In contrast, the TE-PAI algorithm with constant measurement overhead achieves gate complexity $\mathcal{O}(\|c\|_1^2 T^2 / Q)$ for a tunable parameter $Q$, incurring a constant overhead factor of $\exp(Q)$. In TE-PAI, the rotation angles are fixed at $\pm\Delta$ and $\pi$, independent of the target accuracy $\epsilon$ (though dependent on $Q$). As a result, TE-PAI can achieve exact simulation with a finite gate count for any fixed $Q$. There are many variants of qDRIFT, such as for time-dependent Hamiltonians~\cite{time-dependent_l1,unified_time-dependent}, which can be compared to TE-PAI in a similar manner.

Next, we compare TE-PAI to a related algorithm developed in ref.~\cite{similar}
which, similar to TE-PAI, simulates the time evolution by
averaging over random quantum circuits.
The approach decomposes the small-angle unitary rotations into linear
combination of unitary matrices which are then randomly sampled to yield an unbiased
estimator. In contrast, TE-PAI obtains an unbiased estimator by
randomising a linear combination of superoperators in \cref{eq:linear_comb}.

This leads to the following differences in practice.
First,  given the randomly generated quantum circuits $U$ and $U^\prime$, and target unitary observable $O$,
the approach of ref.~\cite{similar} proceeds by controlling these quantum circuits on the state of an ancilla qubit
via the Hadamard test as
\[
	\begin{quantikz}
		\lstick{$\ket{0}$} & \gate{H} & \ctrl{1} & \ctrl{1} & \ctrl{1} & \gate{H} & \meter{} & \qw \\
		\lstick{$\ket{\Psi_0}$} & \qw & \gate{U} & \gate{O} & \gate{U^\prime} & \qw & \qw & \qw
	\end{quantikz}\]
In contrast, TE-PAI needs only execute a random circuit $V$ and directly estimate observables
without the need for controlling the circuits as, e.g., in the following implementation
\[
	\begin{quantikz}
		\lstick{$\ket{\Psi_0}$} & \gate{V} & \meter{O}
	\end{quantikz}
\]
For this reason TE-PAI is particularly well suited for near-term applications, such as NISQ or early-FTQC implementations.

Second, the approach of ref.~\cite{similar} requires a different Hadamard-test circuit for each unitary observable $M$.
The significant advantage of TE-PAI is that it is compatible with all advanced measurement techniques and
can thus be used for the simultaneous estimation of multiple observables, e.g., classical shadows, Pauli grouping,
and can naturally be used to directly estimate expected values of non-unitary observables, e.g.,
estimating the probability of a bitstring.

Ref.~\cite{similar} bounded the measurement overhead $\| g\| _{1}^{\infty }$  of the approach
and we find it coincides with the measurement overhead of TE-PAI as
\[\exp\left[ 2\overline{\| c\| _{1}} T\tan\left(\frac{\Delta }{2}\right)\right].\]
Furthermore, the expected number of gates in ref.~\cite{similar}
is $2  \csc (\Delta )\overline{\| c\| _{1}}T$ which is approximately the
same as the number of gates in TE-PAI (assuming small $\Delta$) as
\[\nu _{\infty } =(3-\cos \Delta )\csc (\Delta )\overline{\| c\| _{1}} T.\]

Furthermore, quantum simulation based on truncated Dyson series, as opposed to product formulas, is well-established  in the literature~\cite{dyson-series}. A related work proposed an unbiased random circuit compiler (URCC) based on the Dyson expansion and achieved exact (on average) simulation of time-dependent Hamiltonians using an LCU expansion, continuous unbiased sampling of its terms, and a leading-order rotation to suppress variance~\cite{dyson-base}. Similar to TE-PAI, which uses fixed rotation angles $\Delta$ and $\pi$ (Pauli operator), URCC applies in its leading-order branch a single Pauli rotation with a fixed angle, while its remaining-order branch consists of Pauli operations sampled from the Dyson expansion. Both URCC and TE-PAI achieve comparable finite gate counts and constant sampling overheads that are independent of the target precision. However, URCC requires complex, controlled quantum circuits as
\[
	\begin{quantikz}
		\lstick{$\ket{0}$}      &\gate{H}    &\ctrl{1}            &\octrl{1}                &\gate{H} &\meter{} & \qw \\
		\lstick{$\ket{\Psi_0}$} &\qw         &\gate{u(s)}   &\gate{u(s^\prime)}   &\qw  &\meter{O} & \qw
	\end{quantikz}
\]
where \(u(s)\) and \(u(s^\prime)\) are time-evolution subcircuits sampled from the Dyson series and \(O\) is the observable of interest.

In \cref{sec:impl}, we develop fault-tolerant implementations of TE-PAI that efficiently prepare multiple copies of the resource states required for the fixed-angle rotations $R_Z(\Delta)$. These methods apply immediately to other techniques, such as qDRIFT and URCC, since they likewise rely on the same fixed-angle Pauli rotations discussed above. However, the number of rotation gates required by qDRIFT is $2 \|c\|_1^2 T^2 / \epsilon$ for a target accuracy $\epsilon$.
Assuming a time-independent Hamiltonian with $\|c\|_1 = 241.3$ and $T = 1$ (time-independent version of \cref{sec:impl}), then for a target accuracy of $\epsilon = 10^{-3}$, qDRIFT would require approximately $1.16 \times 10^{8}$ rotation gates, which is about three orders of magnitude more than TE-PAI in a single circuit.

Recent advances in scalable quantum simulation for constrained hardware have focused on trading quantum circuit complexity for increased classical post-processing. Similar to TE-PAI, Harrow and Lowe~\cite{circuit_cutting} propose \emph{circuit cutting} techniques that simulate large circuits by partitioning them into smaller, independent fragments. Their key idea is that entangling unitaries can be replaced by ensembles of local unitaries, expressed via quasiprobabilistic decompositions $U(\rho) = \sum_j p_j \, V_j^{(A)} \otimes W_j^{(B)}(\rho),$ where \( \{p_j\} \) are quasiprobabilities and \( V_j^{(A)}, W_j^{(B)} \) are local unitaries. The reconstructed observable expectation values are given by \( \langle O \rangle = \sum_j p_j \, \mathbb{E}[O_j] \), with sample variance scaling as \( \|p\|_1^2 \). To minimize this overhead, they introduce the \emph{product extent}, and construct optimal protocols using \emph{two-copy Hadamard tests}. One key application is the simulation of \emph{clustered Hamiltonians} with weak inter-cluster coupling.

\section{Summary of Probabilistic Angle Interpolation\label{app:PAI}}

We assume a quantum system comprising of $N$ qubits, and consider parameterised quantum gates
$R(\theta )=e^{-i\theta G/2}$, where $G$ is a Pauli string as $G \in \{\openone ,X,Y,Z\}^{\otimes N}$.
These gates are fundamental to quantum technologies given single and two-qubit rotation gates
are typically engineered as Pauli gates.
Here we briefly review Probabilistic Angle Interpolation (PAI)~\cite{PAI},
which enables these gates to operate at discrete angular settings $\Theta _{k}$ determined by $B$ bits, defined as
\begin{equation*}
	\Theta _{k} =k\Delta ,\ \Delta =\frac{2\pi }{2^{B}} ,\ k\in \{0,1,\dotsc ,2^{B} -1\}.
\end{equation*}
The PAI method effectively allows for any continuous rotation angle to be achieved by overrotating from one of the discrete settings, selecting from three potential notch settings for each gate in a circuit. This approach not only ensures the desired rotation but also maintains a probability distribution centered around the same mean value as would be achieved with infinite angular resolution.
The trade-off, however, is an increased number of circuit repetitions, which grows exponentially as $e^{\nu \Delta ^{2} /4}$ with the number of gates $\nu $. Nevertheless, ref~\cite{PAI} found that at a resolution of $B=7$ bits, the overhead
is reasonable  for circuits containing up to a few thousand parametrized gates, as relevant in non-error corrected machines.

We introduce the following notation for the superoperators of the aforementioned discrete-angle rotation gates as
\begin{equation*}
	R_{1} :=R( \Theta _{k}) ,R_{2} :=R( \Theta _{k+1}) ,R_{3} :=R( \Theta _{k} +\pi ),
\end{equation*}
then any overrotation $R( \Theta _{k} +\theta )$ by a continuous angle $\theta < \Delta$ can be expressed
as a linear combination of the discrete gates as
\begin{equation*}
	\mathcal{R}( \Theta _{k} +\theta ) =\gamma _{1} (\theta )\mathcal{R}_{1} +\gamma _{2} (\theta )\mathcal{R}_{2} +\gamma _{3} (\theta )\mathcal{R}_{3} .
\end{equation*}
By solving a system of trigonometric equations, ref~\cite{PAI} obtained
the analytic form of the coefficients $\gamma _{l} (\theta )$ as
\begin{equation*}
	\begin{aligned}
		 & \gamma _{1} (\theta )=\csc\left(\frac{\Delta }{2}\right)\cos\left(\frac{\theta }{2}\right)\sin\left(\frac{\Delta }{2} -\frac{\theta }{2}\right) , \\
		 & \gamma _{2} (\theta )=\csc (\Delta )\sin (\theta ),                                                                                               \\
		 & \gamma _{3} (\theta )=-\sec\left(\frac{\Delta }{2}\right)\sin\left(\frac{\theta }{2}\right)\sin\left(\frac{\Delta }{2} -\frac{\theta }{2}\right),
	\end{aligned}
\end{equation*}
as a function of the continuous-angle $\theta $.
We can also analytically compute the vector norm as
\begin{equation*}
	\| \gamma \| _{1} =\sec\left(\frac{\Delta }{2}\right)\cos\left(\frac{\Delta }{2} -\theta \right).
\end{equation*}

Analogously to quasiprobability sampling methods
which mitigate non-unitary error effects, PAI randomly samples the discrete rotation gates according to the above
weights. In particular, we randomly choose one of the three discrete gate variants $\{\mathcal{R}( \Theta _{k}) ,\mathcal{R}( \Theta _{k+1}) ,\mathcal{R}( \Theta _{k} +\pi )\}$ according to the probabilities $p_{l} (\theta )=| \gamma _{l} (\theta )| /\| \gamma (\theta )\| _{1}$ which yields the unbiased estimator of the rotation gate as
\begin{equation*}
	\hat{\mathcal{R}}( \Theta _{k} +\theta ) =\| \gamma (\theta )\| _{1}\operatorname{sign}[ \gamma _{l} (\theta )] \mathcal{R}_{l} ,
\end{equation*}
such that $\mathbb{E}[\hat{\mathcal{R}}( \Theta _{k} +\theta )] =\mathcal{R}( \Theta _{k} +\theta )$.
Ref~\cite{PAI} proved that PAI is optimal in the sense that the choice of the three discrete angle settings
globally minimises the measurement overhead characterised by $\| \gamma (\theta )\| _{1}$.

We now briefly summarise Statement 2 of \cite{PAI} which is concerned with applying the PAI protocol
to each continuous-angle rotation in a circuit.
To simplify notations we assume a circuit $\mathcal{U}_{\text{circ}}$ that contains no other gates than $\nu$ parametrised ones as
\begin{equation*}
	\mathcal{U}_{\text{circ}} =\prod _{j=1}^{\nu } \mathcal{R}^{(j)}( \Theta _{k_{j}} +\theta _{j}),
\end{equation*}
however, it is straightforward to generalise to the case when the circuit contains other non-parametrised gates too.
Here $\mathcal{R}^{(j)}$ denotes the $j^{th}$ parametrised gate which is ideally
set to the continuous-angle that we express as an over rotation by an angle
$\theta _{j}$ relative to the notch setting $\Theta _{k_{j}}$.
Let us, denote the discrete rotations as
\begin{equation*}
	\begin{aligned}
		 & \mathcal{R}_{1}^{(j)} :=\mathcal{R}^{(j)}( \Theta _{k_{j}}) ,\ \ \mathcal{R}_{2}^{(j)} :=\mathcal{R}^{(j)}( \Theta _{k_{j} +1}), \\
		 & \mathcal{R}_{3}^{(j)} :=\mathcal{R}^{(j)}( \Theta _{k_{j}} +\pi ) .
	\end{aligned}
\end{equation*}
At each execution of the circuit, we randomly replace a parametrised gate with the corresponding discrete gate variant, i.e, the $j^{th}$ parametrised gate is replaced by one of the discrete gate variants $\mathcal{R}_{l_{j}}^{(j)}$, according to the probability distribution $p_{l_{j}}( \theta _{j})$. Given a circuit $U_{\text{circ}}$ of $\nu $ parametrised gates, we choose a multi index $\underline{l} \in 3^{\nu }$ according to the probability distribution $p(\underline{l} )=| g_{\underline{l}}| /\| g\| _{1}$ where $g_{\underline{l}}$
\begin{equation*}
	g_{\underline{l}} =\prod _{j=1}^{\nu } \gamma _{l_{j}}^{(j)}( \theta _{j}) .
\end{equation*}
We obtain an unbiased estimator of the ideal circuit as
\begin{equation*}
	\hat{\mathcal{U}}_{\text{circ}} :=\| g\| _{1}\operatorname{sign}( g_{\underline{l}}) \mathcal{U}_{\underline{l}} ,
\end{equation*}
where $\| g\| _{1} =\prod _{j=1}^{\nu }\left\Vert \gamma ^{(j)}\right\Vert _{1}$.
PAI then executing the circuit variants $U_{\underline{l}}$ in which all continuously
parametrised gates are replaced by the discrete ones according to the multi index $\underline{l}$.
This yields an unbiased estimator of the ideal circuit in the sense that $\mathbb{E}[\hat{\mathcal{U}}_{\text{circ}}] =\mathcal{U}_{\text{circ}}$.


After performing a measurement, one multiplies the random outcome with a factor $\| g\| _{1}\operatorname{sign}( g_{\underline{l}})$ that can have negative signs. As a consequence, the variance of the estimator is magnified which implies an increased number of circuit repetitions. Applying PAI to the estimation of an expected value results in an unbiased estimator $\hat{o}$ of the expected value of an observable as $\mathbb{E} [\hat{o} ]=\operatorname{Tr}[ O \mathcal{U}_{\text{circ}} |0\rangle \langle 0|] =o$ . The number of repetitions required to determine the expected value o to accuracy $\epsilon $ scales as
\begin{equation*}
	N_{s} \leq \epsilon ^{-2} \| g\| _{1}^{2},
\end{equation*}
which is increased by the measurement overhead factor $\| g\| _{1}^{2}$
compared to when having access to the ideal unitary.

\section{Classical pre-processing resources \label{app:classical_cost}}
We have finite $N$ in classical pre-processing, and we want to bound the classical pre-processing resources.
An important feature of TE-PAI is that the quantum circuit depth is independent of the Trotter number $N$; increasing $N$ only affects the classical pre-processing resources required to generate the unbiased estimators. In this subsection, we rigorously bound these classical resources.

For a fixed $N$, the additive Trotter error of a first-order product formula is bounded as in \cref{prop:trotter_error},
\begin{equation*}
	\epsilon_T \leq \frac{T^2}{2 N}\,\|c\|^2_T ,
\end{equation*}
where $\|c\|_T$ denotes the commutator norm defined in \cref{eq:commutator_norm}. This error quantifies the algorithmic discrepancy between the exact unitary and the product formula at finite $N$. We recall that the first-order Trotter error involves a sum of commutators. Since for any pair of Pauli operators $\|[P_i,P_j]\|\le 2$, we obtain
\begin{align*}
	\|c\|_T^2
	 & \le 2\sum_{i<j} |c_i|\,|c_j| \nonumber                              \\
	 & = \Bigl(\sum_{j=1}^L |c_j|\Bigr)^2 - \sum_{j=1}^L |c_j|^2 \nonumber \\
	 & = \|c\|_1^2 - \|c\|_2^2 .
\end{align*}
Therefore, a simple universal bound is
\begin{equation*}
	\epsilon_T \le \frac{T^2}{2N}\,\|c\|_1^2,
\end{equation*}
which only requires the easily computable $\ell_1$ norm of the Hamiltonian coefficients.

In practice, we require the algorithmic error $\epsilon_T$ to be negligible compared to the statistical error $\epsilon$ of the estimated expected value. We may demand, for some constant $\kappa \geq 1$ which expresses how much algorithmic errors are below the statistical uncertainty, that
\begin{equation*}
	\epsilon_T \leq \kappa^{-1}\, \epsilon.
\end{equation*}
This condition ensures that algorithmic errors are several orders of magnitude smaller than the statistical uncertainty. Substituting the bound for $\epsilon$ yields a constraint on the Trotter number,
\begin{equation*}
	N  \geq \frac{T^2 \|c\|_1^2}{2 \kappa^{-1} \epsilon}.
\end{equation*}
Thus, a sufficiently large $N$ guarantees that TE-PAI is effectively exact to the desired precision. Therefore, the classical pre-processing cost $\mathcal{C}$, which corresponds to the number of classical samples required to choose an angle from three candidates, becomes
\begin{equation*}
	\mathcal{C} :=  N L N_s\leq \tfrac{1}{2}\kappa \|g\|_{1}^{2} \|c\|_1^2 L T^2  \epsilon^{-3}=\mathcal{O}(\|g\|_{1}^{2} \|c\|_1^2  L T^2\epsilon^{-3}).
\end{equation*}

Let us illustrate this on the example of the spin-ring Hamiltonian we consider in the main text whereby each qubit is touched by seven operators, but only ten pairs per site actually anticommute, giving a maximum commutator weight of $20$. Summing over all sites yields
\begin{equation}\label{eq:trotter_bound}
	\|c\|^2_T \le 2\bigl(4\sum_{k=1}^n|\omega_k|+6n\bigr) \le 20n.
\end{equation}
In our numerical demonstrations, we considered a time evolution of $T = 1$ with $N_s = 1000$ shots. To guarantee that the algorithmic error is within the statistical error, the above bound implies that one must take $N \geq 140 \sqrt{1000}\,\kappa \approx 4427\kappa$. However, in practice the actual Trotter error is often much smaller than the above theoretical bound. Thus, in our simulation in \cref{sec:numeric}, we increased the number of Trotter steps $N$ until the Trotter error became negligible, so that the results matched the exact curve within statistical error bounds. In this way, we found that $N=1000$ is sufficient in practice.

\section{Proof of expected number of gates\label{app:proof2}}
\subsection{Mean value}
\begin{figure}
	\centering
	\includegraphics[width=0.5\textwidth]{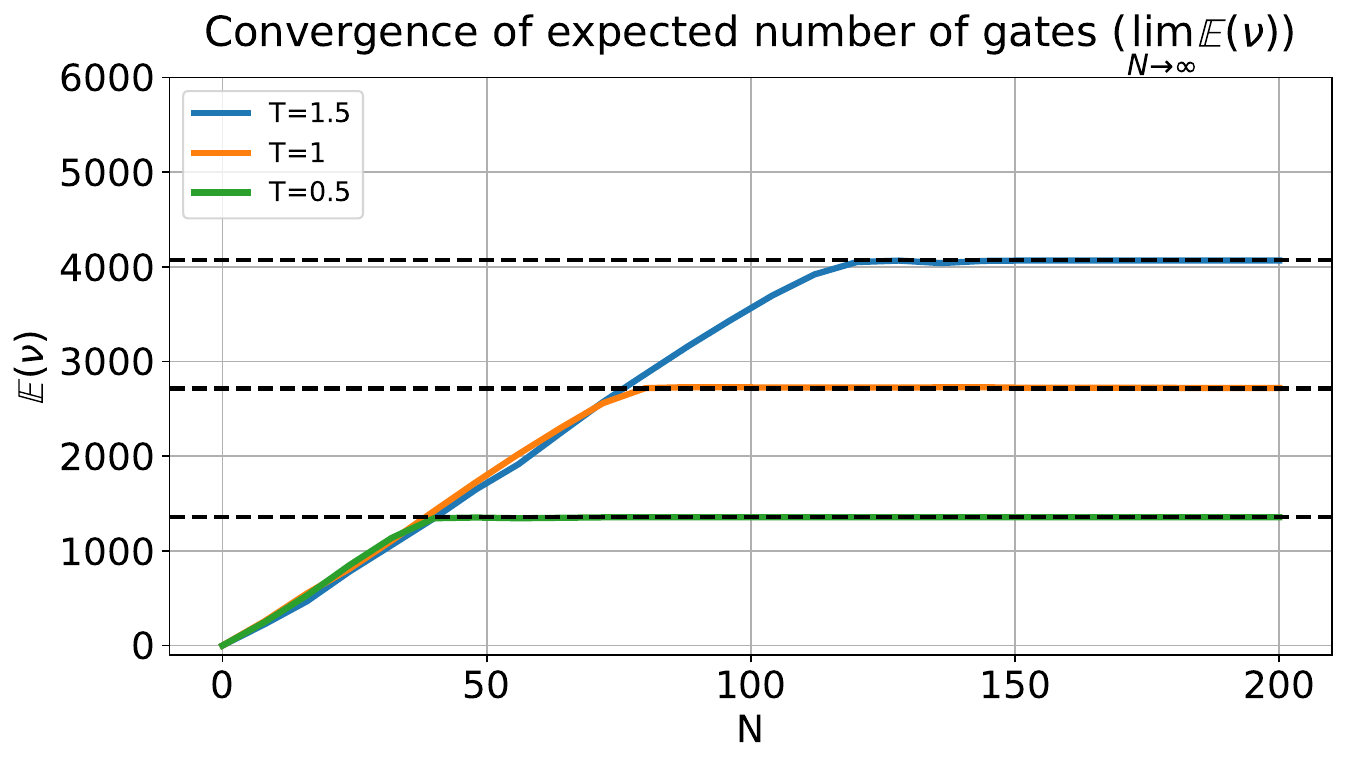}
	\caption{Expected number of gates by \(N\) for the time-dependent Trotter circuit introduced in Section \ref{sec:numeric}, where $\Delta=\pi/2^7$, $T=0.5,\ 1.0,\ 1.5$ and \(\overline{\| c \|_{1}} T \approx 16.654,\ 33.308,\ 49.963\), respectively. We observe that the Expected number of gates approaches its limit value \(\nu_\infty\) as \(N\) increases and converges to it.}
	\label{fig:num_gates_N}
\end{figure}

\begin{proof}
	We now detail our derivation of the expected number $\mathbb{E} (\nu )$
	of gates in TE-PAI circuits.
	As PAI replaces each continuously parametrised gate
	in a circuit with one of three discrete gate variants, and because one of those three options is the identity
	operation which does not increase the number of gates, the number of gates
	is a Bernoulli distributed random variable. More specifically, the operational dynamics of the gates
	are defined such that at position $k$ and $j$, exactly one gate is added to the circuit with probability
	$1-p_{1} (|\theta _{kj}|)$ (either $\pm \Delta$ rotation or $\pi$ rotation ), while the
	identity operation is selected with probability $p_{1} (|\theta _{kj}|)$.
	Given that we implement this selection process across $N$ time steps and $L$ different gate types,
	we effectively conduct $NL$ independent trials. This allows us to compute the expected total number of gates as
	\begin{equation*}
		\mathbb{E} (\nu )=\sum _{j=1}^{N}\sum _{k=1}^{L} (1-p_{1} (|\theta _{kj}|)).
	\end{equation*}
	We express the probability $p_{1}$ in terms of the gate parameters $|\theta _{kj}|$, where:
	\begin{equation*}
		p_{1} (|\theta _{kj}|)=\frac{\sin (\Delta -|\theta _{kj}|)+\sin (\Delta )-\sin (|\theta _{kj}|)}{2(\sin (\Delta -|\theta _{kj}|)+\sin (|\theta _{kj}|))} .
	\end{equation*}
	Expanding these for large $N$ we obtain the series for $p_{1}$ as
	\begin{equation*}
		\begin{aligned}
			p_{1} (|\theta _{kj}|)  = & 1-\frac{1}{2} (3-\cos \Delta )\csc (\Delta )|\theta _{kj}|                        \\
			                          & + \left(\sec ^2 \left(\frac{\Delta}{2}\right)-\frac{1}{4}\right) |\theta _{kj}|^2
			+O\left( |\theta _{kj}|^{3}\right)                                                                            \\
			=                         & 1-(3-\cos \Delta )\csc (\Delta )|c_{k} (t_{j})|\frac{T}{N} + \mathcal{D}_{kj},
		\end{aligned}
	\end{equation*}
	where in the second equation we introduced the error term
	\begin{align*}
		\mathcal{D}_{kj} & := \left(\sec ^2 \left(\frac{\Delta}{2}\right)-\frac{1}{4}\right) |\theta _{kj}|^2
		+ O\left( |\theta _{kj}|^{3}\right)                                                                   \\
		                 & \leq 8 |c_{k} (t_{j})|^2\frac{T^2}{N^2} + O\left( N^{-3}\right)
	\end{align*}
	that we bound using
	$\left(\sec ^2 \left(\frac{\Delta}{2}\right)-\frac{1}{4}\right) \leq 2$ given that  $\Delta \leq \pi/2$,
	and substitute in the rotation angles $|\theta _{kj}| =2 |c_{k} (t_{j})|\frac{T}{N}$.

	With this approximation, the expected number of gates can be expanded for large $N$ as
	\begin{align}\nonumber
		\mathbb{E} (\nu ) & =			\sum _{j=1}^{N}\sum _{k=1}^{L}\left( (3-\cos \Delta )\csc (\Delta )|c_{k} (t_{j} )|\frac{T}{N} - \mathcal{D}_{kj} \right)                 \\
		                  & =\csc (\Delta )(3-\cos \Delta )\sum _{k=1}^{L}\sum _{j=1}^{N}\left( |c_{k} (t_{j} )|\frac{T}{N}\right)    - \mathcal{E} ,\label{eq:summation}
	\end{align}
	where we denoted the sum of individual error terms as
	\begin{equation*}
		\mathcal{E} := \sum _{k=1}^{L}\sum _{j=1}^{N}\mathcal{D}_{kj}
		\leq 8\sum _{k=1}^{L}\sum _{j=1}^{N}  |c_{k} (t_{j})|^2\frac{T^2}{N^2} + O\left( N^{-2}\right).
	\end{equation*}

	We can relate the summation in \cref{eq:summation} to the Riemann integration in \cref{eq:l1-norm}   as
	\begin{equation}\label{eq:riemann_sum_error}
		\int _{0}^{T}\sum _{k=1}^{L}| c_{k} (t)| \, \mathrm{d} t
		= \sum _{k=1}^{L}\sum _{j=1}^{N}\left( |c_{k} (t_{j} )|\frac{T}{N}\right) + \mathcal{Q},
	\end{equation}
	where we can bound the error term as $|\mathcal{Q}| \leq V_{tot}T/N$.
	In the present work we assume that the time-dependent Hamiltonian coefficients
	$c_{k} (t)$ are absolutely continuous functions of time and therefore have bounded variation
	with $V_{tot} = \sum_k V(c_{k}) < \infty$ being a finite constant. In
	case if $c_{k}(t)$ is differentiable, then $V(c_{k})$ is upper bounded by the absolute largest value of the derivative function
	$c'_{k}(t)$, but our results also apply to functions that are not necessarily differentiable everywhere.
	This follows from results of Refs.~\cite{owens2014exploring,tasaki2009convergence} which proved that the error term in the Riemann summation
	$\int_{a}^b f(x) \mathrm{d}x = \sum_{k=1}^{N}f(k \Delta_N) \Delta_N +Q$ is upper bounded as $|Q| \leq V \Delta_N  $  using the notation
	$\Delta_N=(b-a)/N$ and $V$ is the absolute largest variation $V$ of the function.

	Therefore, we finally obtain the explicit expression as
	\begin{align*}
		\mathbb{E} (\nu ) & =\csc (\Delta )(3-\cos \Delta )\sum _{k=1}^{L}\sum _{j=1}^{N}\left( |c_{k} (t_{j} )|\frac{T}{N}\right) - \mathcal{E} \\
		                  & =\csc (\Delta )(3-\cos \Delta )\sum _{k=1}^{L}\int _{0}^{T} |c_{k}(t)| \,\mathrm{d}t + \mathcal{Q}' - \mathcal{E}
		\\
		                  & =\csc (\Delta )(3-\cos \Delta )\overline{\| c\| _{1}} T + \mathcal{Q}' - \mathcal{E}.
	\end{align*}
	Here we can bound the two error terms using the notation  $\mathcal{Q}' = \csc (\Delta )(3-\cos \Delta ) \mathcal{Q} $
	as
	\begin{align*}
		|\mathcal{Q}' - \mathcal{E} | & \leq  |\mathcal{Q}'| + |\mathcal{E}|                                                  \\
		                              & \leq   \frac{3 \csc (\Delta ) T V_{tot}} {N} +
		8 \sum _{k=1}^{L}\sum _{j=1}^{N} |c_{k} (t_{j})|^2\frac{T^2}{N^2} + O\left( N^{-2}\right)                             \\
		                              & =  \frac{ 3 \csc (\Delta ) T V_{tot}} {N} + 	8 \overline{\| c\|^2 _{2}} \frac{T^2}{N}
		+ O\left( N^{-2}\right).
	\end{align*}
	In the last equation, we again replaced the Riemann summation with an integral to obtain the mean
	of the square of the
	$\ell_2$ norm of the coefficient vector $ \overline{\| c\|^2 _{2}}$ up to an error $O\left( N^{-2}\right)$.

	Finally, we conclude our proof as
	\begin{equation}
		\mathbb{E} (\nu ) = \csc (\Delta )(3-\cos \Delta )\overline{\| c\| _{1}} T   + O\left( N^{-1}\right)
	\end{equation}
	In \cref{fig:num_gates_N} we numerically plot  $\mathbb{E} (\nu )$ for increasing \(N\) for the time-dependent Hamiltonian introduced in Section \ref{sec:numeric} and confirm that indeed it converges to the above expression.

\end{proof}

\subsection{Minimal number of gates}
We can also compute the minimum of the number of gates over $\Delta$.
From Theorem \ref{thm:overhead}, we already knew that the sampling cost is upper bounded by $\exp Q$ if we set $\Delta =2\arctan\left(\frac{Q}{2\overline{\| c\| _{1}} T}\right)$. Thus, Substituting this angle yields
\begin{equation*}
	\lim _{N\rightarrow \infty }\mathbb{E} (\nu )=\frac{2(\overline{\| c\| _{1}} T)^{2}}{Q} +Q.
\end{equation*}
Noting that we used the following relations to $(3-\cos \Delta )\csc (\Delta )$.
\begin{equation*}
	\sin (2\arctan (x))=\frac{2x}{1+x^{2}} ,\ \cos (2\arctan (x))=\frac{1-x^{2}}{1+x^{2}} .
\end{equation*}
From the assessment of the arithmetic-geometric mean, the expression $\frac{2(\overline{\| c\| _{1}} T)^{2}}{Q} +Q$ takes its minimum value of $\overline{\| c\| _{1}} T\,2\sqrt{2}$ when $Q=\overline{\| c\| _{1}} T \sqrt{2}$ i.e., $\Delta =2\tan\left(\frac{1}{\sqrt{2}}\right)$.

\begin{figure}
	\centering
	\includegraphics[width=0.5\textwidth]{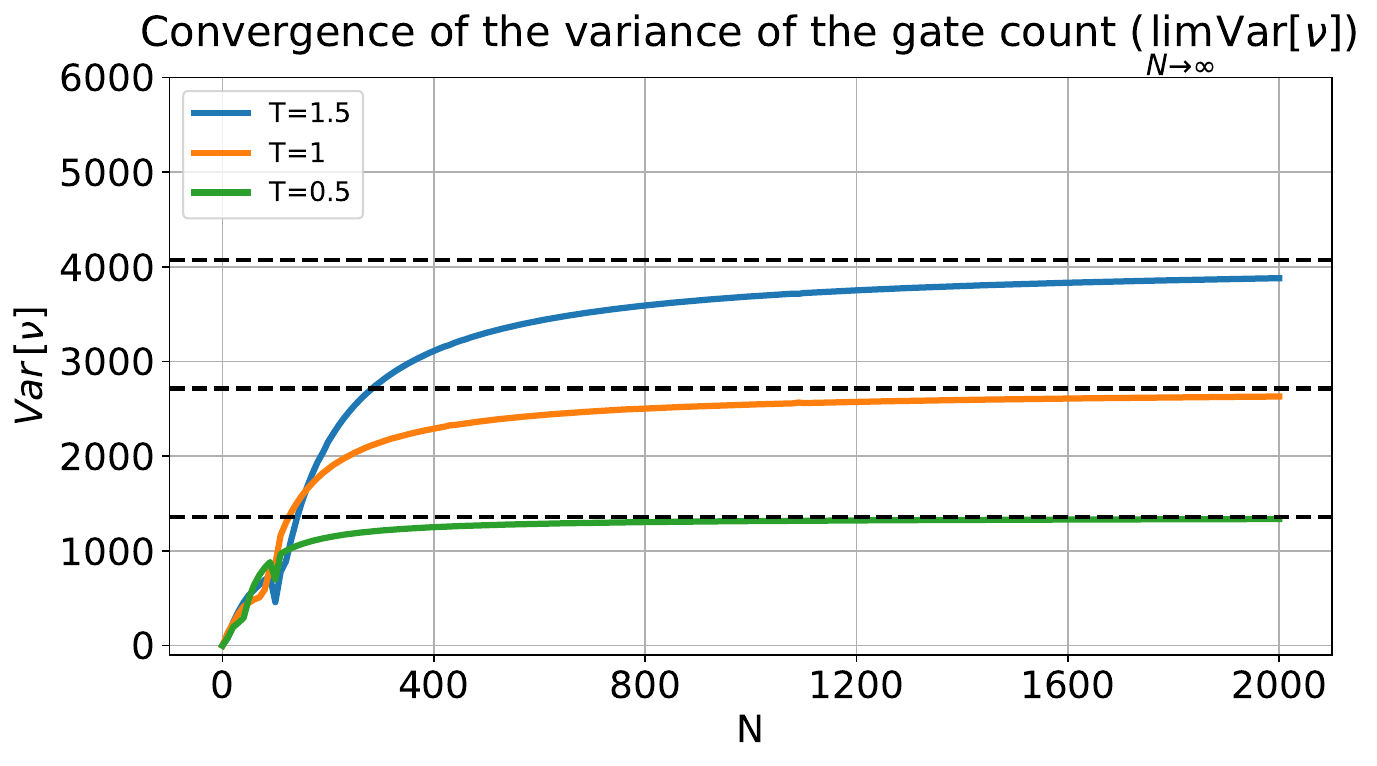}
	\caption{Variance of the gate count by \(N\) for the time-dependent Trotter circuit introduced in Section \ref{sec:numeric}, where $\Delta=\pi/2^7$, $T=0.5,\ 1.0,\ 1.5$ and \(\overline{\| c \|_{1}} T \approx 16.654,\ 33.308,\ 49.963\), respectively. We observe that the variance of the gate count approaches its limit value \(\nu_\infty\) as \(N\) increases and converges to it.}
	\label{fig:var_gates_N}
\end{figure}

\subsection{Variance of the number of gates}
The variance of the number of gates can be computed analyitically given
each position follows a Bernoulli distribution for which the variance is $p(1-p)$ and thus
\begin{equation*}
	\operatorname{Var} [\nu ]=\sum _{j=1}^{N}\sum _{k=1}^{L} p_{1} (|\theta _{kj}|)(1-p_{1} (|\theta _{kj}|)).
\end{equation*}
Here the total variance of the number of gates is upper bounded by the expectation value of the number of gates as
\begin{equation*}
	\operatorname{Var} [\nu ]\leq \sum _{j=1}^{N}\sum _{k=1}^{L} (1-p_{1} (|\theta _{kj}|))=\mathbb{E} [\nu ]
\end{equation*}

For large $N$, the variance can be simplified as follows:
\begin{align*}
	\operatorname{Var} [\nu ] & =\sum _{j=1}^{N}\sum _{k=1}^{L}\csc (\Delta )(3-\cos \Delta )|c_{k} (t_{j} )|\frac{T}{N} +O\left( N^{-2}\right) \\
	                          & =\csc (\Delta )(3-\cos \Delta )\overline{\| c\| _{1}} T + O(N^{-1})
\end{align*}
Above we do not detail all steps as we used the argument as in \cref{eq:riemann_sum_error} to replace the summation with an integral,
and to bound the error term as $O(N^{-1})$. This leads to:

\begin{equation*}
	\lim _{N\rightarrow \infty }\operatorname{Var} [\nu ]=\csc (\Delta )(3-\cos \Delta )\overline{\| c\| _{1}} T=\lim _{N\rightarrow \infty }\mathbb{E} [\nu ].
\end{equation*}
As such, since we consider large $N$, By Lyapunov Central Limit Theorem, we can approximate the distribution of gate numbers well by the normal distribution:
\begin{equation*}
	\mathcal{N} (\mathbb{E} [\nu ],\mathbb{E} [\nu ] )
\end{equation*}

We numerically plot the variance of the number of gates in \cref{fig:var_gates_N} for the time-dependent Hamiltonian introduced in \cref{sec:numeric}, respectively, and confirm
convergence to the above expressions.

\begin{figure}
	\centering
	\includegraphics[width=0.5\textwidth]{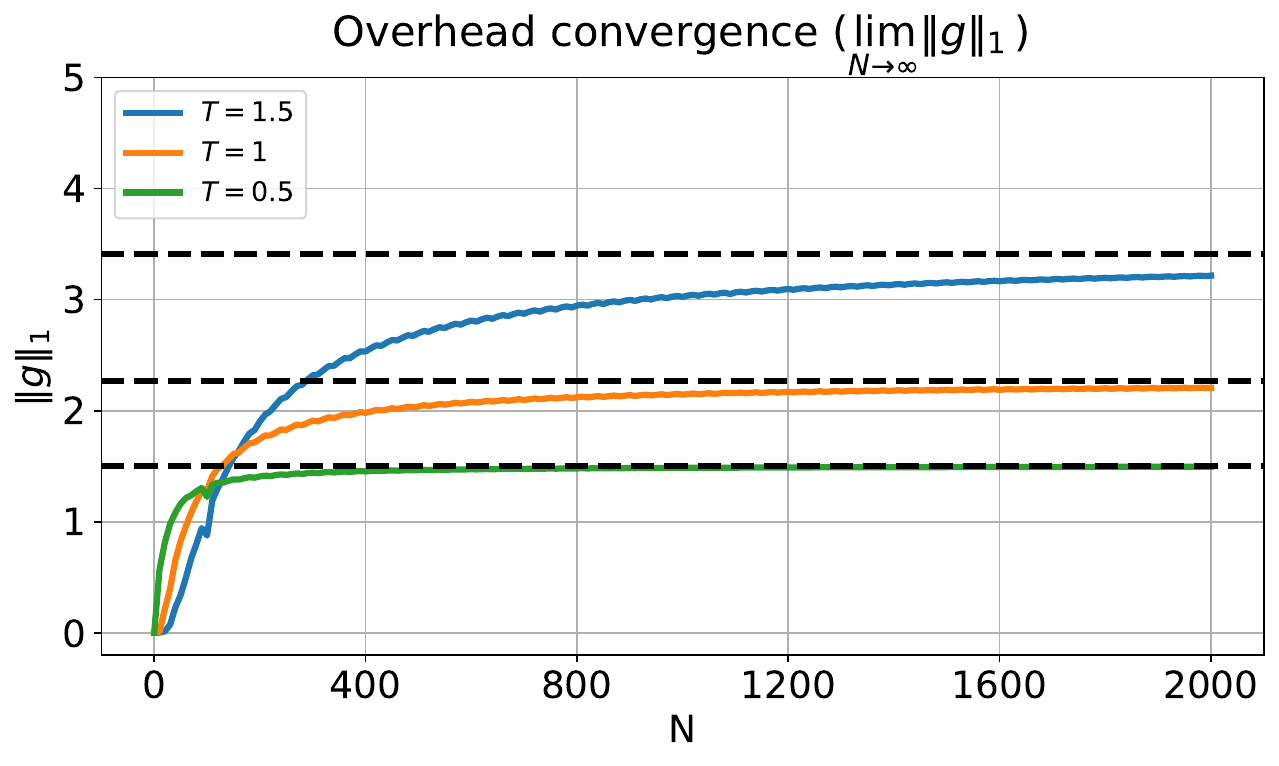}
	\caption{Measurement overhead by \(N\) for the time-dependent Trotter circuit introduced in Section \ref{sec:numeric}, where $\Delta=\pi/2^7$, $T=0.5,\ 1.0,\ 1.5$ and \(\overline{\| c \|_{1}} T \approx 16.654,\ 33.308,\ 49.963\), respectively. We observe that the overhead approaches its limit value \(\|g\|^\infty_1\) as \(N\) increases and converges to it.}
	\label{fig:overhead_N}
\end{figure}

\section{Proof of measurement overhead\label{app:proof1}}
\begin{proof}

	In this section we provide a detailed derivation of the measurement overhead in probabilistic angle interpolation (PAI) applied within Trotter circuits for simulating time-dependent Hamiltonian systems. As we detailed above, the measurement
	overhead of PAI is bounded by the following expression as
	\begin{equation*}
		\| g\| _{1} =\prod _{j=1}^{N}\prod _{k=1}^{L}\Vert \gamma _{k}( |\theta _{kj}|)\Vert _{1},
	\end{equation*}
	which quantifies the cumulative measurement overhead of the circuit by considering the overhead introduced by individual rotation gates.

	The measurement overhead for each gate is given by:
	\begin{equation*}
		\begin{aligned}
			\Vert \gamma ( |\theta _{kj}|)\Vert _{1} & =\sec\left(\frac{\Delta }{2}\right)\cos\left(\frac{\Delta }{2} -|\theta _{kj}|\right) \\
			                                         & =\cos( |\theta _{kj}|) +\sin(| \theta _{kj}|)\tan\left(\frac{\Delta }{2}\right).
		\end{aligned}
	\end{equation*}

	We again consider the limit of large $N$ and thus small angles $\theta _{k}( t)$,
	and obtain the series expansion as
	\begin{equation*}
		\Vert \gamma ( |\theta _{kj}|)\Vert _{1} =1+\tan\left(\frac{\Delta }{2}\right) |\theta _{kj}| + \mathcal{D}_{kj},
	\end{equation*}
	where also substitute in our expression for the angles  $\theta _{kj}$
	and introduce the error term $\mathcal{D}_{kj} \in O(N^{-2})$ as
	\begin{equation*}
		\mathcal{D}_{kj} = - |\theta _{kj}|^2 /2 +O(\theta _{kj}^3) = -2|c_{k}( t_{j})|^2\frac{T^2}{N^2} {+}O\left( N^{-3}\right).
	\end{equation*}

	The total measurement cost $\| g\| _{1}$ can thus be evaluated by expanding each term in the product as
	\begin{equation*}
		\| g\| _{1}  =\prod _{j=1}^{N}\prod _{k=1}^{L}\left[ 1+2\tan\left(\frac{\Delta }{2}\right) |c_{k}( t_{j})|\frac{T}{N} + \mathcal{D}_{kj} \right].
	\end{equation*}

	We now take the logarithm of $\| g\| _{1}$, which allows us to convert the product into a sum as
	\begin{align}
		\nonumber
		\log \| g\| _{1} & =\sum _{j=1}^{N}\sum _{k=1}^{L}\log\left[ 1+2\tan\left(\frac{\Delta }{2}\right) |c_{k}( t_{j})|\frac{T}{N} + \mathcal{D}_{kj}  \right] \\
		\nonumber
		                 & =\sum _{j=1}^{N}\sum _{k=1}^{L}\left[ 2\tan\left(\frac{\Delta }{2}\right) |c_{k}( t_{j})|\frac{T}{N} + \mathcal{D}_{kj} \right]        \\
		                 &
		= 2\tan\left(\frac{\Delta }{2}\right) \sum _{j=1}^{N}\sum _{k=1}^{L}  |c_{k}( t_{j})|\frac{T}{N}
		+ \mathcal{E},         \label{eq:summation_for_variance}
	\end{align}
	where in the second equation we used the expansion $\log(1+ x+\epsilon) = \log(1+x)+\epsilon/(1+x) + O(\epsilon^2)$ and $\log(1+ x) = x - x^2/2 + O(x^3)$
	for small $\epsilon=\mathcal{D}_{kj}\in O(N^{-2})$ and $x = 2\tan\left(\frac{\Delta }{2}\right) |c_{k}( t_{j})|\frac{T}{N} \in O(N^{-1})$, respectively.
	We introduce the error term as
	\begin{equation}
		\mathcal{E} := \sum _{j=1}^{N}\sum _{k=1}^{L} \mathcal{D}_{kj}.
	\end{equation}
	We can again replace the summation in \cref{eq:summation_for_variance} with Riemann integration
	as introduced in  \cref{eq:riemann_sum_error} as
	\begin{equation}
		\sum _{j=1}^{N}\sum _{k=1}^{L}  |c_{k}( t_{j})|\frac{T}{N}  =  \sum _{k=1}^{L}\int _{0}^{T} |c_{k}(t)| \, \mathrm{d}t +\mathcal{Q},
	\end{equation}
	up to the error term 	$|\mathcal{Q}| \leq V_{tot} T/N$.
	This yields
	\begin{equation*}
		\begin{aligned}
			\log \| g\| _{1}  = & 2\tan\left(\frac{\Delta }{2}\right)
			\sum _{k=1}^{L}\int _{0}^{T} |c_{k}(t)| \, \mathrm{d}t  +
			\mathcal{Q}' +  \mathcal{E}  ,
		\end{aligned}
	\end{equation*}
	where we use the notation $\mathcal{Q}' = 2\tan\left(\frac{\Delta }{2}\right) \mathcal{Q} $.
	We finally obtain
	\begin{equation*}
		\begin{aligned}
			\| g\| _{1} & =\exp\left[ 2\tan\left(\frac{\Delta }{2}\right)
			\overline{\| c\| _{1}} T + \mathcal{Q}' + \mathcal{E} \right]  \\
			            & = \exp\left[ 2\tan\left(\frac{\Delta }{2}\right)
			\overline{\| c\| _{1}} T \right] \exp[ \mathcal{Q}' + \mathcal{E}],
		\end{aligned}
	\end{equation*}
	where we used the notation $\overline{\| c\| _{1}} =\frac{1}{T}\int _{0}^{T}\sum _{k=1}^{L}| c_{k} (t)| dt$ from Eq. \ref{eq:l1-norm}, we conclude our proof
	by bounding the error terms as
	\begin{align*}
		|\mathcal{Q}' + \mathcal{E}| & \leq |\mathcal{Q}'| + |\mathcal{E}|                                                                                             \\
		                             & \leq 2\tan\left(\frac{\Delta }{2}\right)    \frac{T V_{tot}}{N} + \sum _{j=1}^{N}\sum _{k=1}^{L} |\mathcal{D}_{kj}|             \\
		                             & = 2\tan\left(\frac{\Delta }{2}\right)    \frac{T V_{tot}}{N} + 2 \overline{\| c\|^2 _{2}} \frac{T^2}{N} + O\left( N^{-2}\right)
	\end{align*}
	which yields the expression
	\begin{equation}
		\| g\| _{1}  =\exp\left[ 2\tan\left(\frac{\Delta }{2}\right)
		\overline{\| c\| _{1}} T \right] + O(N^{-1}).
	\end{equation}
\end{proof}
In \cref{fig:overhead_N} we numerically plot  $\| g\| _{1}$ for increasing \(N\) for the time-dependent Hamiltonian introduced in Section \ref{sec:numeric} and confirm that indeed it converges to the above expression.

\begin{figure*}[ht]
	\centering
	\includegraphics[width=0.9\textwidth]{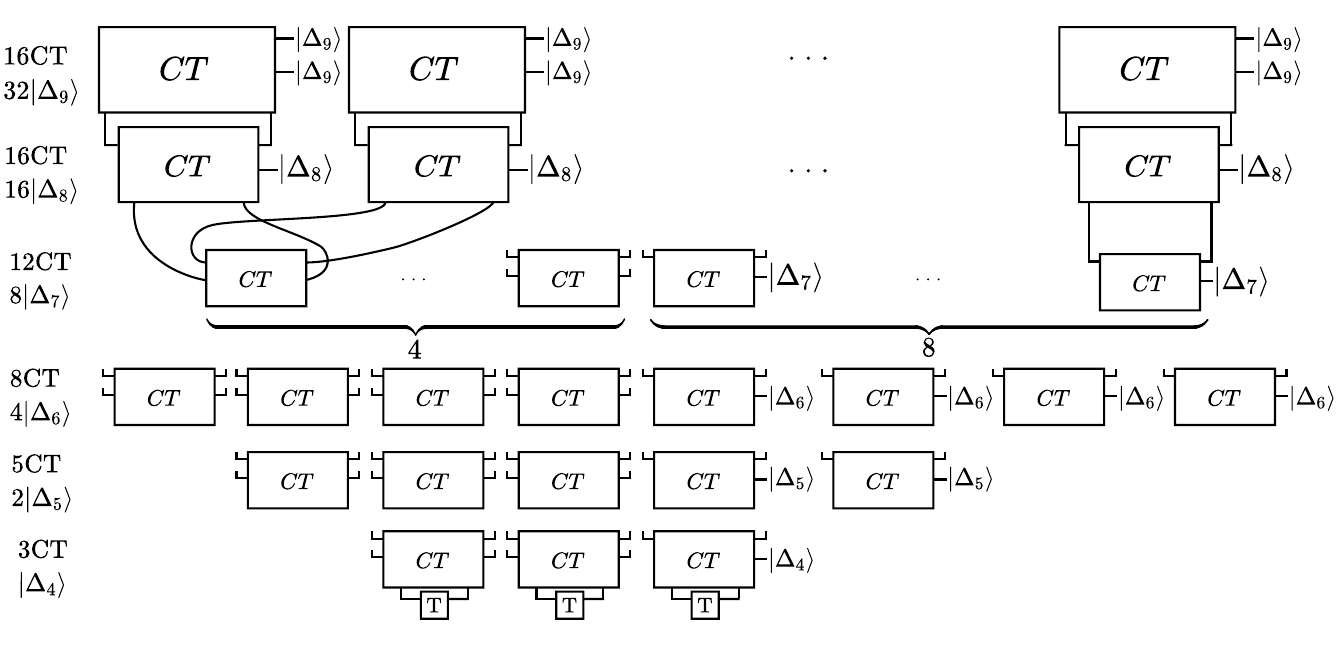}
	\caption{
		The catalyst tower to generate resource states in Clifford hierarchy of $\ell_0=9$ which follows an exponential distribution, i.e., generating $2^{\ell-4}$ resource states of $\ket{\Delta_{\ell}}$ for $\ell = 4, 5, \dots, \ell_0=9$. In the top layer, since each $CT$ circuit generates two $\ket{\Delta_9}$ states, we require $2^{\ell - 5}$ $CT$ circuits. In the second layer, each $CT$ circuit is connected to exactly one $CT$ circuit from the first layer, resulting in $16$ $CT$ circuits to generate $16$ $\ket{\Delta_8}$ states. For the third layer, since we need to generate $8$ $\ket{\Delta_7}$ states, we need $8$ $CT$ circuits. Additionally, to provide $\ket{\Delta_7}$ to the remaining $CT$ circuits in the second layer, we require $(16 - 8)/2 = 4$ extra $CT$ circuits, which do not generate additional $\ket{\Delta_7}$ states. The same approach applies to the following layers. In the final layer, since $\ket{3\Delta}$ corresponds to the T-state, we can directly apply three T-gates for this layer. Therefore, in total, we need 60 $CT$ circuits and 3 T-gates, resulting in a total cost of 243 T-gates and 60 ancilla qubits.
		\label{fig:tower}}
\end{figure*}

\section{Proof of \cref{cor:Q_control_param}}
In Remark \ref{cor:Q_control_param}, we introduced a control trade-off parameter \( Q \) that manages the trade-off between the circuit depth and the number of shots required. The expression for the number of gates \( \nu_\infty \)  is given by:
\[
	\nu_\infty = \frac{2\left(\overline{\| c\| _{1}} T\right)^{2}}{Q} + Q,
\]
Here, we have two competing factors
as the term \( \frac{2\left(\overline{\| c\| _{1}} T\right)^{2}}{Q} \) decreases and the
second term increases as \( Q \) increases. We can rewrite the expression as
\[
	\nu_\infty = \frac{2\left(\overline{\| c\| _{1}} T\right)^{2}}{Q} + \frac{Q^2}{Q}=\frac{2\left(\overline{\| c\| _{1}} T\right)^{2} + Q^2}{Q}.
\]

The lower bound of $\nu _{\infty }$ in \cref{thm:num_gate2} is attained when $Q =  \overline{\| c\| _{1}} T \sqrt{2}$, and now, we substitute this bound \( Q = \overline{\| c\| _{1}} T \sqrt{2} \) into the expression as
\begin{align*}
	\nu _{\infty } & \leq
	\frac{2\left(\overline{\| c\| _{1}} T\right)^{2}    +  \left(\overline{\| c\| _{1}} T \sqrt{2}  \right)^2  }{Q}
	= \frac{4\left(\overline{\| c\| _{1}} T\right)^{2}}{Q}
\end{align*}
gives the upper bound, holds for \( Q \leq \overline{\| c\| _{1}} T \sqrt{2} \).

\begin{figure*}[tb]
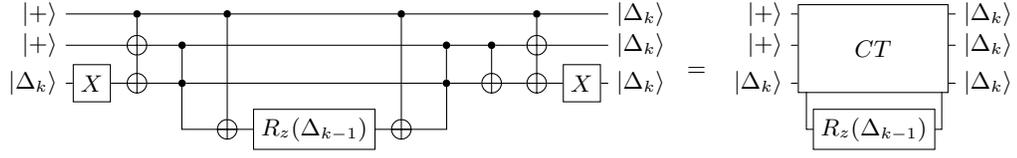

	\centering
	\begin{yquantgroup}
		\registers{
			qubit {} a;
			qubit {} b;
			qubit {} c;
			qubit {} d;}
		\circuit{
		init {$\ket{+}$} a;
		init {$\ket{+}$} b;
		init {$\ket{\Delta_k}$} c;
		discard d;

		x c;
		cnot b,c|a;
		zz (b,c);
		cnot d|a;
		[name=ct]
		box {$R_z(\Delta_{k-1})$} d;
		cnot d|a;
		zz (b,c);
		cnot c|b;
		cnot b,c|a;
		x c;

		output {$\ket{\Delta_k}$} a;
		output {$\ket{\Delta_k}$} b;
		output {$\ket{\Delta_k}$} c;
		\draw (ct)+(-1.76,0.65) |-  (ct) ;
		\draw (ct) -| + (1.76,0.65) ;}
		\equals
		\circuit{
		init {$\ket{+}$} a;
		init {$\ket{+}$} b;
		init {$\ket{\Delta_k}$} c;

		discard d;
		hspace {2mm} d;
		box {$\qquad CT \qquad$} (a,b,c);
		[name=ct]
		box {$R_z(\Delta_{k-1})$} d;

		output {$\ket{\Delta_k}$} a;
		output {$\ket{\Delta_k}$} b;
		output {$\ket{\Delta_k}$} c;

		\draw (ct)+(-0.9,0.5) |-  (ct) ;
		\draw (ct) -| + (0.9,0.5) ;}
	\end{yquantgroup}
	\caption{The catalyst circuit as a tower of height $h=1$. The catalyst towers are built by
		connecting these $CT$ circuits as in \cref{fig:tower} and in \cite{sun2024low}.}
	\label{fig:ct}
\end{figure*}

\section{Details of resource estimation\label{app:resource}}
\subsection{Mapping Pauli rotations to $Z$ rotations}
In fault-tolerant quantum computing (FTQC), Clifford gates are considered relatively cheap to implement
as natural operations in stabiliser codes. A Clifford operator on a quantum system described by $n$ qubits is a unitary operator $U$ such that for any Pauli operator $P$, the operator $UPU^{\dagger }$ is also a Pauli operator. Formally, the Clifford group $\mathcal{C}$ is defined as:
\begin{equation*}
	\mathcal{C} :=\left\{U \mid UPU^{\dagger } \in \mathcal{P} ,\forall P\in \mathcal{P}\right\},
\end{equation*}
where $\mathcal{P}$ denotes the Pauli group. Clifford gates are pivotal in quantum error correction schemes as they are efficiently implementable and largely error-free compared to non-Clifford gates.

Non-Clifford gates, such as T-gates, on the other hand, are more challenging and costly to
implement due to the need for non-trivial additional measures, such as magic state distillation.
Thus, minimizing the number of non-Clifford gates, such as continuos-angle rotations,
is crucial for minimising the resources requirements in early-FTQC.

In our TE-PAI approach, we require only discrete Pauli rotations $R_{\sigma}(\theta)$ of the form $e^{ i \theta \sigma/2}$, where $\sigma$ represents Pauli strings. These rotations can be efficiently mapped to single-qubit $Z$ rotations, interleaved with Clifford gates. More precisely, any Pauli rotation $R_{\sigma}(\theta)$ can be expressed as
\begin{equation*}
	R_{\sigma } (\theta ) = U R_{Z} (\theta ) U^{\dagger },
\end{equation*}
where $U$ is a sequence of Clifford gates determined by the specific Pauli operator $\sigma$ and the rotation angle $\theta$. This formulation allows us to focus on efficiently implementing the $R_Z(\pm\Delta)$ rotation gate
which is then the only non-Clifford resource we require.

\subsection{Clifford hierarchy}

The Clifford hierarchy, denoted as $\mathcal{C}_{\ell}$, is defined recursively, beginning with the Pauli group $\mathcal{P}$ at the first level as $\mathcal{C}_{1}:=\mathcal{P}$ and for $\ell > 1$, the higher levels of the hierarchy are defined as
\begin{equation*}
	\mathcal{C}_{\ell } :=\left\{U\mid UPU^{\dagger } \in \mathcal{C}_{\ell -1} ,\forall P\in \mathcal{P}\right\}.
\end{equation*}
This recursive definition means that the unitary $U$ at level $\ell$ conjugates elements of the Pauli group $\mathcal{P}$ to operators in $\mathcal{C}_{\ell-1}$. Thus, $\mathcal{C}_{2} =\mathcal{C}$ becomes a Clifford group.

\subsection{Catalyst towers for Clifford hierarchy rotations\label{app:tower}}

Here, we explain how we construct a catalyst tower to generate resource states
in Clifford hierarchy for repeat-until-success method. The construction of the
catalyst tower is based on \cite{sun2024low}.
In \cref{fig:tower}, we give the constructive example for $\ell_0=9$, but our
construction is straightforwardly generalises to higher $\ell_0$.

The white boxes in \cref{fig:tower} indicate the catalyst circuits that were introduced in \cite{sun2024low}
and which we denote as $CT$ and we define them explicitly in \cref{fig:ct}.
While \cite{sun2024low} concatenated these circuits to yield a catalyst tower that outputs an approximately
equal number of $\ket{\Delta_\ell}$ resource states, our construction in \cref{fig:tower} outputs resource states $\ket{\Delta_\ell}$
according to an exponential distribution as required for the repeat-until-success implementation of
the rotation gate $R_z(\Delta_\ell)$.
Specifically, our catalyst tower outputs $2^{\ell-4}$ resource states of $\ket{\Delta_{\ell}}$ for $\ell = 4, 5, \dots, \ell_0$, and we explicitly demonstrate the case of $\ell_0 = 9$. $\ket{\Delta_{3}}$ is a T-state, and we assume that T-states are natively produced
by the fault-tolerant quantum hardware, e.g., via magic state distillation.

The tower is constructed as follows:
\begin{itemize}
	\item\textbf{Top Layer}: In the top layer, each $CT$ circuit generates two $\ket{\Delta_{\ell_0}}$ states. Therefore, we need $2^{\ell_0 - 5}$ $CT$ circuits at the top layer to generate all required $2^{\ell_0 -4}$ of $\ket{\Delta_{\ell_0}}$ states.

	\item\textbf{Second Layer}: Each $CT$ circuit in the second layer is connected to exactly one $CT$ circuit in the top layer. The total number of $CT$ circuits in the second layer is $2^{\ell_0 - 5}$, which generates $2^{\ell_0 - 5}$ number of $\ket{\Delta_{\ell_0-1}}$ states.

	\item\textbf{Third Layer}:  In the third layer, we aim to generate $2^{\ell_0 - 6}$ number of $\ket{\Delta_{\ell_0-2}}$ states, so $2^{\ell_0 - 6}$ $CT$ circuits are required. To ensure we provide enough $\ket{\Delta_{\ell_0-2}}$ states to the $CT$ circuits in the second layer, we use an additional $(2^{\ell_0 - 5} - 2^{\ell_0 - 6}) / 2 = 2^{\ell_0 - 7}$ extra $CT$ circuits, which do not generate additional resource states but are used solely to support the generation of $\ket{\Delta_{\ell_0-2}}$ states for the second layer. Thus in total, we need $3\times2^{\ell_0 - 7}$ $CT$ circuits.

	\item\textbf{Remaining Layers}:
	      The same process continues for subsequent layers, where we progressively halve the number of $\ket{\Delta_{\ell}}$ states generated at each layer. At every step, additional $CT$ circuits are used to provide resource states to the above layer, following the same pattern.
	\item\textbf{Final Layer}:
	      In the final layer, the resource state $\ket{\Delta_3}$ corresponds to the T-gate, so we can directly apply T-gates to complete the process.
\end{itemize}

Based on the above, by using mathematical induction, we can calculate that $(\ell_0-\ell+1) 2^{\ell-5}$ $CT$ circuits are required at each layer $\ell$. Note that if $\ell_0$ is an even number, this expression for the final layer $\ell=4$ is not an integer. Thus, it will require one extra T-gate and $CT$ circuit that generate second $\ket{\Delta_4}$. Therefore, the total number of $CT$ circuits required is calculated as follows:
\vspace{-7pt}
\begin{equation*}
	\left\lceil\sum_{\ell=4}^{\ell_0}(\ell_0-\ell+1) 2^{\ell-5}\right\rceil=\left\lceil\frac{2^{\ell_0 -2} -\ell_0 +1}{2}\right\rceil.
\end{equation*}
The number of ancillary qubits is equal to the number of $CT$ circuits. Additionally, $\lceil(\ell_0 -3) /2\rceil$ T-gates are applied to the final layer. Since each $CT$ circuit requires 4 T gates, the total T cost of the entire process becomes $\left( 2^{\ell_0 } -3\ell_0 +1\right) /2$ for odd $\ell_0$ and $\left( 2^{\ell_0 } -3\ell_0 +6\right) /2$ for even $\ell_0$.


\begin{thebibliography}{66}%
	\makeatletter
	\providecommand \@ifxundefined [1]{%
		\@ifx{#1\undefined}
	}%
	\providecommand \@ifnum [1]{%
		\ifnum #1\expandafter \@firstoftwo
		\else \expandafter \@secondoftwo
		\fi
	}%
	\providecommand \@ifx [1]{%
		\ifx #1\expandafter \@firstoftwo
		\else \expandafter \@secondoftwo
		\fi
	}%
	\providecommand \natexlab [1]{#1}%
	\providecommand \enquote  [1]{``#1''}%
	\providecommand \bibnamefont  [1]{#1}%
	\providecommand \bibfnamefont [1]{#1}%
	\providecommand \citenamefont [1]{#1}%
	\providecommand \href@noop [0]{\@secondoftwo}%
	\providecommand \href [0]{\begingroup \@sanitize@url \@href}%
	\providecommand \@href[1]{\@@startlink{#1}\@@href}%
	\providecommand \@@href[1]{\endgroup#1\@@endlink}%
	\providecommand \@sanitize@url [0]{\catcode `\\12\catcode `\$12\catcode `\&12\catcode `\#12\catcode `\^12\catcode `\_12\catcode `\%12\relax}%
	\providecommand \@@startlink[1]{}%
	\providecommand \@@endlink[0]{}%
	\providecommand \url  [0]{\begingroup\@sanitize@url \@url }%
	\providecommand \@url [1]{\endgroup\@href {#1}{\urlprefix }}%
	\providecommand \urlprefix  [0]{URL }%
	\providecommand \Eprint [0]{\href }%
	\providecommand \doibase [0]{https://doi.org/}%
	\providecommand \selectlanguage [0]{\@gobble}%
	\providecommand \bibinfo  [0]{\@secondoftwo}%
	\providecommand \bibfield  [0]{\@secondoftwo}%
	\providecommand \translation [1]{[#1]}%
	\providecommand \BibitemOpen [0]{}%
	\providecommand \bibitemStop [0]{}%
	\providecommand \bibitemNoStop [0]{.\EOS\space}%
	\providecommand \EOS [0]{\spacefactor3000\relax}%
	\providecommand \BibitemShut  [1]{\csname bibitem#1\endcsname}%
	\let\auto@bib@innerbib\@empty
	\bibitem [{\citenamefont {Feynman}(1982)}]{Feynman}%
	\BibitemOpen
	\bibfield  {author} {\bibinfo {author} {\bibfnamefont {R.~P.}\ \bibnamefont {Feynman}},\ }\bibfield  {title} {\bibinfo {title} {Simulating physics with computers},\ }\href {https://doi.org/10.1007/bf02650179} {\bibfield  {journal} {\bibinfo  {journal} {International Journal of Theoretical Physics}\ }\textbf {\bibinfo {volume} {21}},\ \bibinfo {pages} {467–488} (\bibinfo {year} {1982})}\BibitemShut {NoStop}%
	\bibitem [{\citenamefont {Lloyd}(1996)}]{Lloyd}%
	\BibitemOpen
	\bibfield  {author} {\bibinfo {author} {\bibfnamefont {S.}~\bibnamefont {Lloyd}},\ }\bibfield  {title} {\bibinfo {title} {Universal quantum simulators},\ }\href {https://doi.org/10.1126/science.273.5278.1073} {\bibfield  {journal} {\bibinfo  {journal} {Science}\ }\textbf {\bibinfo {volume} {273}},\ \bibinfo {pages} {1073–1078} (\bibinfo {year} {1996})}\BibitemShut {NoStop}%
	\bibitem [{\citenamefont {Suzuki}(1990)}]{Suzuki1}%
	\BibitemOpen
	\bibfield  {author} {\bibinfo {author} {\bibfnamefont {M.}~\bibnamefont {Suzuki}},\ }\bibfield  {title} {\bibinfo {title} {Fractal decomposition of exponential operators with applications to many-body theories and monte carlo simulations},\ }\href {https://doi.org/https://doi.org/10.1016/0375-9601(90)90962-N} {\bibfield  {journal} {\bibinfo  {journal} {Physics Letters A}\ }\textbf {\bibinfo {volume} {146}},\ \bibinfo {pages} {319} (\bibinfo {year} {1990})}\BibitemShut {NoStop}%
	\bibitem [{\citenamefont {Suzuki}(1991)}]{Suzuki2}%
	\BibitemOpen
	\bibfield  {author} {\bibinfo {author} {\bibfnamefont {M.}~\bibnamefont {Suzuki}},\ }\bibfield  {title} {\bibinfo {title} {General theory of fractal path integrals with applications to many-body theories and statistical physics},\ }\href {https://doi.org/10.1063/1.529425} {\bibfield  {journal} {\bibinfo  {journal} {Journal of Mathematical Physics}\ }\textbf {\bibinfo {volume} {32}},\ \bibinfo {pages} {400–407} (\bibinfo {year} {1991})}\BibitemShut {NoStop}%
	\bibitem [{\citenamefont {Reiher}\ \emph {et~al.}(2017)\citenamefont {Reiher}, \citenamefont {Wiebe}, \citenamefont {Svore}, \citenamefont {Wecker},\ and\ \citenamefont {Troyer}}]{chemistry1}%
	\BibitemOpen
	\bibfield  {author} {\bibinfo {author} {\bibfnamefont {M.}~\bibnamefont {Reiher}}, \bibinfo {author} {\bibfnamefont {N.}~\bibnamefont {Wiebe}}, \bibinfo {author} {\bibfnamefont {K.~M.}\ \bibnamefont {Svore}}, \bibinfo {author} {\bibfnamefont {D.}~\bibnamefont {Wecker}},\ and\ \bibinfo {author} {\bibfnamefont {M.}~\bibnamefont {Troyer}},\ }\bibfield  {title} {\bibinfo {title} {Elucidating reaction mechanisms on quantum computers},\ }\href {https://doi.org/10.1073/pnas.1619152114} {\bibfield  {journal} {\bibinfo  {journal} {Proceedings of the National Academy of Sciences}\ }\textbf {\bibinfo {volume} {114}},\ \bibinfo {pages} {7555–7560} (\bibinfo {year} {2017})}\BibitemShut {NoStop}%
	\bibitem [{\citenamefont {McArdle}\ \emph {et~al.}(2020)\citenamefont {McArdle}, \citenamefont {Endo}, \citenamefont {Aspuru-Guzik}, \citenamefont {Benjamin},\ and\ \citenamefont {Yuan}}]{chemistry2}%
	\BibitemOpen
	\bibfield  {author} {\bibinfo {author} {\bibfnamefont {S.}~\bibnamefont {McArdle}}, \bibinfo {author} {\bibfnamefont {S.}~\bibnamefont {Endo}}, \bibinfo {author} {\bibfnamefont {A.}~\bibnamefont {Aspuru-Guzik}}, \bibinfo {author} {\bibfnamefont {S.~C.}\ \bibnamefont {Benjamin}},\ and\ \bibinfo {author} {\bibfnamefont {X.}~\bibnamefont {Yuan}},\ }\bibfield  {title} {\bibinfo {title} {Quantum computational chemistry},\ }\href {https://doi.org/10.1103/RevModPhys.92.015003} {\bibfield  {journal} {\bibinfo  {journal} {Rev. Mod. Phys.}\ }\textbf {\bibinfo {volume} {92}},\ \bibinfo {pages} {015003} (\bibinfo {year} {2020})}\BibitemShut {NoStop}%
	\bibitem [{\citenamefont {Bauer}\ \emph {et~al.}(2020)\citenamefont {Bauer}, \citenamefont {Bravyi}, \citenamefont {Motta},\ and\ \citenamefont {Chan}}]{chemistry3}%
	\BibitemOpen
	\bibfield  {author} {\bibinfo {author} {\bibfnamefont {B.}~\bibnamefont {Bauer}}, \bibinfo {author} {\bibfnamefont {S.}~\bibnamefont {Bravyi}}, \bibinfo {author} {\bibfnamefont {M.}~\bibnamefont {Motta}},\ and\ \bibinfo {author} {\bibfnamefont {G.~K.-L.}\ \bibnamefont {Chan}},\ }\bibfield  {title} {\bibinfo {title} {Quantum algorithms for quantum chemistry and quantum materials science},\ }\href {https://doi.org/10.1021/acs.chemrev.9b00829} {\bibfield  {journal} {\bibinfo  {journal} {Chemical Reviews}\ }\textbf {\bibinfo {volume} {120}},\ \bibinfo {pages} {12685–12717} (\bibinfo {year} {2020})}\BibitemShut {NoStop}%
	\bibitem [{\citenamefont {Su}\ \emph {et~al.}(2021)\citenamefont {Su}, \citenamefont {Berry}, \citenamefont {Wiebe}, \citenamefont {Rubin},\ and\ \citenamefont {Babbush}}]{chemistry4}%
	\BibitemOpen
	\bibfield  {author} {\bibinfo {author} {\bibfnamefont {Y.}~\bibnamefont {Su}}, \bibinfo {author} {\bibfnamefont {D.~W.}\ \bibnamefont {Berry}}, \bibinfo {author} {\bibfnamefont {N.}~\bibnamefont {Wiebe}}, \bibinfo {author} {\bibfnamefont {N.}~\bibnamefont {Rubin}},\ and\ \bibinfo {author} {\bibfnamefont {R.}~\bibnamefont {Babbush}},\ }\bibfield  {title} {\bibinfo {title} {Fault-tolerant quantum simulations of chemistry in first quantization},\ }\href {https://doi.org/10.1103/PRXQuantum.2.040332} {\bibfield  {journal} {\bibinfo  {journal} {PRX Quantum}\ }\textbf {\bibinfo {volume} {2}},\ \bibinfo {pages} {040332} (\bibinfo {year} {2021})}\BibitemShut {NoStop}%
	\bibitem [{\citenamefont {Childs}\ and\ \citenamefont {Wiebe}(2012)}]{LCU1}%
	\BibitemOpen
	\bibfield  {author} {\bibinfo {author} {\bibfnamefont {A.~M.}\ \bibnamefont {Childs}}\ and\ \bibinfo {author} {\bibfnamefont {N.}~\bibnamefont {Wiebe}},\ }\bibfield  {title} {\bibinfo {title} {Hamiltonian simulation using linear combinations of unitary operations},\ }\href {https://doi.org/10.26421/qic12.11-12-1} {\bibfield  {journal} {\bibinfo  {journal} {Quantum Information and Computation}\ }\textbf {\bibinfo {volume} {12}},\ \bibinfo {pages} {901–924} (\bibinfo {year} {2012})}\BibitemShut {NoStop}%
	\bibitem [{\citenamefont {Berry}\ \emph {et~al.}(2014)\citenamefont {Berry}, \citenamefont {Childs}, \citenamefont {Cleve}, \citenamefont {Kothari},\ and\ \citenamefont {Somma}}]{LCU2}%
	\BibitemOpen
	\bibfield  {author} {\bibinfo {author} {\bibfnamefont {D.~W.}\ \bibnamefont {Berry}}, \bibinfo {author} {\bibfnamefont {A.~M.}\ \bibnamefont {Childs}}, \bibinfo {author} {\bibfnamefont {R.}~\bibnamefont {Cleve}}, \bibinfo {author} {\bibfnamefont {R.}~\bibnamefont {Kothari}},\ and\ \bibinfo {author} {\bibfnamefont {R.~D.}\ \bibnamefont {Somma}},\ }\bibfield  {title} {\bibinfo {title} {Exponential improvement in precision for simulating sparse hamiltonians},\ }in\ \href {https://doi.org/10.1145/2591796.2591854} {\emph {\bibinfo {booktitle} {Proceedings of the forty-sixth annual ACM symposium on Theory of computing}}},\ \bibinfo {series and number} {STOC ’14}\ (\bibinfo  {publisher} {ACM},\ \bibinfo {year} {2014})\BibitemShut {NoStop}%
	\bibitem [{\citenamefont {Berry}\ \emph {et~al.}(2015)\citenamefont {Berry}, \citenamefont {Childs}, \citenamefont {Cleve}, \citenamefont {Kothari},\ and\ \citenamefont {Somma}}]{LCU3}%
	\BibitemOpen
	\bibfield  {author} {\bibinfo {author} {\bibfnamefont {D.~W.}\ \bibnamefont {Berry}}, \bibinfo {author} {\bibfnamefont {A.~M.}\ \bibnamefont {Childs}}, \bibinfo {author} {\bibfnamefont {R.}~\bibnamefont {Cleve}}, \bibinfo {author} {\bibfnamefont {R.}~\bibnamefont {Kothari}},\ and\ \bibinfo {author} {\bibfnamefont {R.~D.}\ \bibnamefont {Somma}},\ }\bibfield  {title} {\bibinfo {title} {Simulating hamiltonian dynamics with a truncated taylor series},\ }\bibfield  {journal} {\bibinfo  {journal} {Physical Review Letters}\ }\textbf {\bibinfo {volume} {114}},\ \href {https://doi.org/10.1103/physrevlett.114.090502} {10.1103/physrevlett.114.090502} (\bibinfo {year} {2015})\BibitemShut {NoStop}%
	\bibitem [{\citenamefont {Low}\ and\ \citenamefont {Chuang}(2017)}]{QSP1}%
	\BibitemOpen
	\bibfield  {author} {\bibinfo {author} {\bibfnamefont {G.~H.}\ \bibnamefont {Low}}\ and\ \bibinfo {author} {\bibfnamefont {I.~L.}\ \bibnamefont {Chuang}},\ }\bibfield  {title} {\bibinfo {title} {Optimal hamiltonian simulation by quantum signal processing},\ }\bibfield  {journal} {\bibinfo  {journal} {Physical Review Letters}\ }\textbf {\bibinfo {volume} {118}},\ \href {https://doi.org/10.1103/physrevlett.118.010501} {10.1103/physrevlett.118.010501} (\bibinfo {year} {2017})\BibitemShut {NoStop}%
	\bibitem [{\citenamefont {Low}\ and\ \citenamefont {Chuang}(2019)}]{QSP2}%
	\BibitemOpen
	\bibfield  {author} {\bibinfo {author} {\bibfnamefont {G.~H.}\ \bibnamefont {Low}}\ and\ \bibinfo {author} {\bibfnamefont {I.~L.}\ \bibnamefont {Chuang}},\ }\bibfield  {title} {\bibinfo {title} {Hamiltonian simulation by qubitization},\ }\href {https://doi.org/10.22331/q-2019-07-12-163} {\bibfield  {journal} {\bibinfo  {journal} {Quantum}\ }\textbf {\bibinfo {volume} {3}},\ \bibinfo {pages} {163} (\bibinfo {year} {2019})}\BibitemShut {NoStop}%
	\bibitem [{\citenamefont {Childs}\ and\ \citenamefont {Berry}(2012)}]{qw1}%
	\BibitemOpen
	\bibfield  {author} {\bibinfo {author} {\bibfnamefont {A.~M.}\ \bibnamefont {Childs}}\ and\ \bibinfo {author} {\bibfnamefont {D.~W.}\ \bibnamefont {Berry}},\ }\bibfield  {title} {\bibinfo {title} {Black-box hamiltonian simulation and unitary implementation},\ }\href {https://doi.org/10.26421/qic12.1-2-4} {\bibfield  {journal} {\bibinfo  {journal} {Quantum Information and Computation}\ }\textbf {\bibinfo {volume} {12}},\ \bibinfo {pages} {29–62} (\bibinfo {year} {2012})}\BibitemShut {NoStop}%
	\bibitem [{\citenamefont {Childs}(2009)}]{qw2}%
	\BibitemOpen
	\bibfield  {author} {\bibinfo {author} {\bibfnamefont {A.~M.}\ \bibnamefont {Childs}},\ }\bibfield  {title} {\bibinfo {title} {On the relationship between continuous- and discrete-time quantum walk},\ }\href {https://doi.org/10.1007/s00220-009-0930-1} {\bibfield  {journal} {\bibinfo  {journal} {Communications in Mathematical Physics}\ }\textbf {\bibinfo {volume} {294}},\ \bibinfo {pages} {581–603} (\bibinfo {year} {2009})}\BibitemShut {NoStop}%
	\bibitem [{\citenamefont {Jnane}\ \emph {et~al.}(2024)\citenamefont {Jnane}, \citenamefont {Steinberg}, \citenamefont {Cai}, \citenamefont {Nguyen},\ and\ \citenamefont {Koczor}}]{mitigation_shadow}%
	\BibitemOpen
	\bibfield  {author} {\bibinfo {author} {\bibfnamefont {H.}~\bibnamefont {Jnane}}, \bibinfo {author} {\bibfnamefont {J.}~\bibnamefont {Steinberg}}, \bibinfo {author} {\bibfnamefont {Z.}~\bibnamefont {Cai}}, \bibinfo {author} {\bibfnamefont {H.~C.}\ \bibnamefont {Nguyen}},\ and\ \bibinfo {author} {\bibfnamefont {B.}~\bibnamefont {Koczor}},\ }\bibfield  {title} {\bibinfo {title} {Quantum error mitigated classical shadows},\ }\href {https://doi.org/10.1103/prxquantum.5.010324} {\bibfield  {journal} {\bibinfo  {journal} {PRX Quantum}\ }\textbf {\bibinfo {volume} {5}},\ \bibinfo {pages} {010324} (\bibinfo {year} {2024})}\BibitemShut {NoStop}%
	\bibitem [{\citenamefont {Huang}\ \emph {et~al.}(2020)\citenamefont {Huang}, \citenamefont {Kueng},\ and\ \citenamefont {Preskill}}]{shadow}%
	\BibitemOpen
	\bibfield  {author} {\bibinfo {author} {\bibfnamefont {H.-Y.}\ \bibnamefont {Huang}}, \bibinfo {author} {\bibfnamefont {R.}~\bibnamefont {Kueng}},\ and\ \bibinfo {author} {\bibfnamefont {J.}~\bibnamefont {Preskill}},\ }\bibfield  {title} {\bibinfo {title} {Predicting many properties of a quantum system from very few measurements},\ }\href {https://doi.org/10.1038/s41567-020-0932-7} {\bibfield  {journal} {\bibinfo  {journal} {Nature Physics}\ }\textbf {\bibinfo {volume} {16}},\ \bibinfo {pages} {1050–1057} (\bibinfo {year} {2020})}\BibitemShut {NoStop}%
	\bibitem [{\citenamefont {Crawford}\ \emph {et~al.}(2021)\citenamefont {Crawford}, \citenamefont {van Straaten}, \citenamefont {Wang}, \citenamefont {Parks}, \citenamefont {Campbell},\ and\ \citenamefont {Brierley}}]{Crawford2019}%
	\BibitemOpen
	\bibfield  {author} {\bibinfo {author} {\bibfnamefont {O.}~\bibnamefont {Crawford}}, \bibinfo {author} {\bibfnamefont {B.}~\bibnamefont {van Straaten}}, \bibinfo {author} {\bibfnamefont {D.}~\bibnamefont {Wang}}, \bibinfo {author} {\bibfnamefont {T.}~\bibnamefont {Parks}}, \bibinfo {author} {\bibfnamefont {E.}~\bibnamefont {Campbell}},\ and\ \bibinfo {author} {\bibfnamefont {S.}~\bibnamefont {Brierley}},\ }\bibfield  {title} {\bibinfo {title} {{Efficient quantum measurement of Pauli operators in the presence of finite sampling error}},\ }\href {https://doi.org/https://doi.org/10.22331/q-2021-01-20-385} {\bibfield  {journal} {\bibinfo  {journal} {Quantum}\ }\textbf {\bibinfo {volume} {5}},\ \bibinfo {pages} {385} (\bibinfo {year} {2021})}\BibitemShut {NoStop}%
	\bibitem [{\citenamefont {Jena}\ \emph {et~al.}(2019)\citenamefont {Jena}, \citenamefont {Genin},\ and\ \citenamefont {Mosca}}]{jena2019pauli}%
	\BibitemOpen
	\bibfield  {author} {\bibinfo {author} {\bibfnamefont {A.}~\bibnamefont {Jena}}, \bibinfo {author} {\bibfnamefont {S.}~\bibnamefont {Genin}},\ and\ \bibinfo {author} {\bibfnamefont {M.}~\bibnamefont {Mosca}},\ }\bibfield  {title} {\bibinfo {title} {{Pauli partitioning with respect to gate sets}},\ }\href@noop {} {\bibfield  {journal} {\bibinfo  {journal} {arXiv preprint arXiv:1907.07859}\ } (\bibinfo {year} {2019})}\BibitemShut {NoStop}%
	\bibitem [{\citenamefont {Chan}\ \emph {et~al.}(2025)\citenamefont {Chan}, \citenamefont {Meister}, \citenamefont {Goh},\ and\ \citenamefont {Koczor}}]{chan2022algorithmic}%
	\BibitemOpen
	\bibfield  {author} {\bibinfo {author} {\bibfnamefont {H.~H.~S.}\ \bibnamefont {Chan}}, \bibinfo {author} {\bibfnamefont {R.}~\bibnamefont {Meister}}, \bibinfo {author} {\bibfnamefont {M.~L.}\ \bibnamefont {Goh}},\ and\ \bibinfo {author} {\bibfnamefont {B.}~\bibnamefont {Koczor}},\ }\bibfield  {title} {\bibinfo {title} {Algorithmic shadow spectroscopy},\ }\href {https://doi.org/10.1103/PRXQuantum.6.010352} {\bibfield  {journal} {\bibinfo  {journal} {PRX Quantum}\ }\textbf {\bibinfo {volume} {6}},\ \bibinfo {pages} {010352} (\bibinfo {year} {2025})}\BibitemShut {NoStop}%
	\bibitem [{\citenamefont {Wang}\ \emph {et~al.}(2023)\citenamefont {Wang}, \citenamefont {França}, \citenamefont {Zhang}, \citenamefont {Zhu},\ and\ \citenamefont {Johnson}}]{Wang_2023}%
	\BibitemOpen
	\bibfield  {author} {\bibinfo {author} {\bibfnamefont {G.}~\bibnamefont {Wang}}, \bibinfo {author} {\bibfnamefont {D.~S.}\ \bibnamefont {França}}, \bibinfo {author} {\bibfnamefont {R.}~\bibnamefont {Zhang}}, \bibinfo {author} {\bibfnamefont {S.}~\bibnamefont {Zhu}},\ and\ \bibinfo {author} {\bibfnamefont {P.~D.}\ \bibnamefont {Johnson}},\ }\bibfield  {title} {\bibinfo {title} {Quantum algorithm for ground state energy estimation using circuit depth with exponentially improved dependence on precision},\ }\href {https://doi.org/10.22331/q-2023-11-06-1167} {\bibfield  {journal} {\bibinfo  {journal} {Quantum}\ }\textbf {\bibinfo {volume} {7}},\ \bibinfo {pages} {1167} (\bibinfo {year} {2023})}\BibitemShut {NoStop}%
	\bibitem [{\citenamefont {Lin}\ and\ \citenamefont {Tong}(2022)}]{Lin_2022}%
	\BibitemOpen
	\bibfield  {author} {\bibinfo {author} {\bibfnamefont {L.}~\bibnamefont {Lin}}\ and\ \bibinfo {author} {\bibfnamefont {Y.}~\bibnamefont {Tong}},\ }\bibfield  {title} {\bibinfo {title} {Heisenberg-limited ground-state energy estimation for early fault-tolerant quantum computers},\ }\bibfield  {journal} {\bibinfo  {journal} {{PRX} Quantum}\ }\textbf {\bibinfo {volume} {3}},\ \href {https://doi.org/10.1103/prxquantum.3.010318} {10.1103/prxquantum.3.010318} (\bibinfo {year} {2022})\BibitemShut {NoStop}%
	\bibitem [{\citenamefont {Ding}\ and\ \citenamefont {Lin}(2023)}]{Ding2023simultaneous}%
	\BibitemOpen
	\bibfield  {author} {\bibinfo {author} {\bibfnamefont {Z.}~\bibnamefont {Ding}}\ and\ \bibinfo {author} {\bibfnamefont {L.}~\bibnamefont {Lin}},\ }\bibfield  {title} {\bibinfo {title} {Simultaneous estimation of multiple eigenvalues with short-depth quantum circuit on early fault-tolerant quantum computers},\ }\href {https://doi.org/10.22331/q-2023-10-11-1136} {\bibfield  {journal} {\bibinfo  {journal} {{Quantum}}\ }\textbf {\bibinfo {volume} {7}},\ \bibinfo {pages} {1136} (\bibinfo {year} {2023})}\BibitemShut {NoStop}%
	\bibitem [{\citenamefont {Goh}\ and\ \citenamefont {Koczor}(2024)}]{goh2024direct}%
	\BibitemOpen
	\bibfield  {author} {\bibinfo {author} {\bibfnamefont {M.~L.}\ \bibnamefont {Goh}}\ and\ \bibinfo {author} {\bibfnamefont {B.}~\bibnamefont {Koczor}},\ }\bibfield  {title} {\bibinfo {title} {Direct estimation of the density of states for fermionic systems},\ }\href@noop {} {\bibfield  {journal} {\bibinfo  {journal} {arXiv preprint arXiv:2407.03414}\ } (\bibinfo {year} {2024})}\BibitemShut {NoStop}%
	\bibitem [{\citenamefont {Yang}\ \emph {et~al.}(2021)\citenamefont {Yang}, \citenamefont {Lu},\ and\ \citenamefont {Li}}]{PRXQuantum.2.040361}%
	\BibitemOpen
	\bibfield  {author} {\bibinfo {author} {\bibfnamefont {Y.}~\bibnamefont {Yang}}, \bibinfo {author} {\bibfnamefont {B.-N.}\ \bibnamefont {Lu}},\ and\ \bibinfo {author} {\bibfnamefont {Y.}~\bibnamefont {Li}},\ }\bibfield  {title} {\bibinfo {title} {Accelerated quantum monte carlo with mitigated error on noisy quantum computer},\ }\href {https://doi.org/10.1103/PRXQuantum.2.040361} {\bibfield  {journal} {\bibinfo  {journal} {PRX Quantum}\ }\textbf {\bibinfo {volume} {2}},\ \bibinfo {pages} {040361} (\bibinfo {year} {2021})}\BibitemShut {NoStop}%
	\bibitem [{\citenamefont {Campbell}(2019)}]{qDRIFT1}%
	\BibitemOpen
	\bibfield  {author} {\bibinfo {author} {\bibfnamefont {E.}~\bibnamefont {Campbell}},\ }\bibfield  {title} {\bibinfo {title} {Random compiler for fast hamiltonian simulation},\ }\href {https://doi.org/10.1103/physrevlett.123.070503} {\bibfield  {journal} {\bibinfo  {journal} {Physical Review Letters}\ }\textbf {\bibinfo {volume} {123}},\ \bibinfo {pages} {070503} (\bibinfo {year} {2019})}\BibitemShut {NoStop}%
	\bibitem [{\citenamefont {Ouyang}\ \emph {et~al.}(2020)\citenamefont {Ouyang}, \citenamefont {White},\ and\ \citenamefont {Campbell}}]{qDRIFT2}%
	\BibitemOpen
	\bibfield  {author} {\bibinfo {author} {\bibfnamefont {Y.}~\bibnamefont {Ouyang}}, \bibinfo {author} {\bibfnamefont {D.~R.}\ \bibnamefont {White}},\ and\ \bibinfo {author} {\bibfnamefont {E.~T.}\ \bibnamefont {Campbell}},\ }\bibfield  {title} {\bibinfo {title} {Compilation by stochastic hamiltonian sparsification},\ }\href {https://doi.org/10.22331/q-2020-02-27-235} {\bibfield  {journal} {\bibinfo  {journal} {Quantum}\ }\textbf {\bibinfo {volume} {4}},\ \bibinfo {pages} {235} (\bibinfo {year} {2020})}\BibitemShut {NoStop}%
	\bibitem [{\citenamefont {Chen}\ \emph {et~al.}(2021)\citenamefont {Chen}, \citenamefont {Huang}, \citenamefont {Kueng},\ and\ \citenamefont {Tropp}}]{qDRIFT3}%
	\BibitemOpen
	\bibfield  {author} {\bibinfo {author} {\bibfnamefont {C.-F.}\ \bibnamefont {Chen}}, \bibinfo {author} {\bibfnamefont {H.-Y.}\ \bibnamefont {Huang}}, \bibinfo {author} {\bibfnamefont {R.}~\bibnamefont {Kueng}},\ and\ \bibinfo {author} {\bibfnamefont {J.~A.}\ \bibnamefont {Tropp}},\ }\bibfield  {title} {\bibinfo {title} {Concentration for random product formulas},\ }\href {https://doi.org/10.1103/prxquantum.2.040305} {\bibfield  {journal} {\bibinfo  {journal} {PRX Quantum}\ }\textbf {\bibinfo {volume} {2}},\ \bibinfo {pages} {040305} (\bibinfo {year} {2021})}\BibitemShut {NoStop}%
	\bibitem [{\citenamefont {Zhang}\ \emph {et~al.}(2022)\citenamefont {Zhang}, \citenamefont {Huo}, \citenamefont {Liu}, \citenamefont {Li},\ and\ \citenamefont {Yuan}}]{dyson-base}%
	\BibitemOpen
	\bibfield  {author} {\bibinfo {author} {\bibfnamefont {X.-M.}\ \bibnamefont {Zhang}}, \bibinfo {author} {\bibfnamefont {Z.}~\bibnamefont {Huo}}, \bibinfo {author} {\bibfnamefont {K.}~\bibnamefont {Liu}}, \bibinfo {author} {\bibfnamefont {Y.}~\bibnamefont {Li}},\ and\ \bibinfo {author} {\bibfnamefont {X.}~\bibnamefont {Yuan}},\ }\href {https://doi.org/10.48550/ARXIV.2212.09445} {\bibinfo {title} {Unbiased random circuit compiler for time-dependent hamiltonian simulation}} (\bibinfo {year} {2022})\BibitemShut {NoStop}%
	\bibitem [{\citenamefont {Granet}\ and\ \citenamefont {Dreyer}(2024)}]{similar}%
	\BibitemOpen
	\bibfield  {author} {\bibinfo {author} {\bibfnamefont {E.}~\bibnamefont {Granet}}\ and\ \bibinfo {author} {\bibfnamefont {H.}~\bibnamefont {Dreyer}},\ }\bibfield  {title} {\bibinfo {title} {Continuous hamiltonian dynamics on digital quantum computers without discretization error},\ }\href {https://doi.org/10.1038/s41534-024-00745-6} {\bibfield  {journal} {\bibinfo  {journal} {npj Quantum Information}\ }\textbf {\bibinfo {volume} {10}},\ \bibinfo {pages} {82} (\bibinfo {year} {2024})}\BibitemShut {NoStop}%
	\bibitem [{\citenamefont {Koczor}\ \emph {et~al.}(2024)\citenamefont {Koczor}, \citenamefont {Morton},\ and\ \citenamefont {Benjamin}}]{PAI}%
	\BibitemOpen
	\bibfield  {author} {\bibinfo {author} {\bibfnamefont {B.}~\bibnamefont {Koczor}}, \bibinfo {author} {\bibfnamefont {J.~J.~L.}\ \bibnamefont {Morton}},\ and\ \bibinfo {author} {\bibfnamefont {S.~C.}\ \bibnamefont {Benjamin}},\ }\bibfield  {title} {\bibinfo {title} {Probabilistic interpolation of quantum rotation angles},\ }\href {https://doi.org/10.1103/PhysRevLett.132.130602} {\bibfield  {journal} {\bibinfo  {journal} {Phys. Rev. Lett.}\ }\textbf {\bibinfo {volume} {132}},\ \bibinfo {pages} {130602} (\bibinfo {year} {2024})}\BibitemShut {NoStop}%
	\bibitem [{\citenamefont {Lieb}\ and\ \citenamefont {Robinson}(1972)}]{LR}%
	\BibitemOpen
	\bibfield  {author} {\bibinfo {author} {\bibfnamefont {E.~H.}\ \bibnamefont {Lieb}}\ and\ \bibinfo {author} {\bibfnamefont {D.~W.}\ \bibnamefont {Robinson}},\ }\bibfield  {title} {\bibinfo {title} {The finite group velocity of quantum spin systems},\ }\href {https://doi.org/10.1007/bf01645779} {\bibfield  {journal} {\bibinfo  {journal} {Communications in Mathematical Physics}\ }\textbf {\bibinfo {volume} {28}},\ \bibinfo {pages} {251–257} (\bibinfo {year} {1972})}\BibitemShut {NoStop}%
	\bibitem [{\citenamefont {Haah}\ \emph {et~al.}(2021)\citenamefont {Haah}, \citenamefont {Hastings}, \citenamefont {Kothari},\ and\ \citenamefont {Low}}]{Haah2021}%
	\BibitemOpen
	\bibfield  {author} {\bibinfo {author} {\bibfnamefont {J.}~\bibnamefont {Haah}}, \bibinfo {author} {\bibfnamefont {M.~B.}\ \bibnamefont {Hastings}}, \bibinfo {author} {\bibfnamefont {R.}~\bibnamefont {Kothari}},\ and\ \bibinfo {author} {\bibfnamefont {G.~H.}\ \bibnamefont {Low}},\ }\bibfield  {title} {\bibinfo {title} {Quantum algorithm for simulating real time evolution of lattice hamiltonians},\ }\href {https://doi.org/10.1137/18m1231511} {\bibfield  {journal} {\bibinfo  {journal} {SIAM Journal on Computing}\ }\textbf {\bibinfo {volume} {52}},\ \bibinfo {pages} {FOCS18} (\bibinfo {year} {2021})}\BibitemShut {NoStop}%
	\bibitem [{\citenamefont {Lao}\ and\ \citenamefont {Browne}(2022)}]{2qan}%
	\BibitemOpen
	\bibfield  {author} {\bibinfo {author} {\bibfnamefont {L.}~\bibnamefont {Lao}}\ and\ \bibinfo {author} {\bibfnamefont {D.~E.}\ \bibnamefont {Browne}},\ }\bibfield  {title} {\bibinfo {title} {{2QAN: a quantum compiler for 2-local qubit hamiltonian simulation algorithms}},\ }in\ \href {https://doi.org/10.1145/3470496.3527394} {\emph {\bibinfo {booktitle} {Proceedings of the 49th Annual International Symposium on Computer Architecture}}},\ \bibinfo {series and number} {ISCA '22}\ (\bibinfo  {publisher} {Association for Computing Machinery},\ \bibinfo {address} {New York, NY, USA},\ \bibinfo {year} {2022})\ p.\ \bibinfo {pages} {351–365}\BibitemShut {NoStop}%
	\bibitem [{\citenamefont {Childs}\ \emph {et~al.}(2021)\citenamefont {Childs}, \citenamefont {Su}, \citenamefont {Tran}, \citenamefont {Wiebe},\ and\ \citenamefont {Zhu}}]{trotter_err}%
	\BibitemOpen
	\bibfield  {author} {\bibinfo {author} {\bibfnamefont {A.~M.}\ \bibnamefont {Childs}}, \bibinfo {author} {\bibfnamefont {Y.}~\bibnamefont {Su}}, \bibinfo {author} {\bibfnamefont {M.~C.}\ \bibnamefont {Tran}}, \bibinfo {author} {\bibfnamefont {N.}~\bibnamefont {Wiebe}},\ and\ \bibinfo {author} {\bibfnamefont {S.}~\bibnamefont {Zhu}},\ }\bibfield  {title} {\bibinfo {title} {Theory of trotter error with commutator scaling},\ }\href {https://doi.org/10.1103/PhysRevX.11.011020} {\bibfield  {journal} {\bibinfo  {journal} {Phys. Rev. X}\ }\textbf {\bibinfo {volume} {11}},\ \bibinfo {pages} {011020} (\bibinfo {year} {2021})}\BibitemShut {NoStop}%
	\bibitem [{\citenamefont {An}\ \emph {et~al.}(2022)\citenamefont {An}, \citenamefont {Fang},\ and\ \citenamefont {Lin}}]{oscillate}%
	\BibitemOpen
	\bibfield  {author} {\bibinfo {author} {\bibfnamefont {D.}~\bibnamefont {An}}, \bibinfo {author} {\bibfnamefont {D.}~\bibnamefont {Fang}},\ and\ \bibinfo {author} {\bibfnamefont {L.}~\bibnamefont {Lin}},\ }\bibfield  {title} {\bibinfo {title} {Time-dependent {H}amiltonian simulation of highly oscillatory dynamics and superconvergence for {S}chr{\"{o}}dinger equation},\ }\href {https://doi.org/10.22331/q-2022-04-15-690} {\bibfield  {journal} {\bibinfo  {journal} {{Quantum}}\ }\textbf {\bibinfo {volume} {6}},\ \bibinfo {pages} {690} (\bibinfo {year} {2022})}\BibitemShut {NoStop}%
	\bibitem [{\citenamefont {Koch}\ \emph {et~al.}(2022)\citenamefont {Koch}, \citenamefont {Boscain}, \citenamefont {Calarco}, \citenamefont {Dirr}, \citenamefont {Filipp}, \citenamefont {Glaser}, \citenamefont {Kosloff}, \citenamefont {Montangero}, \citenamefont {Schulte-Herbr{\"u}ggen}, \citenamefont {Sugny} \emph {et~al.}}]{koch2022quantum}%
	\BibitemOpen
	\bibfield  {author} {\bibinfo {author} {\bibfnamefont {C.~P.}\ \bibnamefont {Koch}}, \bibinfo {author} {\bibfnamefont {U.}~\bibnamefont {Boscain}}, \bibinfo {author} {\bibfnamefont {T.}~\bibnamefont {Calarco}}, \bibinfo {author} {\bibfnamefont {G.}~\bibnamefont {Dirr}}, \bibinfo {author} {\bibfnamefont {S.}~\bibnamefont {Filipp}}, \bibinfo {author} {\bibfnamefont {S.~J.}\ \bibnamefont {Glaser}}, \bibinfo {author} {\bibfnamefont {R.}~\bibnamefont {Kosloff}}, \bibinfo {author} {\bibfnamefont {S.}~\bibnamefont {Montangero}}, \bibinfo {author} {\bibfnamefont {T.}~\bibnamefont {Schulte-Herbr{\"u}ggen}}, \bibinfo {author} {\bibfnamefont {D.}~\bibnamefont {Sugny}}, \emph {et~al.},\ }\bibfield  {title} {\bibinfo {title} {Quantum optimal control in quantum technologies. strategic report on current status, visions and goals for research in europe},\ }\href@noop {} {\bibfield  {journal} {\bibinfo  {journal} {EPJ Quantum Technology}\ }\textbf {\bibinfo {volume} {9}},\ \bibinfo {pages} {19} (\bibinfo {year}
		{2022})}\BibitemShut {NoStop}%
	\bibitem [{\citenamefont {Blanes}\ \emph {et~al.}(2009)\citenamefont {Blanes}, \citenamefont {Casas}, \citenamefont {Oteo},\ and\ \citenamefont {Ros}}]{magnus}%
	\BibitemOpen
	\bibfield  {author} {\bibinfo {author} {\bibfnamefont {S.}~\bibnamefont {Blanes}}, \bibinfo {author} {\bibfnamefont {F.}~\bibnamefont {Casas}}, \bibinfo {author} {\bibfnamefont {J.}~\bibnamefont {Oteo}},\ and\ \bibinfo {author} {\bibfnamefont {J.}~\bibnamefont {Ros}},\ }\bibfield  {title} {\bibinfo {title} {The magnus expansion and some of its applications},\ }\href {https://doi.org/https://doi.org/10.1016/j.physrep.2008.11.001} {\bibfield  {journal} {\bibinfo  {journal} {Physics Reports}\ }\textbf {\bibinfo {volume} {470}},\ \bibinfo {pages} {151} (\bibinfo {year} {2009})}\BibitemShut {NoStop}%
	\bibitem [{\citenamefont {Koczor}(2024)}]{koczor2024sparse}%
	\BibitemOpen
	\bibfield  {author} {\bibinfo {author} {\bibfnamefont {B.}~\bibnamefont {Koczor}},\ }\bibfield  {title} {\bibinfo {title} {Sparse probabilistic synthesis of quantum operations},\ }\href {https://doi.org/10.1103/PRXQuantum.5.040352} {\bibfield  {journal} {\bibinfo  {journal} {PRX Quantum}\ }\textbf {\bibinfo {volume} {5}},\ \bibinfo {pages} {040352} (\bibinfo {year} {2024})}\BibitemShut {NoStop}%
	\bibitem [{\citenamefont {Lee}\ \emph {et~al.}(2021)\citenamefont {Lee}, \citenamefont {Berry}, \citenamefont {Gidney}, \citenamefont {Huggins}, \citenamefont {McClean}, \citenamefont {Wiebe},\ and\ \citenamefont {Babbush}}]{hyper_contraction}%
	\BibitemOpen
	\bibfield  {author} {\bibinfo {author} {\bibfnamefont {J.}~\bibnamefont {Lee}}, \bibinfo {author} {\bibfnamefont {D.~W.}\ \bibnamefont {Berry}}, \bibinfo {author} {\bibfnamefont {C.}~\bibnamefont {Gidney}}, \bibinfo {author} {\bibfnamefont {W.~J.}\ \bibnamefont {Huggins}}, \bibinfo {author} {\bibfnamefont {J.~R.}\ \bibnamefont {McClean}}, \bibinfo {author} {\bibfnamefont {N.}~\bibnamefont {Wiebe}},\ and\ \bibinfo {author} {\bibfnamefont {R.}~\bibnamefont {Babbush}},\ }\bibfield  {title} {\bibinfo {title} {Even more efficient quantum computations of chemistry through tensor hypercontraction},\ }\href {https://doi.org/10.1103/PRXQuantum.2.030305} {\bibfield  {journal} {\bibinfo  {journal} {PRX Quantum}\ }\textbf {\bibinfo {volume} {2}},\ \bibinfo {pages} {030305} (\bibinfo {year} {2021})}\BibitemShut {NoStop}%
	\bibitem [{\citenamefont {Labib}\ \emph {et~al.}(2024)\citenamefont {Labib}, \citenamefont {Clader}, \citenamefont {Stamatopoulos},\ and\ \citenamefont {Zeng}}]{labib2024quantum}%
	\BibitemOpen
	\bibfield  {author} {\bibinfo {author} {\bibfnamefont {F.}~\bibnamefont {Labib}}, \bibinfo {author} {\bibfnamefont {B.~D.}\ \bibnamefont {Clader}}, \bibinfo {author} {\bibfnamefont {N.}~\bibnamefont {Stamatopoulos}},\ and\ \bibinfo {author} {\bibfnamefont {W.~J.}\ \bibnamefont {Zeng}},\ }\bibfield  {title} {\bibinfo {title} {Quantum amplitude estimation from classical signal processing},\ }\href@noop {} {\bibfield  {journal} {\bibinfo  {journal} {arXiv preprint arXiv:2405.14697}\ } (\bibinfo {year} {2024})}\BibitemShut {NoStop}%
	\bibitem [{\citenamefont {Gilyén}\ \emph {et~al.}(2019)\citenamefont {Gilyén}, \citenamefont {Su}, \citenamefont {Low},\ and\ \citenamefont {Wiebe}}]{QSVT}%
	\BibitemOpen
	\bibfield  {author} {\bibinfo {author} {\bibfnamefont {A.}~\bibnamefont {Gilyén}}, \bibinfo {author} {\bibfnamefont {Y.}~\bibnamefont {Su}}, \bibinfo {author} {\bibfnamefont {G.~H.}\ \bibnamefont {Low}},\ and\ \bibinfo {author} {\bibfnamefont {N.}~\bibnamefont {Wiebe}},\ }\bibfield  {title} {\bibinfo {title} {Quantum singular value transformation and beyond: exponential improvements for quantum matrix arithmetics},\ }in\ \href {https://doi.org/10.1145/3313276.3316366} {\emph {\bibinfo {booktitle} {Proceedings of the 51st Annual ACM SIGACT Symposium on Theory of Computing}}},\ \bibinfo {series and number} {STOC ’19}\ (\bibinfo  {publisher} {ACM},\ \bibinfo {year} {2019})\ p.\ \bibinfo {pages} {193–204}\BibitemShut {NoStop}%
	\bibitem [{\citenamefont {Luitz}\ \emph {et~al.}(2015)\citenamefont {Luitz}, \citenamefont {Laflorencie},\ and\ \citenamefont {Alet}}]{PhysRevB.91.081103}%
	\BibitemOpen
	\bibfield  {author} {\bibinfo {author} {\bibfnamefont {D.~J.}\ \bibnamefont {Luitz}}, \bibinfo {author} {\bibfnamefont {N.}~\bibnamefont {Laflorencie}},\ and\ \bibinfo {author} {\bibfnamefont {F.}~\bibnamefont {Alet}},\ }\bibfield  {title} {\bibinfo {title} {Many-body localization edge in the random-field heisenberg chain},\ }\href {https://doi.org/10.1103/PhysRevB.91.081103} {\bibfield  {journal} {\bibinfo  {journal} {Phys. Rev. B}\ }\textbf {\bibinfo {volume} {91}},\ \bibinfo {pages} {081103} (\bibinfo {year} {2015})}\BibitemShut {NoStop}%
	\bibitem [{\citenamefont {Childs}\ \emph {et~al.}(2018)\citenamefont {Childs}, \citenamefont {Maslov}, \citenamefont {Nam}, \citenamefont {Ross},\ and\ \citenamefont {Su}}]{Childs_2018}%
	\BibitemOpen
	\bibfield  {author} {\bibinfo {author} {\bibfnamefont {A.~M.}\ \bibnamefont {Childs}}, \bibinfo {author} {\bibfnamefont {D.}~\bibnamefont {Maslov}}, \bibinfo {author} {\bibfnamefont {Y.}~\bibnamefont {Nam}}, \bibinfo {author} {\bibfnamefont {N.~J.}\ \bibnamefont {Ross}},\ and\ \bibinfo {author} {\bibfnamefont {Y.}~\bibnamefont {Su}},\ }\bibfield  {title} {\bibinfo {title} {Toward the first quantum simulation with quantum speedup},\ }\href {https://doi.org/10.1073/pnas.1801723115} {\bibfield  {journal} {\bibinfo  {journal} {Proceedings of the National Academy of Sciences}\ }\textbf {\bibinfo {volume} {115}},\ \bibinfo {pages} {9456–9461} (\bibinfo {year} {2018})}\BibitemShut {NoStop}%
	\bibitem [{\citenamefont {Cai}\ \emph {et~al.}(2023)\citenamefont {Cai}, \citenamefont {Babbush}, \citenamefont {Benjamin}, \citenamefont {Endo}, \citenamefont {Huggins}, \citenamefont {Li}, \citenamefont {McClean},\ and\ \citenamefont {O'Brien}}]{RevModPhys.95.045005}%
	\BibitemOpen
	\bibfield  {author} {\bibinfo {author} {\bibfnamefont {Z.}~\bibnamefont {Cai}}, \bibinfo {author} {\bibfnamefont {R.}~\bibnamefont {Babbush}}, \bibinfo {author} {\bibfnamefont {S.~C.}\ \bibnamefont {Benjamin}}, \bibinfo {author} {\bibfnamefont {S.}~\bibnamefont {Endo}}, \bibinfo {author} {\bibfnamefont {W.~J.}\ \bibnamefont {Huggins}}, \bibinfo {author} {\bibfnamefont {Y.}~\bibnamefont {Li}}, \bibinfo {author} {\bibfnamefont {J.~R.}\ \bibnamefont {McClean}},\ and\ \bibinfo {author} {\bibfnamefont {T.~E.}\ \bibnamefont {O'Brien}},\ }\bibfield  {title} {\bibinfo {title} {Quantum error mitigation},\ }\href {https://doi.org/10.1103/RevModPhys.95.045005} {\bibfield  {journal} {\bibinfo  {journal} {Rev. Mod. Phys.}\ }\textbf {\bibinfo {volume} {95}},\ \bibinfo {pages} {045005} (\bibinfo {year} {2023})}\BibitemShut {NoStop}%
	\bibitem [{\citenamefont {Dalzell}\ \emph {et~al.}(2024)\citenamefont {Dalzell}, \citenamefont {Hunter-Jones},\ and\ \citenamefont {Brand{\~a}o}}]{dalzell2024random}%
	\BibitemOpen
	\bibfield  {author} {\bibinfo {author} {\bibfnamefont {A.~M.}\ \bibnamefont {Dalzell}}, \bibinfo {author} {\bibfnamefont {N.}~\bibnamefont {Hunter-Jones}},\ and\ \bibinfo {author} {\bibfnamefont {F.~G.}\ \bibnamefont {Brand{\~a}o}},\ }\bibfield  {title} {\bibinfo {title} {Random quantum circuits transform local noise into global white noise},\ }\href@noop {} {\bibfield  {journal} {\bibinfo  {journal} {Communications in Mathematical Physics}\ }\textbf {\bibinfo {volume} {405}},\ \bibinfo {pages} {78} (\bibinfo {year} {2024})}\BibitemShut {NoStop}%
	\bibitem [{\citenamefont {Foldager}\ and\ \citenamefont {Koczor}(2023)}]{white_noise}%
	\BibitemOpen
	\bibfield  {author} {\bibinfo {author} {\bibfnamefont {J.}~\bibnamefont {Foldager}}\ and\ \bibinfo {author} {\bibfnamefont {B.}~\bibnamefont {Koczor}},\ }\bibfield  {title} {\bibinfo {title} {Can shallow quantum circuits scramble local noise into global white noise?},\ }\href {https://doi.org/10.1088/1751-8121/ad0ac7} {\bibfield  {journal} {\bibinfo  {journal} {Journal of Physics A: Mathematical and Theoretical}\ }\textbf {\bibinfo {volume} {57}},\ \bibinfo {pages} {015306} (\bibinfo {year} {2023})}\BibitemShut {NoStop}%
	\bibitem [{\citenamefont {Ross}\ and\ \citenamefont {Selinger}(2016)}]{synthesis0}%
	\BibitemOpen
	\bibfield  {author} {\bibinfo {author} {\bibfnamefont {N.~J.}\ \bibnamefont {Ross}}\ and\ \bibinfo {author} {\bibfnamefont {P.}~\bibnamefont {Selinger}},\ }\bibfield  {title} {\bibinfo {title} {Optimal ancilla-free clifford+t approximation of z-rotations},\ }\href {https://doi.org/10.26421/qic16.11-12-1} {\bibfield  {journal} {\bibinfo  {journal} {Quantum Information and Computation}\ }\textbf {\bibinfo {volume} {16}},\ \bibinfo {pages} {901–953} (\bibinfo {year} {2016})}\BibitemShut {NoStop}%
	\bibitem [{\citenamefont {Bocharov}\ \emph {et~al.}(2015{\natexlab{a}})\citenamefont {Bocharov}, \citenamefont {Roetteler},\ and\ \citenamefont {Svore}}]{synthesis2}%
	\BibitemOpen
	\bibfield  {author} {\bibinfo {author} {\bibfnamefont {A.}~\bibnamefont {Bocharov}}, \bibinfo {author} {\bibfnamefont {M.}~\bibnamefont {Roetteler}},\ and\ \bibinfo {author} {\bibfnamefont {K.~M.}\ \bibnamefont {Svore}},\ }\bibfield  {title} {\bibinfo {title} {Efficient synthesis of probabilistic quantum circuits with fallback},\ }\href {https://doi.org/10.1103/PhysRevA.91.052317} {\bibfield  {journal} {\bibinfo  {journal} {Phys. Rev. A}\ }\textbf {\bibinfo {volume} {91}},\ \bibinfo {pages} {052317} (\bibinfo {year} {2015}{\natexlab{a}})}\BibitemShut {NoStop}%
	\bibitem [{\citenamefont {Bocharov}\ \emph {et~al.}(2015{\natexlab{b}})\citenamefont {Bocharov}, \citenamefont {Roetteler},\ and\ \citenamefont {Svore}}]{synthesis}%
	\BibitemOpen
	\bibfield  {author} {\bibinfo {author} {\bibfnamefont {A.}~\bibnamefont {Bocharov}}, \bibinfo {author} {\bibfnamefont {M.}~\bibnamefont {Roetteler}},\ and\ \bibinfo {author} {\bibfnamefont {K.~M.}\ \bibnamefont {Svore}},\ }\bibfield  {title} {\bibinfo {title} {Efficient synthesis of universal repeat-until-success quantum circuits},\ }\href {https://doi.org/10.1103/PhysRevLett.114.080502} {\bibfield  {journal} {\bibinfo  {journal} {Phys. Rev. Lett.}\ }\textbf {\bibinfo {volume} {114}},\ \bibinfo {pages} {080502} (\bibinfo {year} {2015}{\natexlab{b}})}\BibitemShut {NoStop}%
	\bibitem [{\citenamefont {Kliuchnikov}\ \emph {et~al.}(2023)\citenamefont {Kliuchnikov}, \citenamefont {Lauter}, \citenamefont {Minko}, \citenamefont {Paetznick},\ and\ \citenamefont {Petit}}]{synthesis_summary}%
	\BibitemOpen
	\bibfield  {author} {\bibinfo {author} {\bibfnamefont {V.}~\bibnamefont {Kliuchnikov}}, \bibinfo {author} {\bibfnamefont {K.}~\bibnamefont {Lauter}}, \bibinfo {author} {\bibfnamefont {R.}~\bibnamefont {Minko}}, \bibinfo {author} {\bibfnamefont {A.}~\bibnamefont {Paetznick}},\ and\ \bibinfo {author} {\bibfnamefont {C.}~\bibnamefont {Petit}},\ }\bibfield  {title} {\bibinfo {title} {Shorter quantum circuits via single-qubit gate approximation},\ }\href {https://doi.org/10.22331/q-2023-12-18-1208} {\bibfield  {journal} {\bibinfo  {journal} {{Quantum}}\ }\textbf {\bibinfo {volume} {7}},\ \bibinfo {pages} {1208} (\bibinfo {year} {2023})}\BibitemShut {NoStop}%
	\bibitem [{\citenamefont {Landahl}\ and\ \citenamefont {Cesare}(2013)}]{reed}%
	\BibitemOpen
	\bibfield  {author} {\bibinfo {author} {\bibfnamefont {A.~J.}\ \bibnamefont {Landahl}}\ and\ \bibinfo {author} {\bibfnamefont {C.}~\bibnamefont {Cesare}},\ }\bibfield  {title} {\bibinfo {title} {Complex instruction set computing architecture for performing accurate quantum {$Z$} rotations with less magic},\ }\href@noop {} {\bibfield  {journal} {\bibinfo  {journal} {arXiv preprint arXiv:1302.3240}\ } (\bibinfo {year} {2013})}\BibitemShut {NoStop}%
	\bibitem [{\citenamefont {Gidney}(2018)}]{hamming-Gidney}%
	\BibitemOpen
	\bibfield  {author} {\bibinfo {author} {\bibfnamefont {C.}~\bibnamefont {Gidney}},\ }\bibfield  {title} {\bibinfo {title} {Halving the cost of quantum addition},\ }\href {https://doi.org/10.22331/q-2018-06-18-74} {\bibfield  {journal} {\bibinfo  {journal} {Quantum}\ }\textbf {\bibinfo {volume} {2}},\ \bibinfo {pages} {74} (\bibinfo {year} {2018})}\BibitemShut {NoStop}%
	\bibitem [{\citenamefont {Kivlichan}\ \emph {et~al.}(2020)\citenamefont {Kivlichan}, \citenamefont {Gidney}, \citenamefont {Berry}, \citenamefont {Wiebe}, \citenamefont {McClean}, \citenamefont {Sun}, \citenamefont {Jiang}, \citenamefont {Rubin}, \citenamefont {Fowler}, \citenamefont {Aspuru-Guzik}, \citenamefont {Neven},\ and\ \citenamefont {Babbush}}]{hamming}%
	\BibitemOpen
	\bibfield  {author} {\bibinfo {author} {\bibfnamefont {I.~D.}\ \bibnamefont {Kivlichan}}, \bibinfo {author} {\bibfnamefont {C.}~\bibnamefont {Gidney}}, \bibinfo {author} {\bibfnamefont {D.~W.}\ \bibnamefont {Berry}}, \bibinfo {author} {\bibfnamefont {N.}~\bibnamefont {Wiebe}}, \bibinfo {author} {\bibfnamefont {J.}~\bibnamefont {McClean}}, \bibinfo {author} {\bibfnamefont {W.}~\bibnamefont {Sun}}, \bibinfo {author} {\bibfnamefont {Z.}~\bibnamefont {Jiang}}, \bibinfo {author} {\bibfnamefont {N.}~\bibnamefont {Rubin}}, \bibinfo {author} {\bibfnamefont {A.}~\bibnamefont {Fowler}}, \bibinfo {author} {\bibfnamefont {A.}~\bibnamefont {Aspuru-Guzik}}, \bibinfo {author} {\bibfnamefont {H.}~\bibnamefont {Neven}},\ and\ \bibinfo {author} {\bibfnamefont {R.}~\bibnamefont {Babbush}},\ }\bibfield  {title} {\bibinfo {title} {Improved fault-tolerant quantum simulation of condensed-phase correlated electrons via {T}rotterization},\ }\href {https://doi.org/10.22331/q-2020-07-16-296} {\bibfield  {journal} {\bibinfo  {journal}
			{{Quantum}}\ }\textbf {\bibinfo {volume} {4}},\ \bibinfo {pages} {296} (\bibinfo {year} {2020})}\BibitemShut {NoStop}%
	\bibitem [{\citenamefont {Sun}\ \emph {et~al.}(2024)\citenamefont {Sun}, \citenamefont {Boyd}, \citenamefont {Cai}, \citenamefont {Jnane}, \citenamefont {Koczor}, \citenamefont {Meister}, \citenamefont {Minko}, \citenamefont {Pring}, \citenamefont {Benjamin},\ and\ \citenamefont {Stamatopoulos}}]{sun2024low}%
	\BibitemOpen
	\bibfield  {author} {\bibinfo {author} {\bibfnamefont {Z.}~\bibnamefont {Sun}}, \bibinfo {author} {\bibfnamefont {G.}~\bibnamefont {Boyd}}, \bibinfo {author} {\bibfnamefont {Z.}~\bibnamefont {Cai}}, \bibinfo {author} {\bibfnamefont {H.}~\bibnamefont {Jnane}}, \bibinfo {author} {\bibfnamefont {B.}~\bibnamefont {Koczor}}, \bibinfo {author} {\bibfnamefont {R.}~\bibnamefont {Meister}}, \bibinfo {author} {\bibfnamefont {R.}~\bibnamefont {Minko}}, \bibinfo {author} {\bibfnamefont {B.}~\bibnamefont {Pring}}, \bibinfo {author} {\bibfnamefont {S.~C.}\ \bibnamefont {Benjamin}},\ and\ \bibinfo {author} {\bibfnamefont {N.}~\bibnamefont {Stamatopoulos}},\ }\bibfield  {title} {\bibinfo {title} {Low depth phase oracle using a parallel piecewise circuit},\ }\href@noop {} {\bibfield  {journal} {\bibinfo  {journal} {arXiv preprint arXiv:2409.04587}\ } (\bibinfo {year} {2024})}\BibitemShut {NoStop}%
	\bibitem [{\citenamefont {Gidney}\ and\ \citenamefont {Fowler}(2019)}]{gidney2019efficient}%
	\BibitemOpen
	\bibfield  {author} {\bibinfo {author} {\bibfnamefont {C.}~\bibnamefont {Gidney}}\ and\ \bibinfo {author} {\bibfnamefont {A.~G.}\ \bibnamefont {Fowler}},\ }\bibfield  {title} {\bibinfo {title} {Efficient magic state factories with a catalyzed {CCZ} to {2T} transformation},\ }\href {https://doi.org/https://doi.org/10.22331/q-2019-04-30-135} {\bibfield  {journal} {\bibinfo  {journal} {Quantum}\ }\textbf {\bibinfo {volume} {3}},\ \bibinfo {pages} {135} (\bibinfo {year} {2019})}\BibitemShut {NoStop}%
	\bibitem [{\citenamefont {Harrow}\ and\ \citenamefont {Lowe}(2025)}]{circuit_cutting}%
	\BibitemOpen
	\bibfield  {author} {\bibinfo {author} {\bibfnamefont {A.~W.}\ \bibnamefont {Harrow}}\ and\ \bibinfo {author} {\bibfnamefont {A.}~\bibnamefont {Lowe}},\ }\bibfield  {title} {\bibinfo {title} {Optimal quantum circuit cuts with application to clustered hamiltonian simulation},\ }\href {https://doi.org/10.1103/PRXQuantum.6.010316} {\bibfield  {journal} {\bibinfo  {journal} {PRX Quantum}\ }\textbf {\bibinfo {volume} {6}},\ \bibinfo {pages} {010316} (\bibinfo {year} {2025})}\BibitemShut {NoStop}%
	\bibitem [{\citenamefont {Wan}\ \emph {et~al.}(2022)\citenamefont {Wan}, \citenamefont {Berta},\ and\ \citenamefont {Campbell}}]{statistical-phase-estimation}%
	\BibitemOpen
	\bibfield  {author} {\bibinfo {author} {\bibfnamefont {K.}~\bibnamefont {Wan}}, \bibinfo {author} {\bibfnamefont {M.}~\bibnamefont {Berta}},\ and\ \bibinfo {author} {\bibfnamefont {E.~T.}\ \bibnamefont {Campbell}},\ }\bibfield  {title} {\bibinfo {title} {Randomized quantum algorithm for statistical phase estimation},\ }\href {https://doi.org/10.1103/PhysRevLett.129.030503} {\bibfield  {journal} {\bibinfo  {journal} {Phys. Rev. Lett.}\ }\textbf {\bibinfo {volume} {129}},\ \bibinfo {pages} {030503} (\bibinfo {year} {2022})}\BibitemShut {NoStop}%
	\bibitem [{\citenamefont {Boyd}\ \emph {et~al.}(2024)\citenamefont {Boyd}, \citenamefont {Koczor},\ and\ \citenamefont {Cai}}]{boyd2024high}%
	\BibitemOpen
	\bibfield  {author} {\bibinfo {author} {\bibfnamefont {G.}~\bibnamefont {Boyd}}, \bibinfo {author} {\bibfnamefont {B.}~\bibnamefont {Koczor}},\ and\ \bibinfo {author} {\bibfnamefont {Z.}~\bibnamefont {Cai}},\ }\bibfield  {title} {\bibinfo {title} {High-dimensional subspace expansion using classical shadows},\ }\href@noop {} {\bibfield  {journal} {\bibinfo  {journal} {arXiv preprint arXiv:2406.11533}\ } (\bibinfo {year} {2024})}\BibitemShut {NoStop}%
	\bibitem [{\citenamefont {Jones}\ \emph {et~al.}(2019)\citenamefont {Jones}, \citenamefont {Brown}, \citenamefont {Bush},\ and\ \citenamefont {Benjamin}}]{quest}%
	\BibitemOpen
	\bibfield  {author} {\bibinfo {author} {\bibfnamefont {T.}~\bibnamefont {Jones}}, \bibinfo {author} {\bibfnamefont {A.}~\bibnamefont {Brown}}, \bibinfo {author} {\bibfnamefont {I.}~\bibnamefont {Bush}},\ and\ \bibinfo {author} {\bibfnamefont {S.~C.}\ \bibnamefont {Benjamin}},\ }\bibfield  {title} {\bibinfo {title} {Quest and high performance simulation of quantum computers},\ }\href {https://doi.org/10.1038/s41598-019-47174-9} {\bibfield  {journal} {\bibinfo  {journal} {Scientific Reports}\ }\textbf {\bibinfo {volume} {9}},\ \bibinfo {pages} {10736} (\bibinfo {year} {2019})}\BibitemShut {NoStop}%
	\bibitem [{\citenamefont {Jones}\ and\ \citenamefont {Benjamin}(2020)}]{questlink}%
	\BibitemOpen
	\bibfield  {author} {\bibinfo {author} {\bibfnamefont {T.}~\bibnamefont {Jones}}\ and\ \bibinfo {author} {\bibfnamefont {S.}~\bibnamefont {Benjamin}},\ }\bibfield  {title} {\bibinfo {title} {Questlink—mathematica embiggened by a hardware-optimised quantum emulator},\ }\href {https://doi.org/10.1088/2058-9565/ab8506} {\bibfield  {journal} {\bibinfo  {journal} {Quantum Science and Technology}\ }\textbf {\bibinfo {volume} {5}},\ \bibinfo {pages} {034012} (\bibinfo {year} {2020})}\BibitemShut {NoStop}%
	\bibitem [{\citenamefont {Berry}\ \emph {et~al.}(2020)\citenamefont {Berry}, \citenamefont {Childs}, \citenamefont {Su}, \citenamefont {Wang},\ and\ \citenamefont {Wiebe}}]{time-dependent_l1}%
	\BibitemOpen
	\bibfield  {author} {\bibinfo {author} {\bibfnamefont {D.~W.}\ \bibnamefont {Berry}}, \bibinfo {author} {\bibfnamefont {A.~M.}\ \bibnamefont {Childs}}, \bibinfo {author} {\bibfnamefont {Y.}~\bibnamefont {Su}}, \bibinfo {author} {\bibfnamefont {X.}~\bibnamefont {Wang}},\ and\ \bibinfo {author} {\bibfnamefont {N.}~\bibnamefont {Wiebe}},\ }\bibfield  {title} {\bibinfo {title} {Time-dependent {H}amiltonian simulation with {$L^1$}-norm scaling},\ }\href {https://doi.org/10.22331/q-2020-04-20-254} {\bibfield  {journal} {\bibinfo  {journal} {{Quantum}}\ }\textbf {\bibinfo {volume} {4}},\ \bibinfo {pages} {254} (\bibinfo {year} {2020})}\BibitemShut {NoStop}%
	\bibitem [{\citenamefont {Cao}\ \emph {et~al.}(2024)\citenamefont {Cao}, \citenamefont {Jin},\ and\ \citenamefont {Liu}}]{unified_time-dependent}%
	\BibitemOpen
	\bibfield  {author} {\bibinfo {author} {\bibfnamefont {Y.}~\bibnamefont {Cao}}, \bibinfo {author} {\bibfnamefont {S.}~\bibnamefont {Jin}},\ and\ \bibinfo {author} {\bibfnamefont {N.}~\bibnamefont {Liu}},\ }\bibfield  {title} {\bibinfo {title} {A unifying framework for quantum simulation algorithms for time-dependent hamiltonian dynamics},\ }\bibfield  {journal} {\bibinfo  {journal} {arXiv preprint arXiv:2411.03180}\ }\href {https://doi.org/10.48550/ARXIV.2411.03180} {10.48550/ARXIV.2411.03180} (\bibinfo {year} {2024})\BibitemShut {NoStop}%
	\bibitem [{\citenamefont {Kieferov\'a}\ \emph {et~al.}(2019)\citenamefont {Kieferov\'a}, \citenamefont {Scherer},\ and\ \citenamefont {Berry}}]{dyson-series}%
	\BibitemOpen
	\bibfield  {author} {\bibinfo {author} {\bibfnamefont {M.}~\bibnamefont {Kieferov\'a}}, \bibinfo {author} {\bibfnamefont {A.}~\bibnamefont {Scherer}},\ and\ \bibinfo {author} {\bibfnamefont {D.~W.}\ \bibnamefont {Berry}},\ }\bibfield  {title} {\bibinfo {title} {Simulating the dynamics of time-dependent hamiltonians with a truncated dyson series},\ }\href {https://doi.org/10.1103/PhysRevA.99.042314} {\bibfield  {journal} {\bibinfo  {journal} {Phys. Rev. A}\ }\textbf {\bibinfo {volume} {99}},\ \bibinfo {pages} {042314} (\bibinfo {year} {2019})}\BibitemShut {NoStop}%
	\bibitem [{\citenamefont {Owens}(2014)}]{owens2014exploring}%
	\BibitemOpen
	\bibfield  {author} {\bibinfo {author} {\bibfnamefont {L.}~\bibnamefont {Owens}},\ }\bibfield  {title} {\bibinfo {title} {Exploring the rate of convergence of approximations to the riemann integral},\ }\href@noop {} {\bibfield  {journal} {\bibinfo  {journal} {Preprint}\ } (\bibinfo {year} {2014})}\BibitemShut {NoStop}%
	\bibitem [{\citenamefont {Tasaki}(2009)}]{tasaki2009convergence}%
	\BibitemOpen
	\bibfield  {author} {\bibinfo {author} {\bibfnamefont {H.}~\bibnamefont {Tasaki}},\ }\bibfield  {title} {\bibinfo {title} {Convergence rates of approximate sums of riemann integrals},\ }\href@noop {} {\bibfield  {journal} {\bibinfo  {journal} {Journal of approximation theory}\ }\textbf {\bibinfo {volume} {161}},\ \bibinfo {pages} {477} (\bibinfo {year} {2009})}\BibitemShut {NoStop}%
\end{thebibliography}
%

\end{document}